\documentclass[superscriptaddress,prx,twocolumn,nofootinbib,citeautoscript,longbibliography,notitlepage]{revtex4-1}
\usepackage{graphicx}% Include figure files
\usepackage{dcolumn}% Align table columns on decimal point
\usepackage{bm}% bold math
\usepackage{color}
\usepackage{amsmath}
\usepackage{tabularx,graphicx}
\usepackage{epstopdf}
\usepackage{latexsym}
\usepackage{amssymb}
\usepackage{amsmath}
\usepackage{color, colortbl}
\usepackage{psfrag}
\usepackage{bbm}
\usepackage{booktabs}
\usepackage{bm}
\usepackage{titlesec}
\usepackage{dsfont}
\usepackage{feynmp}
\usepackage{slashed}
\usepackage{multirow}
\usepackage[tight]{subfigure}
\usepackage{comment}
\usepackage{float}
\usepackage{diagbox} %²åÈë±í¸ñºê°ü
\usepackage{makecell}
\usepackage{array}
\usepackage[papersize={8.5in,11in}]{geometry}
\usepackage{color}
\usepackage{fmtcount} %ÂÞÂí×Öĸ°ü
\definecolor{darkblue}{rgb}{0.,0.,0.4}
\definecolor{darkred}{rgb}{0.5,0.,0.}
\definecolor{BlueViolet}{RGB}{138,43,226}
\definecolor{SkyBlue}{RGB}{30,144,255}
\definecolor{DarkGreen}{RGB}{0,100,0}

\usepackage[pdftex,colorlinks=true,linkcolor=darkblue,citecolor=blue,urlcolor=darkred]{hyperref}

\renewcommand{\epsilon}{\varepsilon}

\allowdisplaybreaks[3]

\usepackage{cases}

\allowdisplaybreaks[4]

\UseRawInputEncoding

\begin{document}

\title{Interaction-driven quantum criticality in two-dimensional quadratic band crossing
semimetals with time-reversal symmetry breaking}

\author{Yi-Kun Fang}
\affiliation{Department of Physics, Tianjin University, Tianjin 300072, P.R. China}

\author{Jing Wang}
\altaffiliation{Corresponding author: jing$\textunderscore$wang@tju.edu.cn}
\affiliation{Department of Physics, Tianjin University, Tianjin 300072, P.R. China}
\affiliation{Tianjin Key Laboratory of Low Dimensional Materials Physics and
Preparing Technology, Tianjin University, Tianjin 300072, P.R. China}

\date{\today}

%%%%%%%%%%%%%%%%%%%%%%%%%%%%%%%%%%%%%%%%%%%%%%%%%%%%%%%%%%%%%%%%

\begin{abstract}

We present a systematic investigation of all 16 marginally relevant fermion-fermion interactions
in two-dimensional (2D) time-reversal symmetry-breaking kagom\'{e} semimetals hosting a quadratic band crossing point (QBCP).
Employing a momentum-shell renormalization group approach that treats every interaction on an equal footing, we
derive energy-dependent flow equations that capture the hierarchical evolutions of interaction parameters.
%%Deciphering these flows reveals several intriguing results.
Our analysis begins by tracking the energy-dependent flows of fermion-fermion interactions.
The interaction couplings evolve towards divergence at a critical energy scale, signaling quantum critical behavior
governed by a certain fixed point (FP). The character of this FP depends intimately on structural
parameters $d_{0,1,2,3}$ which classify the microscopic model into rotationally symmetric and asymmetric cases.
Then, we identify two stable FPs in the rotationally symmetric and nine additional FPs in the asymmetric case
dubbed FP$_{1-10}$. Their boundary conditions are approximately demarcated and established by linear and plane-fitting
techniques in the structural parameter space. Furthermore, we examine distinct interaction-driven instabilities near
these FPs by incorporating the relevant external source terms and computing their susceptibilities.
It indicates that the charge density wave and superconductivity become dominant at FP$_{2,4,5,6,8}$ and FP$_{1,9,10}$,
while the $x$-current and bond density prevail at FP$_3$ and FP$_7$, respectively.
In addition to these leading states, several underlying subordinate instabilities are presented as well.
These results would be helpful to further study the low-energy critical behavior in a 2D kagom\'{e} QBCP
and related materials.

\end{abstract}

\maketitle

\section{Introduction}

Semimetals constitute a cornerstone of contemporary condensed matter physics,
exhibiting unique electronic properties that bridge conventional metals and insulators~\cite{Lee2005Nature,
Neto2009RMP,Fu2007PRL,Roy2009PRB,Moore2010Nature,Hasan2010RMP,Qi2011RMP,
Sheng2012book,Bernevig2013book,Herbut2018Science,Armitage2018RMP,Roy2018PRX}.
Over the past two decades, Dirac and Weyl semimetals that are characterized by discrete
band-touching points with linear dispersion have been extensively investigated both theoretically and experimentally~\cite{Wang2012PRB,Young2012PRL,Steinberg2014PRL,Liu2014NM,Liu2014Science,
Xiong2015Science,Neto2009RMP,Burkov2011PRL,Yang2011PRB,
Wan2011PRB,Huang2015PRX,Xu2015Science,Xu2015NP,Lv2015NP,Weng2015PRX,
Lee2005Nature,Fu2007PRL,Roy2009PRB,Moore2010Nature,Hasan2010RMP,
Qi2011RMP,Korshunov2014PRB,Hung2016PRB,Nandkishore2013PRB,Potirniche2014PRB,
Nandkishore2017PRB,Sarma2016PRB,Herbut2018Science}. More recently, research focus has shifted
toward the quadratic-dispersion materials~\cite{Chong2008PRB,Fradkin2008PRB,
Fradkin2009PRL,Vafek2012PRB,Vafek2014PRB,Herbut2012PRB,Mandal2019CMP,Zhu2016PRL,
Vafek2010PRB,Yang2010PRB,Wang2017PRB,Wang2020-arxiv,Roy2020-arxiv,Janssen2020PRB,Shah2011.00249,
Luttinger1956PR,Murakami2004PRB,Janssen2015PRB,Boettcher2016PRB,Janssen2017PRB,Boettcher2017PRB,
Mandal2018PRB,Lin2018PRB,Savary2014PRX,Savary2017PRB,Vojta1810.07695,Lai2014arXiv,Goswami2017PRB,
Szabo2018arXiv,Foster2019PRB,Wang1911.09654,Wang2303.10163}, where the conduction and valence
bands touch parabolically in momentum space, forming a quadratic band crossing point (QBCP).
These distinctive properties emerge in geometrically distinct lattice systems, including kagom\'{e}~\cite{Huse2003PRB,Fradkin2009PRL,Janssen2020PRB}, checkerboard
~\cite{Fradkin2008PRB,Vafek2014PRB}, Lieb~\cite{Tsai2015NJP}, and collinear spin-density-wave
lattices~\cite{Chern2012PRL} under various point-group symmetries~\cite{Fradkin2009PRL,Vafek2012PRB,Vafek2014PRB}.
Crucially, unlike Dirac/Weyl counterparts where the density of states (DOS) vanishes at nodal points,
two-dimensional QBCP systems possess a finite DOS at the Fermi level (i.e., the QBCP)
and gapless quasiparticles with parabolic dispersion~\cite{Chong2008PRB,Fradkin2008PRB,
Fradkin2009PRL,Vafek2012PRB,Vafek2014PRB,Herbut2012PRB,Mandal2019CMP,Zhu2016PRL,
Vafek2010PRB,Yang2010PRB,Wang2017PRB,Wang2020-arxiv,Roy2020-arxiv,Janssen2020PRB,Shah2011.00249}.
This finite DOS enables pronounced interaction effects and accordingly renders them ideal platforms for
studying weak coupling-driven quantum phase transitions~\cite{Fradkin2009PRL,Vafek2012PRB,Vafek2014PRB}.
Such 2D QBCPs therefore have become one of the most active frontiers in condensed-matter research.

It is worth highlighting that 2D QBCP systems display fundamentally distinct time-reversal
symmetry (TRS) properties that are dictated by their underlying lattice geometries. In the
checkerboard lattice, the free Hamiltonian protects the QBCP through crystalline symmetries
combined with TRS, enabling topologically nontrivial band structures~\cite{Fradkin2008PRB,Fradkin2009PRL}.
Due to unconventional dispersion relations, short-range fermion-fermion interactions generate
rich weak-coupling phenomena, including interaction-driven topological phase transitions~\cite{Fradkin2009PRL,Vafek2010PRB,Vafek2012PRB,Yang2010PRB,Vafek2014PRB,Venderbos2016PRB,Wu2016PRL,
Zhu2016PRL,Wang2017PRB}. In sharp contrast, the kagom\'{e}-lattice QBCP materials inherently break TRS due to
geometric frustration and complex hopping parameters~\cite{Huse2003PRB}. Despite sharing quadratic dispersion
characteristics, their distinct Hamiltonian structure with TRS breaking yields qualitatively different low-energy
physics compared to their TRS-preserving counterparts~\cite{Herbut2014PRB}. It therefore raises several interesting unexplored inquiries.
How do fermion-fermion interactions renormalize the low-energy properties of the 2D kagom\'{e} QBCP semimetals,
and what kinds of interaction-driven instabilities can emerge, and which quantum-critical phenomena appear
as the system approaches the potential quantum critical points.

To address these significant issues, we systematically investigate the impacts of all 16 marginally relevant fermion-fermion interactions on the low-energy physics of 2D kagom\'{e} QBCP systems hosting a QBCP~\cite{Fradkin2009PRL,Vafek2010PRB,Vafek2012PRB,Yang2010PRB,Vafek2014PRB,Huse2003PRB}. To unbiasedly treat these interactions and their complex interplay, we employ the momentum-shell renormalization group (RG) method~\cite{Shankar1994RMP,Wilson1975RMP,Polchinski9210046}. This yields energy-dependent RG flow equations for the interaction parameters. Analyzing these flows reveals the fixed points governing the low-energy physics and identifies distinct interaction-driven instabilities as these points are approached.

At the outset, our RG analysis of fermion-fermion interactions reveals a coupling divergence at a critical low-energy scale that is associated with quantum criticality~\cite{Metzner2000PRL,Maiti2010PRB,Vafek2014PRB,Chubukov2016PRX,Halboth2000RPB}. In particular, this critical behavior can be characterized by certain fixed points that depend sensitively on the structural parameters of the microscopic model.  Specifically, there exist two distinct fixed points when the system preserves the rotational symmetry, whereas nine appear once the symmetry is broken. In addition, the boundary conditions for all these fixed points
are obtained by performing linear fitting. Table~\ref{table:fixpoint} summarizes
the primary results for the fixed points. Subsequently, we investigate the potential instabilities near these fixed points~\cite{Metzner2000PRL,Halboth2000RPB,Maiti2010PRB,Vafek2014PRB,Chubukov2016PRX,Roy2018PRX}.
After incorporating external source terms associated with candidate states listed in Table~\ref{table:xiangbian} and computing their
susceptibilities near the fixed points, we notice that the charge density wave (CDW) dominates at FP$_{2,4,5,6,8}$,
while chiral SC1 and SC2 prevail at FP$_{9,10}$, and the $x$-current and bond density become the leading states at FP$_3$ and
FP$_7$, respectively. Particularly, $s$-wave SC joins chiral SC2 as the leading instability at FP$_1$.
In addition to the leading states, we also present the subleading instabilities around these FPs.
Specifically, the chiral SC2 is subordinate to the leading CDW at FP$_{2,4,8}$,
and vice versa at FP$_{1,9,10}$. In particular, there exist triply degenerate states including chiral SC2,
bond density, and CDW around FP$_{3}$. The $x$-current and bond density states serve as alternative subleading state
at FP$_{4,5,6}$ and FP$_{4,9,10}$, respectively. As to FP$_7$, a subleading instability corresponds
to chiral SC1. This analysis establishes a theoretical framework for understanding the low-energy
critical behavior in 2D kagom\'{e} QBCP.

The rest of this paper is organized as follows. Section~\ref{Sec_model} introduces the
microscopic model and develops the effective theory for 2D kagom\'{e} QBCP semimetals.
Sec.~\ref{Sec_RG_analysis} is followed to perform the RG analysis to derive coupled flow equations
for all interaction parameters. In Sec.~\ref{Sec_fate_ff}, we systematically investigate the fates
of fermion-fermion interactions, categorizing three distinct cases based on structural parameter
features, and we present all relevant fixed points with their boundary conditions in parameter space.
Sec.~\ref{Sec_phase_transition} further analyzes symmetry breaking and phase transitions near these
fixed points, addressing both leading and subleading instabilities induced by fermion-fermion interactions.
At last, we conclude with the primary results in Sec.~\ref{Sec_summary}.

\section{Effective theory and RG analysis}\label{Sec_model_RG}

\subsection{Microscopic model and effective action}\label{Sec_model}

The microscopic noninteracting model for a two-dimensional (2D) quadratic band crossing point (QBCP) semimetal
with spin-1/2 electrons on a kagom\'{e} lattice is characterized by the Hamiltonian~\cite{Fradkin2008PRB,Fradkin2009PRL,Huse2003PRB}:
\begin{equation}
H_0 = \sum_{|\mathbf{k}| < \Lambda} \Psi^\dagger_{\mathbf{k}} \mathcal{H}_0(\mathbf{k}) \Psi_{\mathbf{k}},
\label{Eq_H0}
\end{equation}
where $\Lambda$ denotes the momentum cutoff associated with the lattice constant. The low-energy quasiparticles are described by the four-component spinor $\Psi_{\mathbf{k}}^T \equiv (c_{A\uparrow}, c_{A\downarrow}, c_{B\uparrow}, c_{B\downarrow})$, with $A$ and $B$ indexing the two effective sublattices of the kagom\'{e} structure.
In addition, the Hamiltonian density $\mathcal{H}_0$ is explicitly expressed as~\cite{Fradkin2008PRB,Fradkin2009PRL,Huse2003PRB},
\begin{equation}
\mathcal{H}_0 \!= \!d_0 \mathbf{k}^2 \Sigma_{00} \!+\! d_1 (k_x^2 \!-\! k_y^2) \Sigma_{01} \!+\!
d_2 k_x k_y \Sigma_{02} \!+\! d_3 \mathbf{k}^2 \Sigma_{03},
\label{Eq_H0_density}
\end{equation}
with $d_{0,1,2,3}$ being microscopic structure parameters that are material-dependent. Hereby, the $4 \times 4$ matrices $\Sigma_{\mu\nu} \equiv \tau_\mu \otimes \sigma_\nu$ act on the combined sublattice ($\tau$) and spin ($\sigma$) spaces, with $\mu,\nu = 0,1,2,3$ indexing the identity ($\tau_0,\sigma_0$) and Pauli matrices ($\tau_{1-3},\sigma_{1-3}$).

The energy dispersion relations can be obtained by diagonalizing the free Hamiltonian in Eq.~(\ref{Eq_H0_density}),
\begin{equation}
E_{\pm}(\mathbf{k}) = \mathbf{k}^2 \!\!\left( \!d_0 \pm \!\sqrt{d_1^2 \cos^2 2\theta + \tfrac{1}{4} d_2^2 \sin^2 2\theta + d_3^2} \!\right),
\label{Eq_E_pm}
\end{equation}
where $E_+$ ($E_-$) corresponds to the upper (lower) band, and $\theta \equiv \arctan(k_y/k_x)$ specifies the orientation of momentum $\mathbf{k}$. It is of particular importance to address several comments on this energy dispersion~(\ref{Eq_E_pm}).
At first, the rotational symmetry of the system requires the dispersion to be isotropic in $\theta$, which occurs when $d_2 = 2d_1$.
Under this condition, the expression under the radical simplifies to $\sqrt{d_1^2 + d_3^2}$, rendering the dispersion
angle-independent. Conversely, the deviations from $d_2 = 2d_1$ explicitly break the rotational symmetry.
Next, on the other hand, the parameter $d_0$ governs particle-hole symmetry. When $d_0$ vanishes, the system exhibits perfect particle-hole symmetry with $E_+(\mathbf{k}) = -E_-(\mathbf{k})$, yielding symmetric energy bands about $E=0$. However, a finite $d_0$ would break such symmetry, shifting the band degeneracy point away from the Fermi level.
Further, the band splitting magnitude is measured by $\Delta(\mathbf{k})\equiv E_+-E_-
=2\mathbf{k}^2 \sqrt{d_1^2 \cos^2 2\theta + \tfrac{1}{4} d_2^2 \sin^2 2\theta + d_3^2}$. A stable QBCP requires gapless excitations at $\mathbf{k} = 0$ with $E_+ > 0$ and $E_- < 0$ for $\mathbf{k} \neq 0$. This leads to the stability criterion of QBCP~\cite{Fradkin2008PRB,Fradkin2009PRL,Huse2003PRB},
\begin{eqnarray}
|d_{0}| < \sqrt{d_{1}^{2}\cos^{2}2\theta + \frac{1}{4}d_{2}^{2}\sin^{2}2\theta + d_{3}^{2}}.\label{Eq_condition_QBCP}
\end{eqnarray}
As a consequence, the conditions $d_2=2d_1$ and $d_0=0$ are directly related to the rotational symmetry
and particle-hole symmetry, while Eq.~(\ref{Eq_condition_QBCP}) serves as a stability criterion for the emergence of QBCP. Without loss of generality, we are going to investigate the low-energy behavior of a general situation including
both symmetric and asymmetric situations.

The influence of fermion-fermion interactions on low-energy physics in 2D quadratic band crossing systems has attracted significant theoretical attention, particularly for checkerboard lattices~\cite{Fradkin2009PRL,Vafek2010PRB,Vafek2014PRB,Wang2017PRB,Wang2018JPCM, Wang2019JPCM,Wang2020-arxiv}. However, despite possessing particle-hole and sixfold rotational symmetry~\cite{Fradkin2008PRB,Vafek2014PRB}, the time-reversal symmetry broken QBCP system described by
Eq.~(\ref{Eq_H0_density}) remains comparatively unexplored relative to its checkerboard counterpart.
It is therefore of particular importance to investigate how interactions modify low-energy
behavior in the 2D kagom\'{e} QBCP systems. To this end, we introduce all marginal short-range four-fermion interactions~\cite{Fradkin2009PRL,Vafek2014PRB,Vafek2010PRB,Yang2010PRB,Wang2020-arxiv,
Fu-Wang2024AP,Zhang-Wang2025PRB,Roy2009.05055,Roy2023PRR},
\begin{widetext}
\begin{eqnarray}
S_{\rm int} = \sum_{\mu,\nu=0}^3 \lambda_{\mu\nu} \prod_{i=1}^3 \int \frac{d\omega_i d^2\mathbf{k}_i}{(2\pi)^3}
\Psi^{\dagger}(\omega_1,\mathbf{k}_1) \Sigma_{\mu\nu} \Psi(\omega_2,\mathbf{k}_2) \Psi^{\dagger}(\omega_3,\mathbf{k}_3) \Sigma_{\mu\nu} \Psi(\omega_1+\omega_2-\omega_3,\mathbf{k}_1+\mathbf{k}_2-\mathbf{k}_3),\label{Eq_S_int}
\end{eqnarray}
\end{widetext}
where $\lambda_{\mu\nu}$ with $\mu, \nu = 0, 1, 2, 3$ are dimensionless coupling constants measuring the strength of 16 distinct interaction channels~\cite{Fradkin2009PRL,Vafek2014PRB,Vafek2010PRB,Yang2010PRB,Wang2020-arxiv}. These interactions are classified by their transformation properties under the vertex matrices $\Sigma_{\mu\nu} \equiv \tau_\mu \otimes \sigma_\nu$.
These interactions are marginal at tree level, but develop significant one-loop corrections that play a significant role in reshaping the low-energy physics. To proceed, the effective action in momentum space including both the noninteracting Hamiltonian (\ref{Eq_H0})
and the marginal four-fermion interactions~(\ref{Eq_S_int}) can be obtained~\cite{Fradkin2009PRL,Vafek2014PRB,Wang2017PRB},
\begin{eqnarray}
S_{\text{eff}}
&=&\!\!\!\int\!\frac{d\omega d^2 \mathbf{k}}{(2\pi)^3} \Psi^\dagger(\omega,\mathbf{k})
\!\!\left[ (-i\omega + d_0 \mathbf{k}^2) \Sigma_{00} + d_1 (\mathbf{k}_x^2 - \mathbf{k}_y^2) \Sigma_{01}\right.\nonumber\\
&&\left. + d_2 \mathbf{k}_x \mathbf{k}_y \Sigma_{02} + d_3 \mathbf{k}^2 \Sigma_{03} \right]\Psi(\omega,\mathbf{k})+ S_{\rm int},\label{Eq_S_eff}
\end{eqnarray}
where the first term presents the noninteracting dynamics and the second encapsulates all allowed interactions. Within this quantum field theory framework, the free fermion propagator can be derived as~\cite{Altland2006Book},
\begin{eqnarray}
G^{-1}_0(i\omega, \mathbf{k})
&=&(-i\omega+d_0\mathbf{k}^2)\Sigma_{00} + d_1 (k_x^2 - k_y^2) \Sigma_{01} \nonumber\\
&&+ d_2 k_x k_y \Sigma_{02}+ d_3 \mathbf{k}^2 \Sigma_{03},
\end{eqnarray}
which exhibits quadratic dispersion near the QBCP.
We hereafter consider the effective action~(\ref{Eq_S_eff}) as the starting point for RG analysis of the low-energy behavior
of spin-1/2 electrons on the kagom\'{e} lattice.

\begin{figure}[H]
\includegraphics[width=3in]{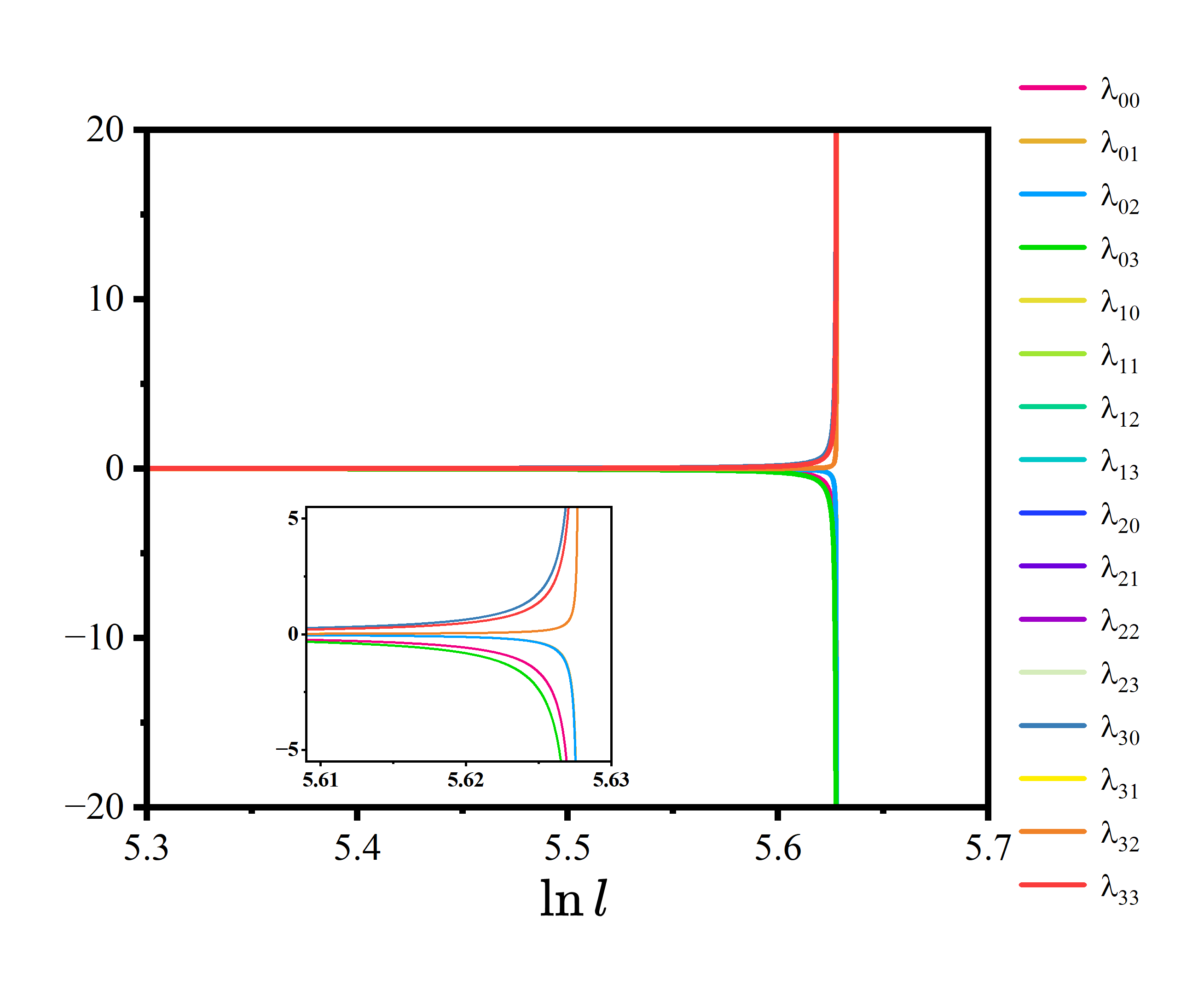}\\
\vspace{-1cm}
\caption{(Color online) Energy-dependent evolutions of fermion-fermion interactions at
fixed structural parameters $d_1 = 0.4$, $d_2 = 0.06$, and $d_3 = 0.7$
(the behavior is insensitive to choices of structural parameters).}
\label{Fig_evolution}
\end{figure}

\subsection{RG analysis and flow equations}\label{Sec_RG_analysis}

To derive the coupled flow equations governing the low-energy behavior of spin-1/2 electrons on kagom\'{e} lattices
with quadratic band crossings, we implement the momentum-shell renormalization group (RG) analysis
at one-loop order~\cite{Shankar1994RMP,Wilson1975RMP,Polchinski9210046}.

To perform the RG analysis, we need to separate the fermionic fields into
slow modes ($|\mathbf{k}| < b\Lambda$) and fast modes ($b\Lambda < |\mathbf{k}| < \Lambda$)~\cite{Shankar1994RMP},
where $\Lambda$ is employed to denote the energy scale and the variable parameter $b$
designated as $b=e^{-l}<1$ with $l>0$ specifying a running energy,
and then compute one-loop corrections by integrating out fast modes~\cite{Vafek2014PRB,Stauber2005PRB, Wang2017PRB, Wang2011PRB,She2010PRB,Huh2008PRB}, which are
provided in Appendix~\ref{Appendix_1L_corrections} for details.
Further, the quantum corrections are absorbed to the old ``slow modes" to yield the new ``slow modes",
which are then rescaled to the new ``fast modes"~\cite{Vafek2012PRB,Vafek2014PRB,Wang2017PRB, Wang2011PRB,Wang2013PRB,She2010PRB,Huh2008PRB,Kim2008PRB,Maiti2010PRB,
She2015PRB,Roy2016PRB}. To this end, we need to adopt the
RG rescaling transformations,
\begin{eqnarray}
k_x&\longrightarrow&k'_xe^{-l},\label{Eq_rescale-k_x}\\
k_y&\longrightarrow&k'_ye^{-l},\label{Eq_rescale-k_y}\\
\omega&\longrightarrow&\omega'e^{-2l},\label{Eq_rescale-omega}\\
\Psi(i\omega,\mathbf{k})&\longrightarrow&\Psi'(i\omega',\mathbf{k}')e^{3l}.
\label{Eq_rescale-Psi}
\end{eqnarray}
These scalings are derived by considering
the noninteracting parts of effective action~(\ref{Eq_S_eff})
as an original fixed point that is invariant during RG processes~\cite{Vafek2014PRB,Wang2011PRB,Huh2008PRB,Wang2019JPCM}.

\begin{figure}[H]
\centering
\includegraphics[width=3in]{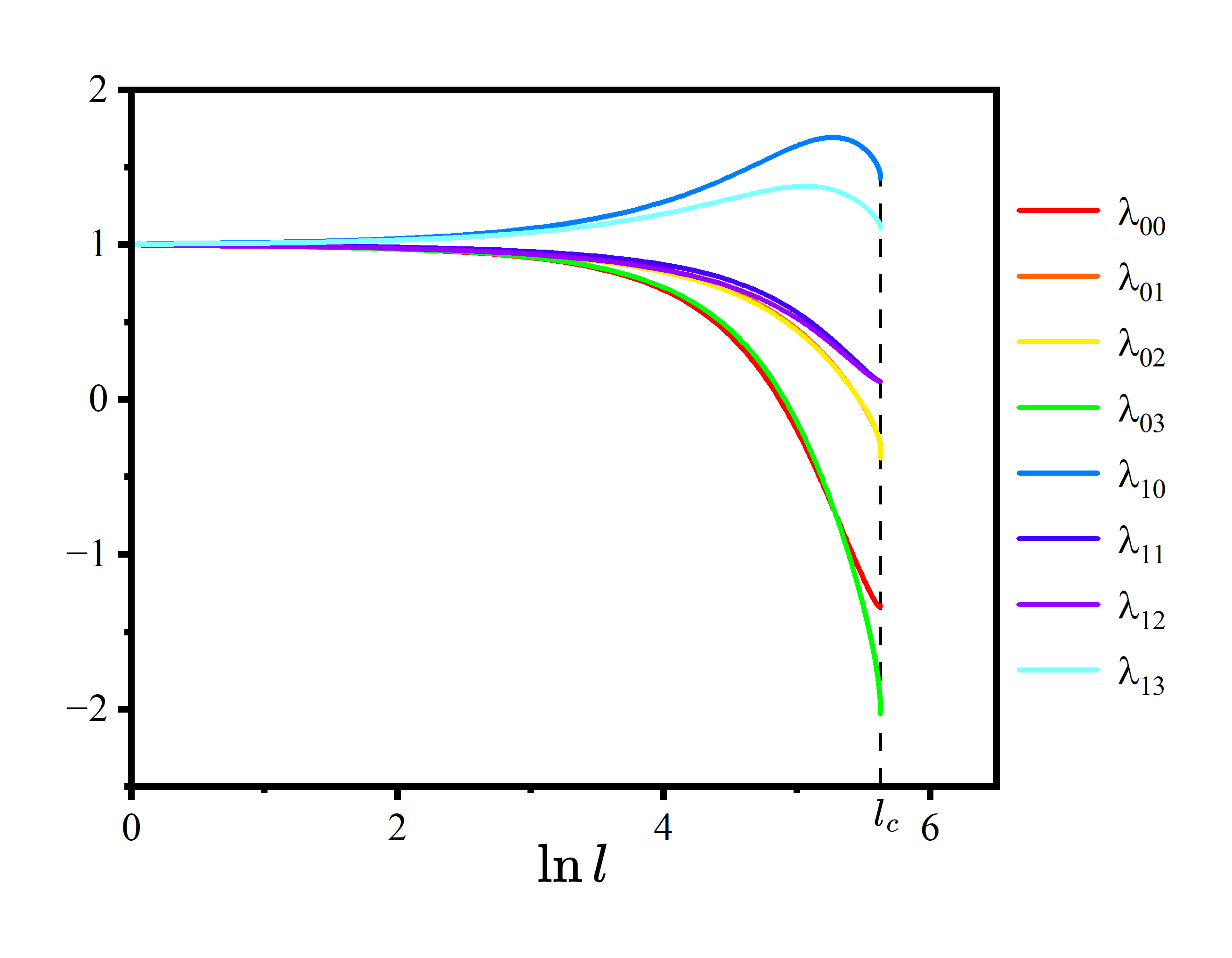}\\
\vspace{-0.6cm}
\caption{(Color online) Energy-dependent rescaled evolutions of eight nondegenerate fermion-fermion interactions
for the same structural parameters as in Fig.~\ref{Fig_evolution}.}
\label{Fig_rescaled_FP}
\end{figure}

On the basis of these RG scalings for the momentum, energy, and the fermion
field~(\ref{Eq_rescale-k_x})-(\ref{Eq_rescale-Psi}), the coupled RG equations of all
fermion-fermion interactions can be derived by combining one-loop corrections~(\ref{Eq_F00})-(\ref{Eq_F33})~\cite{Vafek2014PRB,Wang2011PRB,Huh2008PRB,Wang2019JPCM},
\begin{equation}
\left\{\frac{d\lambda_{\mu\nu}}{dl} = \mathcal{F}_{\mu\nu},\,\,\mathrm{with}\,\, \mu,\nu = 0,1,2,3 \right\},
\label{Eq_RG_flows}
\end{equation}
where $\mathcal{F}_{\mu\nu}$ are presented in Eqs.~(\ref{Eq_F00})-(\ref{Eq_F33}) of Appendix~\ref{Appendix_1L_corrections}.
The RG flow equations~(\ref{Eq_RG_flows}) play a crucial role in unraveling the low-energy behavior of 2D kagom\'{e} QBCP systems.
With the help of these equations, we aim to determine the fixed points that govern the fates of fermion-fermion interactions in the low-energy regime. Additionally, we are going to identify the dominant instabilities and the associated phase transitions,
considering both symmetric and asymmetric cases.

\section{Fates of fermion-fermion interactions}\label{Sec_fate_ff}

In this section, we perform a systematic analysis of the coupled RG flows~(\ref{Eq_RG_flows}) to determine how fermion-fermion interactions evolve under decreasing energy scales. This analysis allows us to identify the fates of these interactions, which are intimately related to the low-energy physics of 2D kagom\'{e} QBCP semimetals.

\begin{figure*}[htbp]
\centering
\subfigure[]{
\includegraphics[width=2.1in]{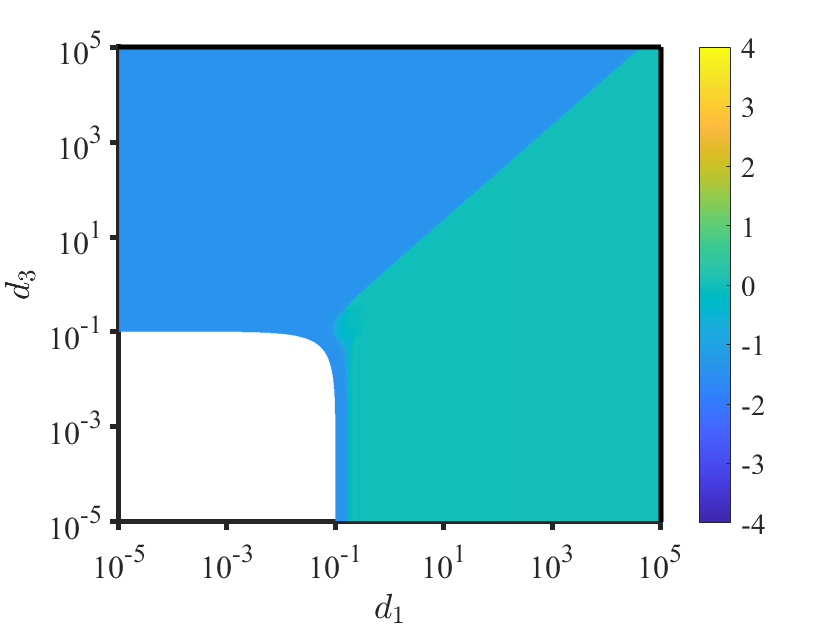}
}
\subfigure[]{
\includegraphics[width=2.1in]{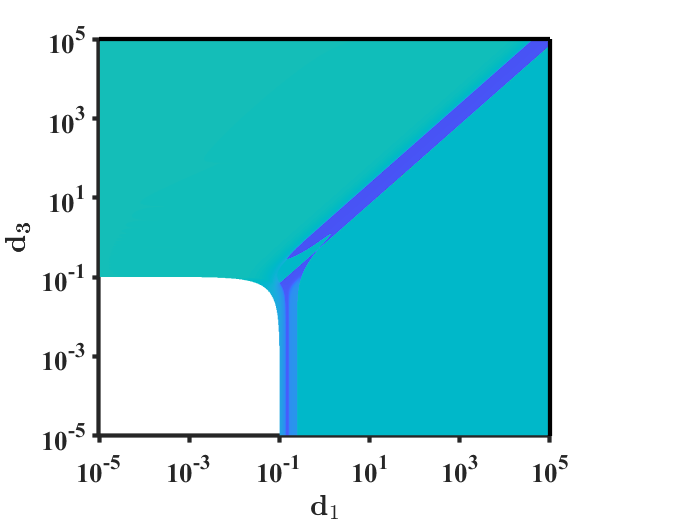}
}
\subfigure[]{
\includegraphics[width=2.1in]{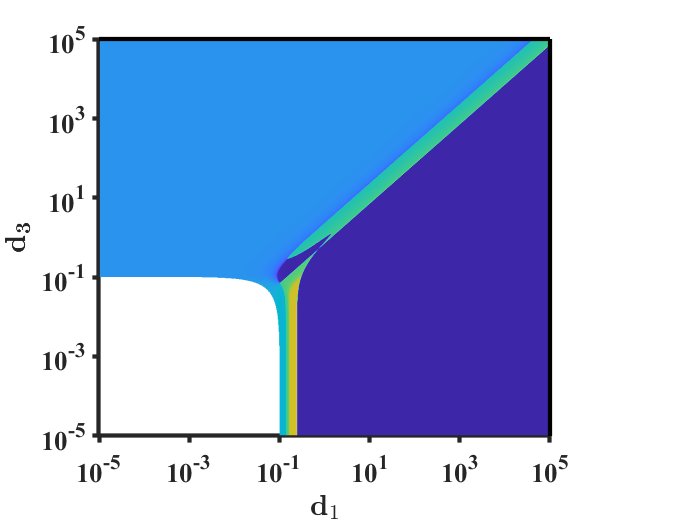}
}\\
\subfigure[]{
\includegraphics[width=2.1in]{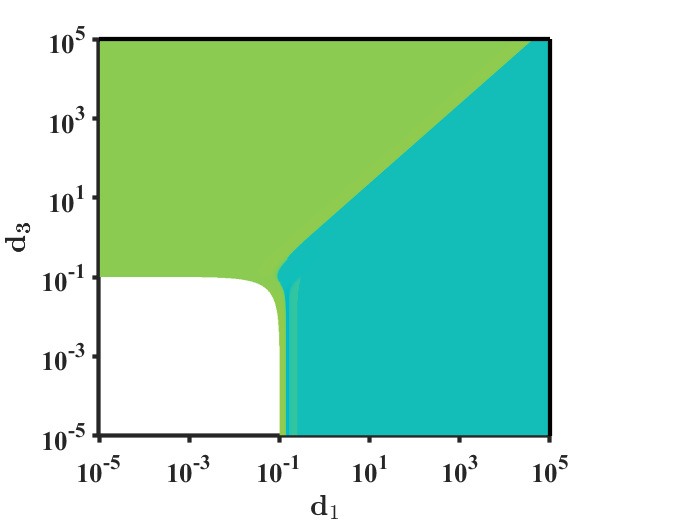}
}
\subfigure[]{
\includegraphics[width=2.1in]{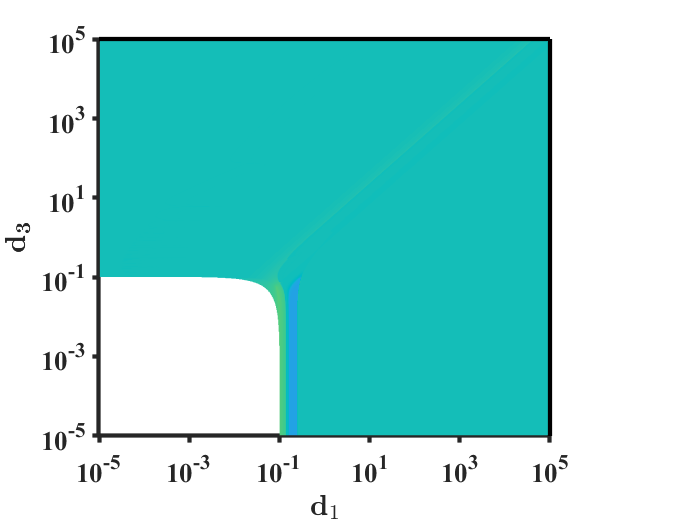}
}
\subfigure[]{
\includegraphics[width=2.1in]{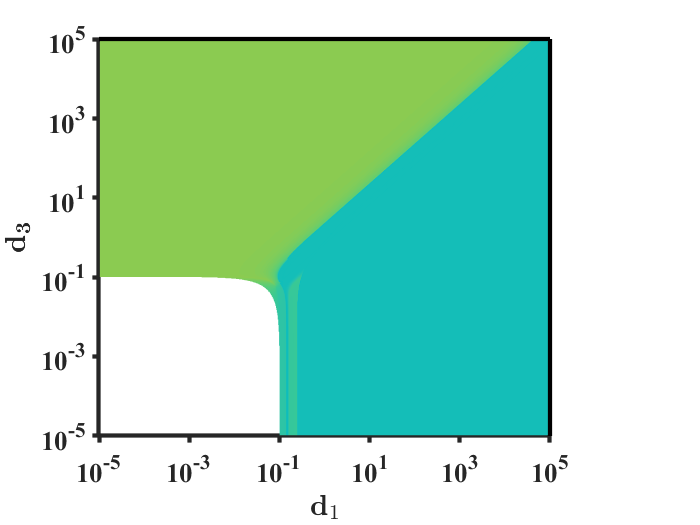}
}\\
\vspace{-0.1cm}
\caption{(Color online) Critical values of fermion-fermion interactions at $l=l_c$ with variations of
the structural parameters $d_1$ and $d_3$ at a fixed $d_0=0.1$ (the basic results are
insensitive to concrete value of $d_0$):
(a) $\lambda_{00}$, (b) $\lambda_{01}$, (c) $\lambda_{03}$, (d) $\lambda_{10}$,
(e) $\lambda_{11}$, and (f) $\lambda_{13}$.}
\label{Fig_case_I_FP}
\end{figure*}

\subsection{Classification of parameters and fixed point characterization}\label{Sec_classification_FPs}

Before performing the numerical analysis, it is essential to classify the initial conditions of the parameters in
RG flows. The coupled RG equations~(\ref{Eq_RG_flows}) depend critically on four structural
parameters $d_0$, $d_1$, $d_2$, and $d_3$ from the Hamiltonian density~(\ref{Eq_H0_density}).
These parameters are constrained by the QBCP stability criterion~(\ref{Eq_condition_QBCP})~\cite{Fradkin2008PRB,Fradkin2009PRL,Huse2003PRB},
\begin{eqnarray}
d_0^2<(d_3^2+d_1^2\cos^2 2\theta+\frac{1}{4}d_2^2\sin^2 2\theta)_{\mathrm{min}}.
\end{eqnarray}
This restriction clusters the parameter space into three distinct cases, namely,
Case I with $\frac{1}{4}d_2^2 = d_1^2$ and $d_3^2+d_1^2>d_0^2$, Case II with $\frac{1}{4}d_2^2 > d_1^2$ and $d_3^2+d_1^2>d_0^2$,
and Case III with $\frac{1}{4}d_2^2 < d_1^2$ and $d_3^2+\frac{1}{4}d_2^2>d_0^2$, respectively.
In addition, the initial values of the fermion-fermion interaction can either be identical or distinct.

As a preliminary step, let us impose identical initial values on all fermion-fermion couplings to
investigate the overall tendencies of these fermionic couplings. Without loss of generality, we
fix $d_0 = 0.1$ and choose  $d_1 = 0.4$, $d_2 = 0.06$,
plus $d_3 = 0.7$ to construct a representative set of asymmetric parameters
that satisfies the QBCP stability criterion. The numerical analysis of the coupled RG
equations~(\ref{Eq_RG_flows}) in Fig.~\ref{Fig_evolution} with $\lambda_{\mu\nu}(l=0)=10^{-2}$ shows that each
coupling exhibits a pronounced energy dependence and diverges at a critical energy scale denoted by $l=l_c$.
It is of particular importance to emphasize that the qualitative trends of flows are
insensitive to the initial magnitude of $\lambda_{\mu\nu}(l=0)$ and thus
$\lambda_{\mu\nu}(l=0)=10^{-2}$ are adopted in the subsequent studies for convenience.

Additionally, we realize from Fig.~\ref{Fig_evolution} that the trajectories within each of the sets
$\{\lambda_{10}, \lambda_{20}, \lambda_{30}\}$, $\{\lambda_{11}, \lambda_{21}, \lambda_{31}\}$,
$\{\lambda_{12}, \lambda_{22}, \lambda_{32}\}$, and
$\{\lambda_{13}, \lambda_{23}, \lambda_{33}\}$ collapse onto a single scaling curve, implying certain
symmetry-related degeneracies. To efficiently capture the essential information, we henceforth retain a
single representative from each degenerate set and
subsequently put the focus on the following eight nondegenerate couplings, namely
$\lambda_{00}$, $\lambda_{01}$, $\lambda_{02}$, $\lambda_{03}$, $\lambda_{10}$, $\lambda_{11}$, $\lambda_{12}$,
and $\lambda_{13}$.

Furthermore, in order to succinctly track the evolution of interactions close to the critical energy scale,
we follow the strategy introduced in Refs.~\cite{Vafek2014PRB,Vafek2010PRB,Yang2010PRB,Roy2018PRX,Chubukov2016PRX,Zhang-Wang2025PRB}
to rescale all the interaction parameters by employing a sign-unchanged parameter,
\begin{eqnarray}
\mathcal{F}(l)\equiv\sqrt{\frac{1}{16}\sum_{\mu\nu}\lambda_{\mu\nu}^2(l)}.
\end{eqnarray}
The rescaled couplings are then defined as
\begin{eqnarray}
\lambda_{\mu\nu}(l)\longrightarrow\frac{\lambda_{\mu\nu}(l)}{\mathcal{F}(l)},
\end{eqnarray}
with $\mu,\nu=0,1,2,3$. For brevity, hereafter we utilize $\lambda_{\mu\nu}$ to denote these rescaled quantities.
Fig.~\ref{Fig_rescaled_FP} displays the rescaled flows of the eight nondegenerate interaction parameters.
There generally exists a low-energy scale dubbed $l_c$ at which the correlation length diverges, marking the emergence of critical behavior~\cite{Vojta2003RPP,Sachdev2011Book}.
Within the RG framework, the very scale $l_c$ is adopted to represent the ``critical energy scale", marking the terminal point of the RG flows where the system approaches a quantum critical regime~\cite{Shankar1994RMP,Sachdev2011Book}.
The values of interaction parameters at the critical scale $l_c$ designate a fixed point (FP) in parameter space~\cite{Vafek2014PRB,Vafek2010PRB,Yang2010PRB,Roy2018PRX,Chubukov2016PRX,Zhang-Wang2025PRB},
which corresponds to
\begin{eqnarray}
\mathrm{FP}
\equiv\left.\{\lambda_{\mu\nu}\,\,\mathrm{with}\,\,\mu,\nu,=0,1,2,3\} \right|_{l=l_c}.
\end{eqnarray}
Principally, these FPs are closely associated with the quantum criticality~\cite{Vojta2003RPP,Sachdev2011Book,Metzner2000PRL,Maiti2010PRB,
Vafek2014PRB,Chubukov2016PRX,Roy2018PRX,Nandkishore2012NP,Vafek2012PRB}
and play an important role in dictating the low-energy physics of 2D QBCP systems.
In the remainder of this section, we are going to seek all the potential FPs allowed
by the RG equations~(\ref{Eq_RG_flows}) and determine the initial conditions that flow into them, and defer their physical
consequences to Sec.~\ref{Sec_phase_transition}.

\begin{figure}[H]
\centering
\includegraphics[width=3in]{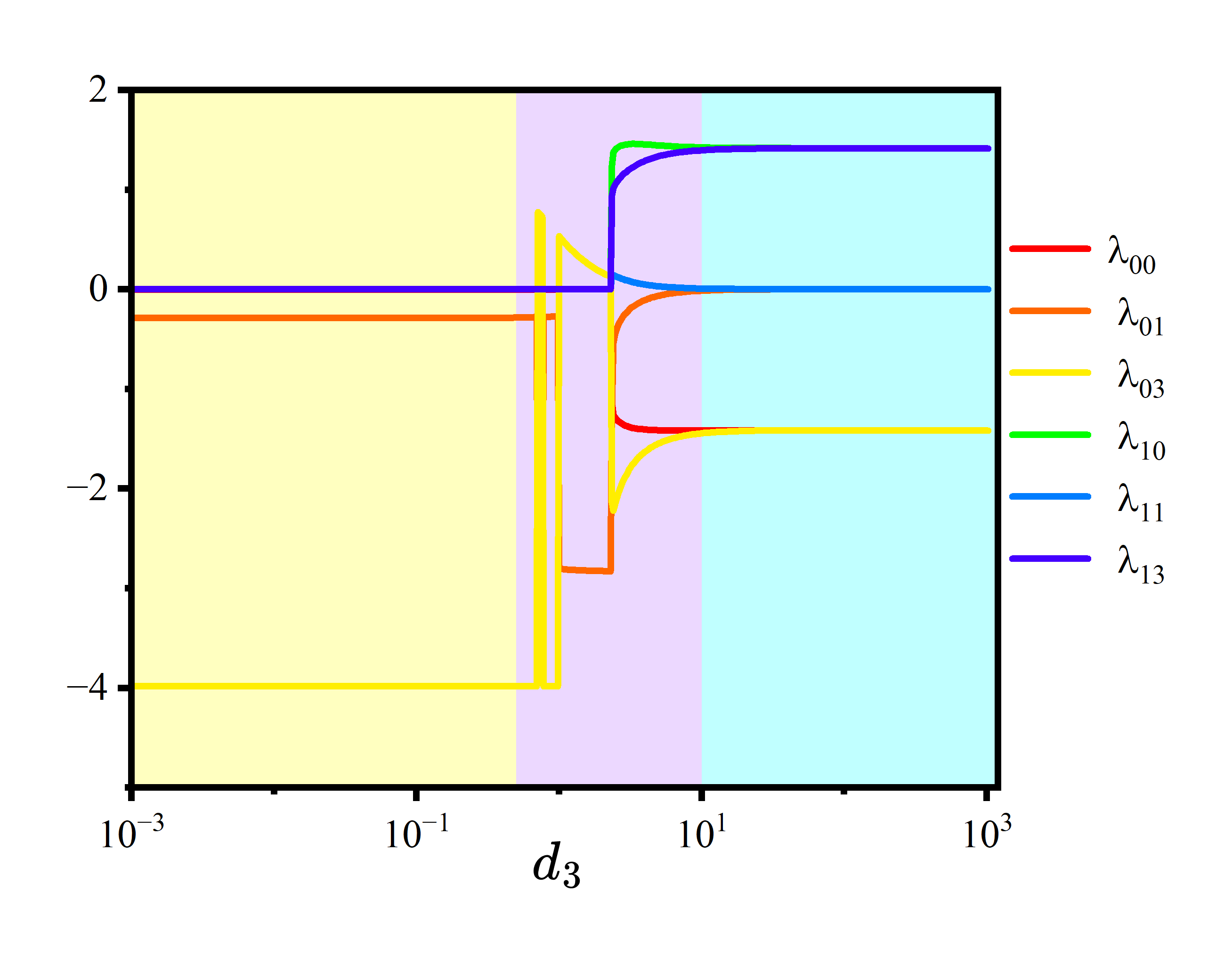}\\
\vspace{-0.35cm}
\caption{(Color online) Critical values of nondegenerate fermion-fermion interactions at $l=l_c$
with variation of the structural parameter $d_3$ at fixed $d_1=1$ and $d_2=2$.
The right (light cyan) and left (light yellow) stable regions correspond to $\mathrm{FP}_1$
and $\mathrm{FP}_2$ in Case $\mathrm{I}$, respectively.}
\label{Fig_FPs1_2}
\end{figure}

\begin{figure*}[htbp]
\centering
\subfigure[]{
\includegraphics[width=2in]{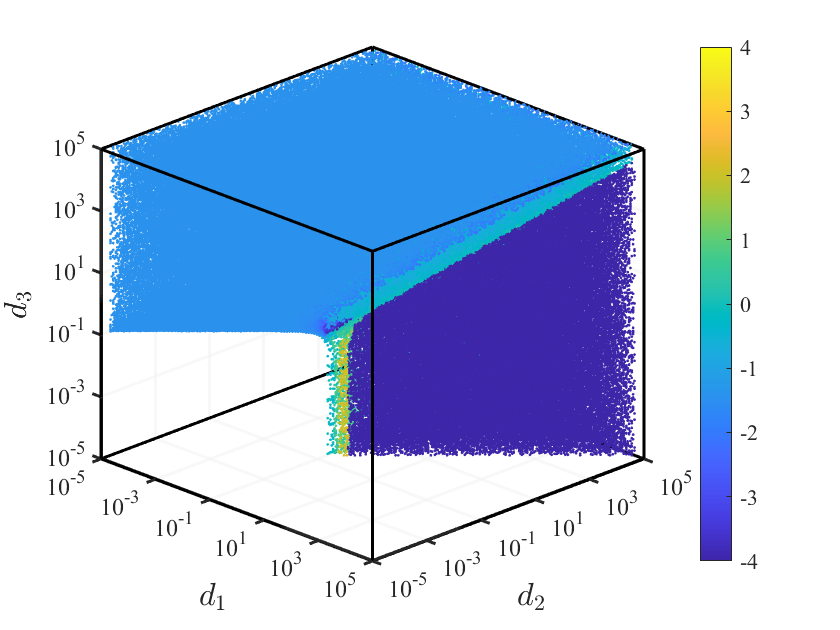}
}
\subfigure[]{
\includegraphics[width=2in]{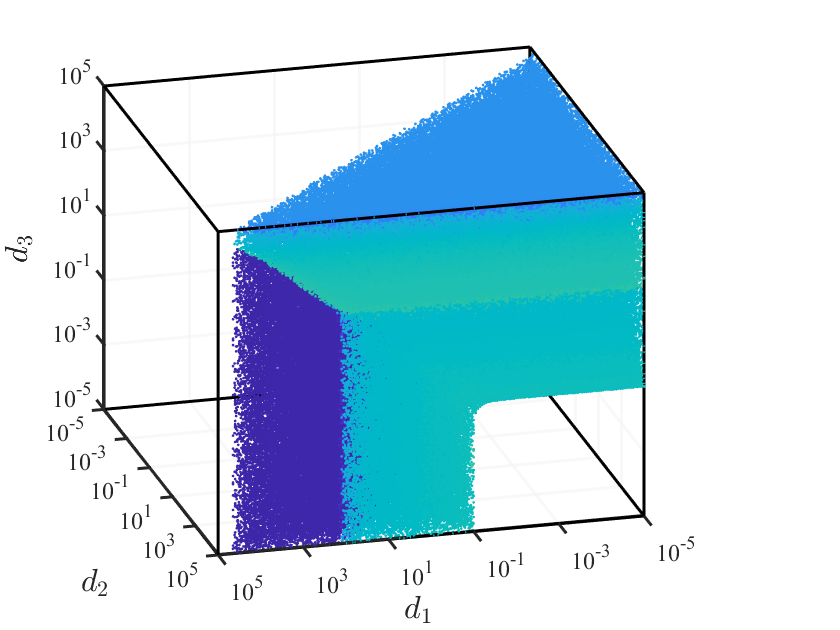}
}
\subfigure[]{
\includegraphics[width=2in]{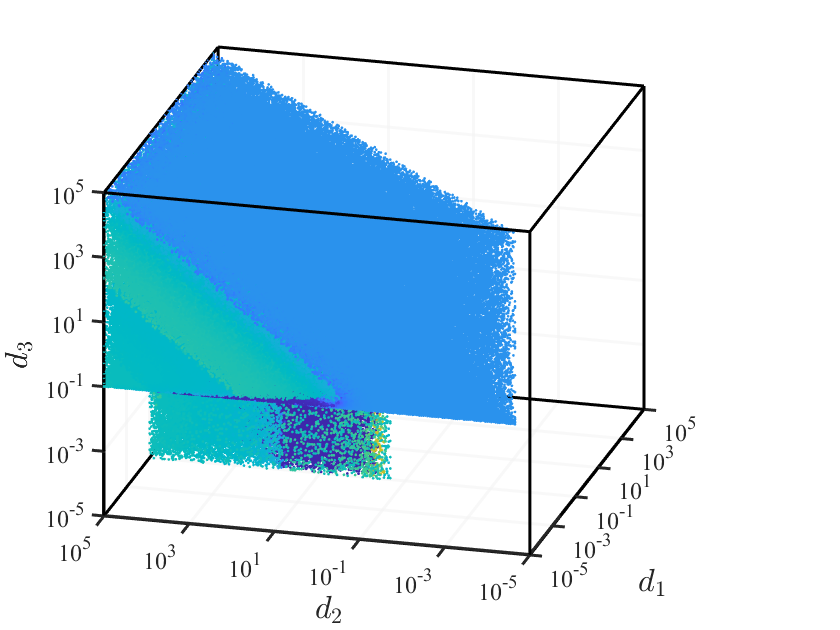}
}
\caption{(Color online) Critical values of fermion-fermion interaction $\lambda_{03}$ at $l=l_c$ as functions of
the structural parameters $d_1$, $d_2$, and $d_3$ at a fixed $d_0=0.1$ (the basic results are
insensitive to concrete value of $d_0$) rotated clockwise by (a) $0^\circ$, (b) $120^\circ$, and (c) $240^\circ$.
The tendencies of the other seven nondegenerate fermion-fermion interaction parameters ($\lambda_{00}$,
$\lambda_{01}$, $\lambda_{02}$, $\lambda_{10}$, $\lambda_{11}$, $\lambda_{12}$, and $\lambda_{13}$) are
qualitatively similar and provided in Appendix~\ref{Sec_Appendix_BC_flows}.}
\label{Fig_lambda03_d123}
\end{figure*}

\subsection{Potential fixed points and required boundaries}

\subsubsection{Case $\mathrm{I}$}\label{Sec_Case-I}

At first, let us consider Case $\mathrm{I}$, defined by the condition $\frac{1}{4}d_2^2 = d_1^2$,
which restores rotational symmetry. Due to this symmetry, the four independent parameters $d_{0,1,2,3}$
collapse to three, namely $d_0$, $d_1$, and $d_3$. In this circumstance, numerical analysis of the RG flows~(\ref{Eq_RG_flows})
indicates that there exist the degeneracies for $\lambda_{01}$ and $\lambda_{02}$
as well as $\lambda_{11}$ and $\lambda_{12}$. This accordingly reduces
the eight interaction parameters discussed in Sec.~\ref{Sec_classification_FPs} for the general case
to six couplings, which can be selected as $\lambda_{00}$, $\lambda_{01}$, $\lambda_{03}$,
$\lambda_{10}$, $\lambda_{11}$, and $\lambda_{13}$.

To proceed, let us focus on how the variations of structural parameters influence the
behavior of interaction parameters, which is of close relevance to the FPs. Since the parameter
$d_0$ primarily rescales the absolute magnitude of the couplings without altering their qualitative behavior,
we hereby set $d_0=0.1$ from now on and put our focus on the impacts of
$d_1$ and $d_3$ on the tendencies of interaction parameters.
Figure~\ref{Fig_case_I_FP} shows that the critical values of these six kinds of interactions are sensitive at $l_c$ with
tuning the structural parameters $d_1$ and $d_3$.
Reading from Fig.~\ref{Fig_FPs1_2} we identify two distinct stable regions where interaction parameters
become insensitive to structural variations and flow towards the same fixed point at the lowest-energy limit.
Specifically, the yellow region with $d_1 \ll d_3$ of Fig.~\ref{Fig_FPs1_2} signals that the system goes
to the first FP,
\begin{eqnarray}
\mathrm{FP}_1
&\equiv&(\lambda_{00}, \lambda_{01}, \lambda_{02}, \lambda_{03},
\lambda_{10}, \lambda_{11}, \lambda_{12},\lambda_{13})|_{l=l_c}\nonumber\\
&\approx&(-1.414, 0, 0, -1.414, 1.414, 0,  0, 1.414),\label{Eq_FP1}
\end{eqnarray}
and the cyan region of Fig.~\ref{Fig_FPs1_2} is characterized by $d_1 \gg d_3$, giving rise to the second FP,
\begin{eqnarray}
\mathrm{FP}_2&\equiv&(\lambda_{00}, \lambda_{01}, \lambda_{02}, \lambda_{03},
\lambda_{10}, \lambda_{11}, \lambda_{12},\lambda_{13})|_{l=l_c}\nonumber\\
&\approx&(0, -0.288, -0.288, -3.979, 0,  0,  0,  0).\label{Eq_FP2}
\end{eqnarray}
The remainder of Fig.~\ref{Fig_FPs1_2} corresponds to a crossover region that fails to support a
stable fixed point and is therefore physically irrelevant to critical behavior~\cite{Vafek2014PRB,Vafek2010PRB,Yang2010PRB,Roy2018PRX,Chubukov2016PRX,Zhang-Wang2025PRB}.

\begin{figure*}[htbp]
\subfigure[]{
\includegraphics[width=1.6in]{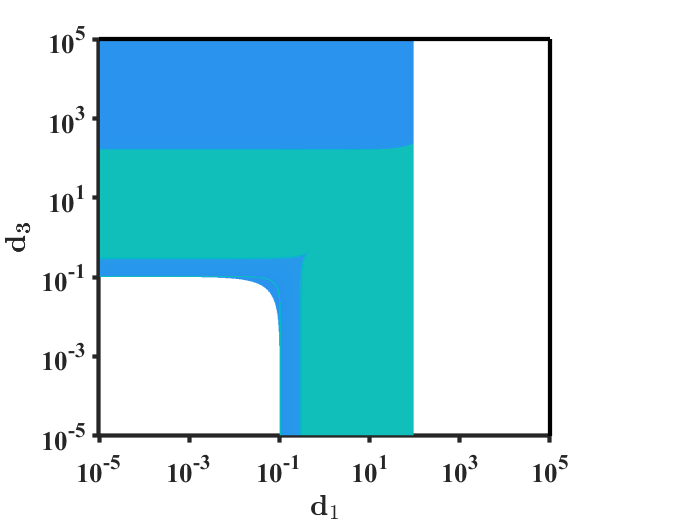}
}
\subfigure[]{
\includegraphics[width=1.6in]{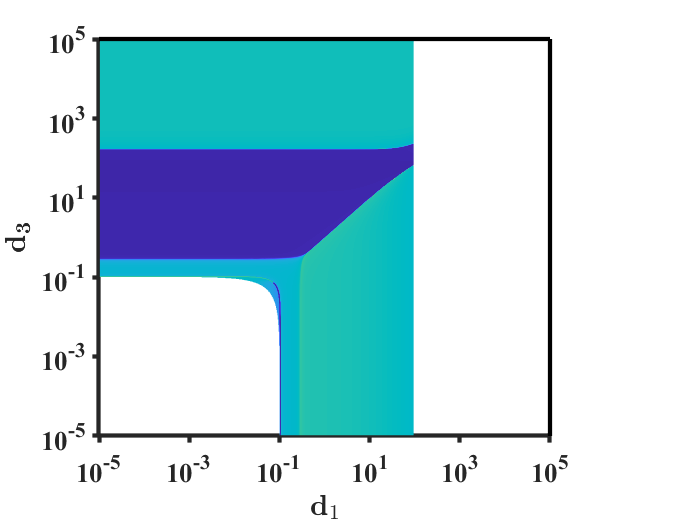}
}
\subfigure[]{
\includegraphics[width=1.6in]{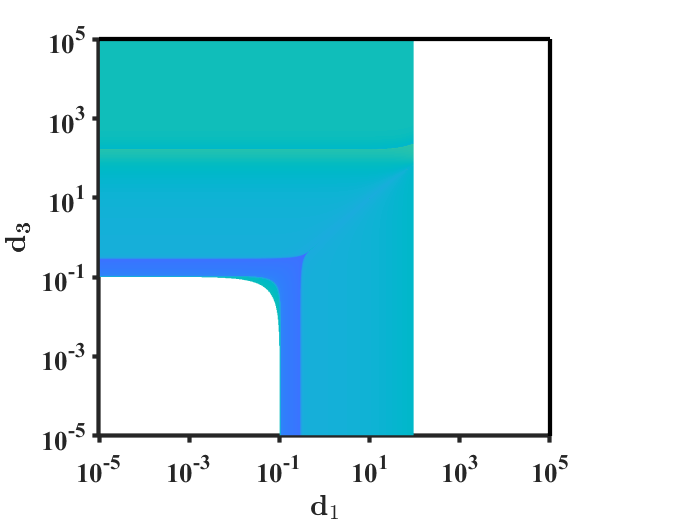}
}
\subfigure[]{
\includegraphics[width=1.6in]{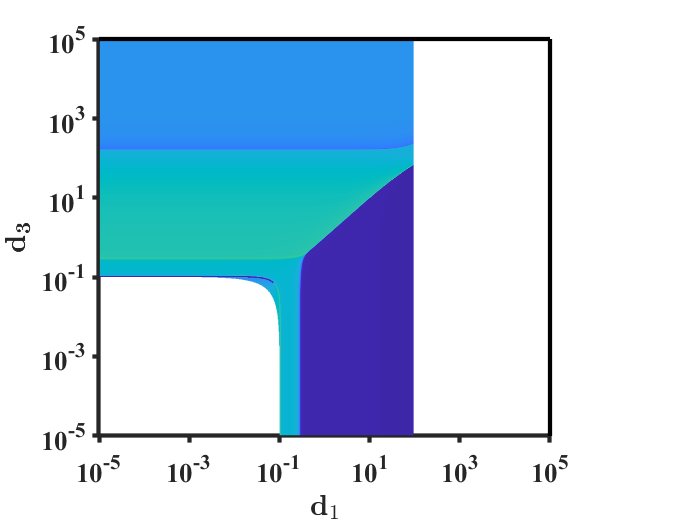}
}\\
\subfigure[]{
\includegraphics[width=1.6in]{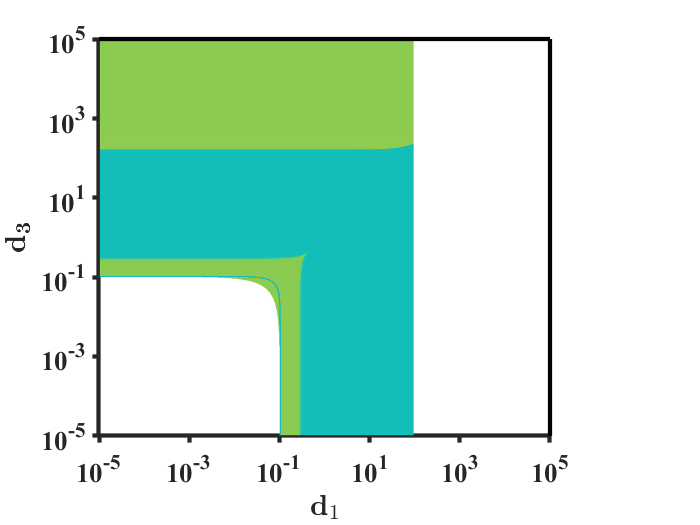}
}
\subfigure[]{
\includegraphics[width=1.6in]{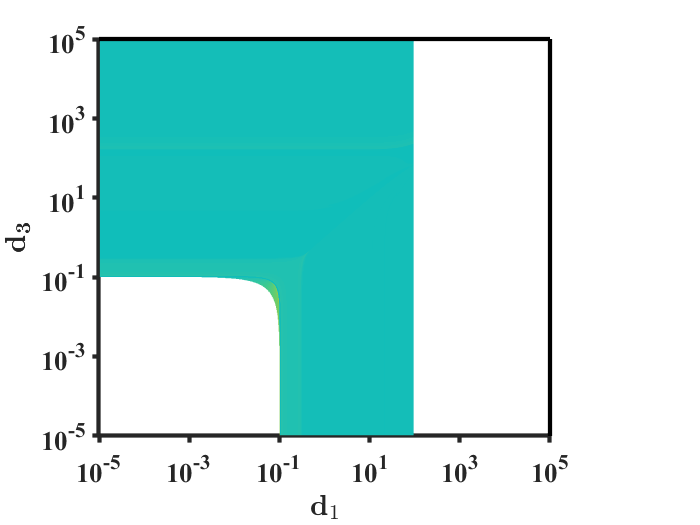}
}
\subfigure[]{
\includegraphics[width=1.6in]{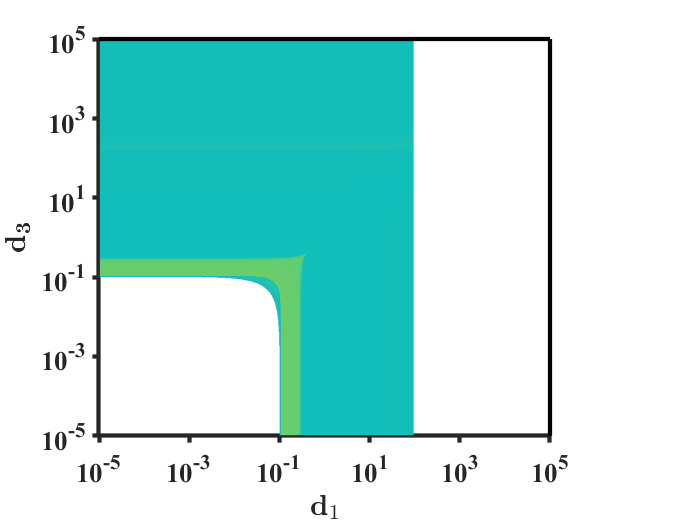}
}
\subfigure[]{
\includegraphics[width=1.6in]{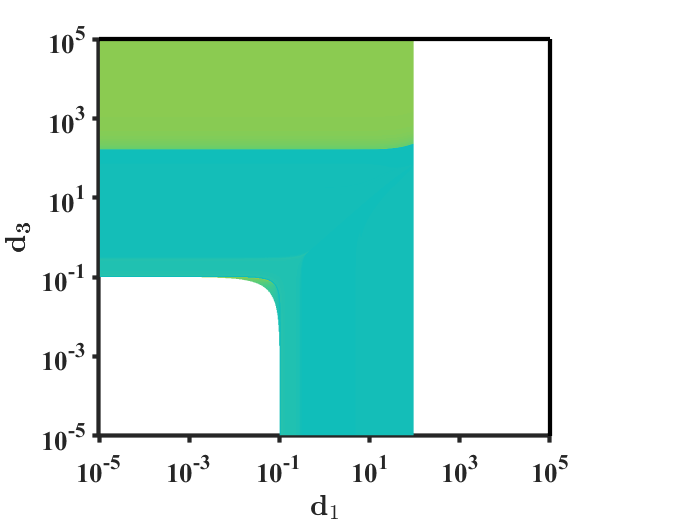}
}
\caption{(Color online) Critical values of fermion-fermion interaction at $l=l_c$ as functions of
the structural parameters $d_1$ and $d_3$ at a fixed $d_0=0.1$ and $d_2=200$
(i.e, a cross profile of Fig.~\ref{Fig_lambda03_d123} and seven other nondegenerate fermion-fermion interactions
at $d_2=200$): (a) $\lambda_{00}$, (b) $\lambda_{01}$, (c) $\lambda_{02}$, (d) $\lambda_{03}$, (e) $\lambda_{10}$,
(f) $\lambda_{11}$, (g) $\lambda_{12}$, and (h) $\lambda_{13}$.}
\label{figure:d2=200}
\end{figure*}

To proceed, it is necessary to establish the boundary conditions in structural parameter space that are required to
achieve these two FPs. For convenience, let us designate $u\equiv\log{d_1}$, $v\equiv\log{d_2}$, and $w\equiv\log{d_3}$.
After combining Fig.~\ref{Fig_case_I_FP} and Fig.~\ref{Fig_FPs1_2}, we perform the linear fitting analysis
and obtain the boundary conditions (BCs) that are approximately bounded by four lines,
namely, $\mathrm{BC}$-$\mathrm{FP}_1$-$\mathrm{case}$ $\mathrm{I}$,
\begin{small}
\begin{numcases}{}
\mathrm{Line1}:\,\,w + 0.2300000 = 0,& \nonumber \\
\mathrm{Line2}:\,\,1.039256u -w + 1.713409 = 0,& \label{Eq_condition_1}\\
\mathrm{Line3}:\,\,u + 5.000000 = 0,& \nonumber \\
\mathrm{Line4}:\,\,w - 5.000000 = 0, \nonumber
\end{numcases}
\end{small}
for $\mathrm{FP}_1$, and  $\mathrm{BC}$-$\mathrm{FP}_2$-$\mathrm{case}$ $\mathrm{I}$
\begin{small}
\begin{numcases}{}
\mathrm{Line1}:\,\,193.500000u - w + 53.050000 = 0,& \nonumber \\
\mathrm{Line2}:\,\,1.037879u - w - 0.839394 = 0,& \label{Eq_condition_2}\\
\mathrm{Line3}:\,\,w + 5.000000= 0,& \nonumber \\
\mathrm{Line4}:\,\,u - 5.000000= 0,\nonumber
\end{numcases}
\end{small}
for $\mathrm{FP}_2$, respectively. These linear conditions delineate the parameter regions
that flow into the corresponding FPs and thus
provide quantitative guidance for analyzing the low-energy critical behavior.

\subsubsection{Case $\mathrm{II}$}

Next, we turn to Case $\mathrm{II}$, defined by $\frac{1}{4}d_2^2 > d_1^2$ and $d_3^2+d_1^2>d_0^2$.
In this situation, the rotational symmetry is explicitly broken and thus all four structural parameters
$d_{0,1,2,3}$ can be tuned independently. As mentioned earlier, the $d_0$ plays a trivial role in determining
the qualitative behavior of fermionic couplings and thus it is significant to examine the
effects of variations of $d_{1,2,3}$. In addition, both degenerate pairs $(\lambda_{01},\lambda_{02})$
and $(\lambda_{11},\lambda_{12})$ for Case $\mathrm{I}$ are no longer overlapped but split from each other.
This yields eight nondegenerate couplings for Case $\mathrm{II}$, i.e., $\lambda_{00}$, $\lambda_{01}$,
$\lambda_{02}$, $\lambda_{03}$, $\lambda_{10}$, $\lambda_{11}$, $\lambda_{12}$,
and $\lambda_{13}$ as addressed in Sec~\ref{Sec_classification_FPs}.

\begin{figure*}[htbp]
\subfigure[]{
\includegraphics[width=1.5in]{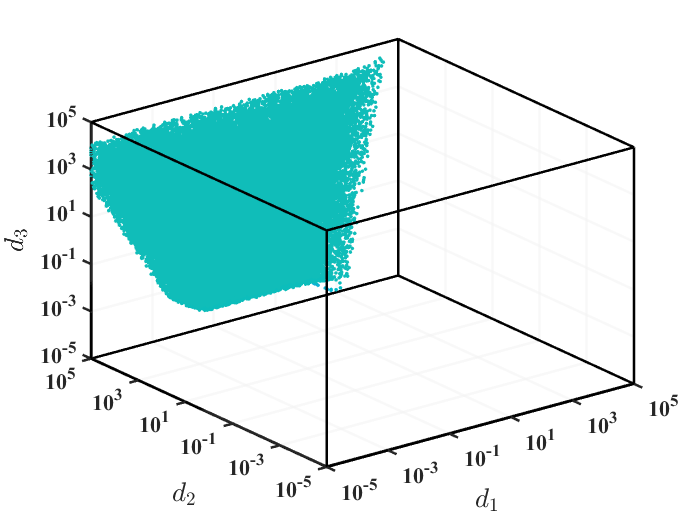}
}
\subfigure[]{
\includegraphics[width=1.5in]{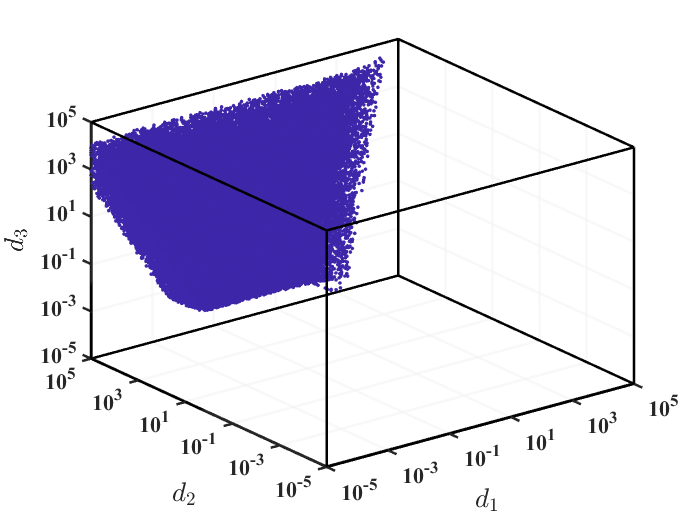}
}
\subfigure[]{
\includegraphics[width=1.5in]{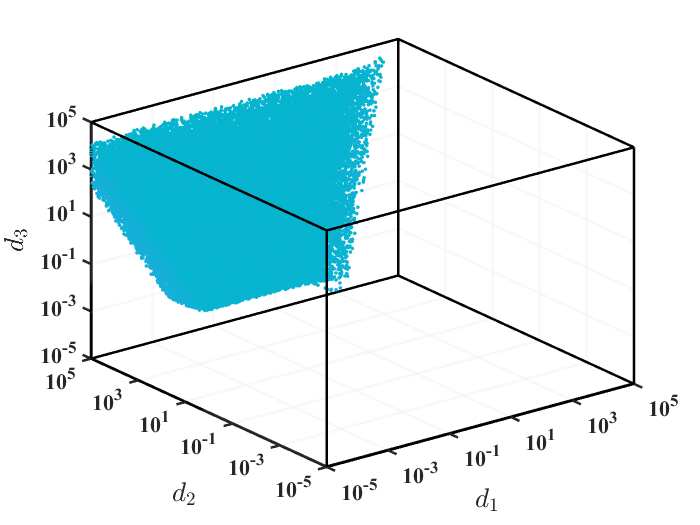}
}
\subfigure[]{
\includegraphics[width=1.5in]{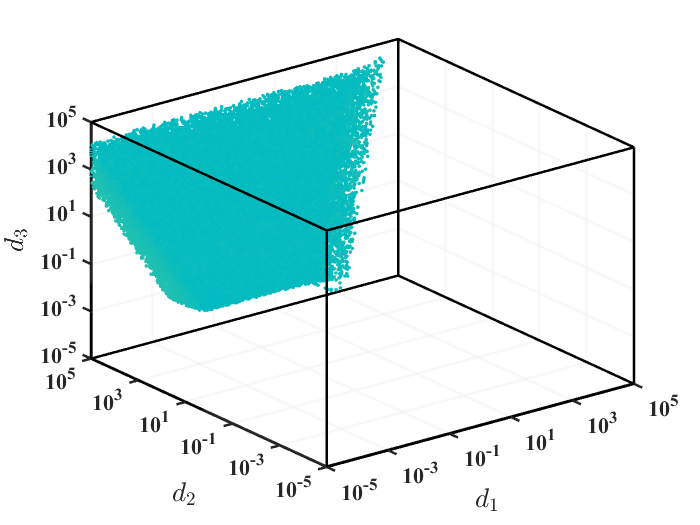}
}\\
\subfigure[]{
\includegraphics[width=1.5in]{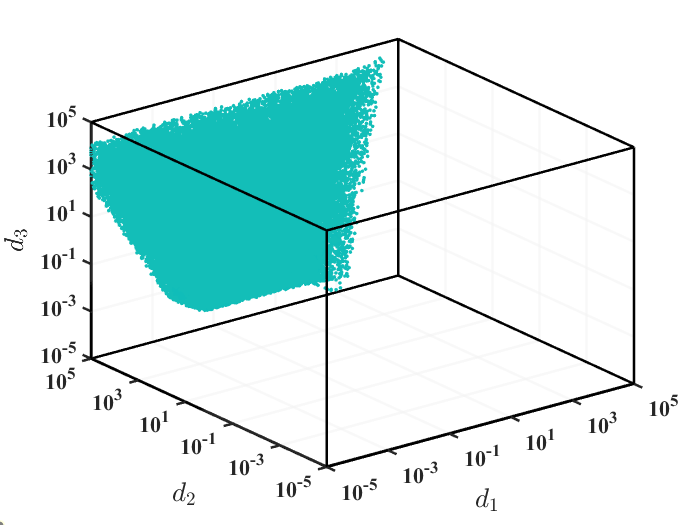}
}
\subfigure[]{
\includegraphics[width=1.5in]{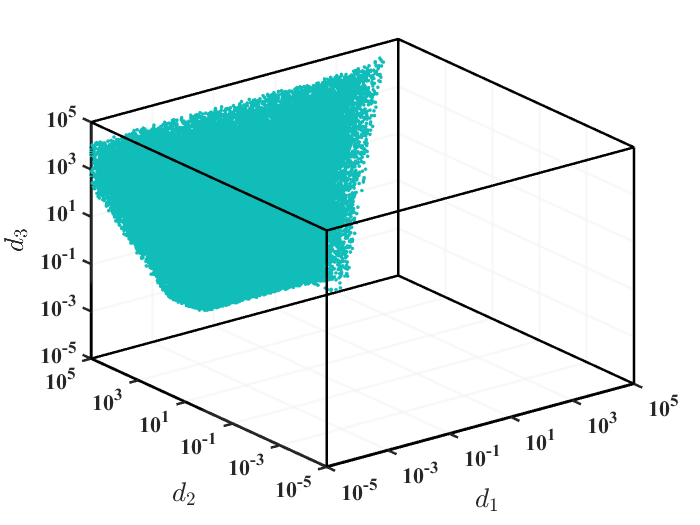}
}
\subfigure[]{
\includegraphics[width=1.5in]{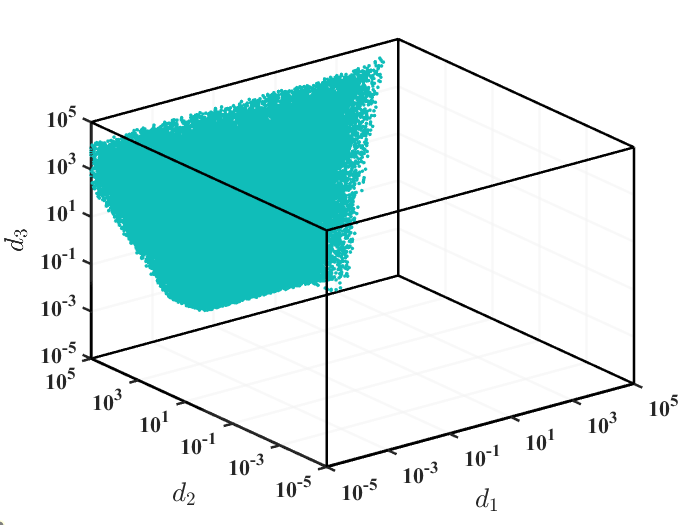}
}
\subfigure[]{
\includegraphics[width=1.5in]{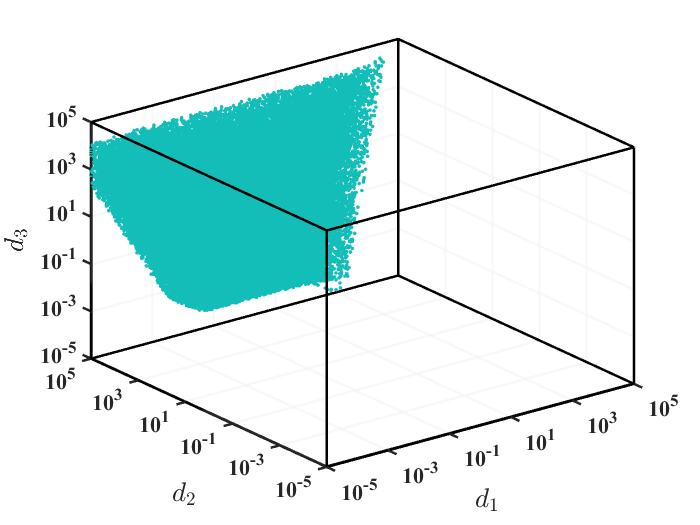}
}
\caption{(Color online) Stable subregions in the $d_1$-$d_2$-$d_3$ structural parameter
space for critical fermion-fermion interaction values at $l=l_c$
[color mapping follows the same scale as Fig.~\ref{Fig_lambda03_d123}(a)]:
(a) $\lambda_{00}$, (b) $\lambda_{01}$, (c)$\lambda_{02}$, (d) $\lambda_{03}$, (e) $\lambda_{10}$, (f)$\lambda_{11}$,
(g) $\lambda_{12}$, and (h)$\lambda_{13}$, which constitute the $\mathrm{FP}_3$~(\ref{Eq_FP3}).
The related subregions for $\mathrm{FP}_{1,4,5,6}$ are provided in
Fig.~\ref{Fig_FP1-4-II} and Fig.~\ref{Fig_FP5-6-II} of Appendix~\ref{Sec_Appendix_BC_flows}.}
\label{Fig_FP3}
\end{figure*}

Although it is more complicated, the asymmetric situation is both generic and physically relevant.
Fig.~\ref{Fig_lambda03_d123} shows the critical values of fermion-fermion interaction $\lambda_{03}$ at $l=l_c$ as functions of
the structural parameters $d_1$, $d_2$, and $d_3$ at a fixed $d_0$ viewed from three distinct perspectives. Analogous results
are obtained for the remaining seven nondegenerate couplings and thus are not shown for brevity.
To delineate the stable regions, we take a cross-section of Fig.~\ref{Fig_lambda03_d123} at $d_2 = 200$ as shown in Fig.~\ref{figure:d2=200}. It clearly displays the well-defined stable regions that are separated by narrow transition zones.
Within each stable region, the critical values of fermion-fermion interactions remain considerably flat and are insensitive
to the structural parameter variations.

\begin{figure}[htbp]
\includegraphics[width=3in]{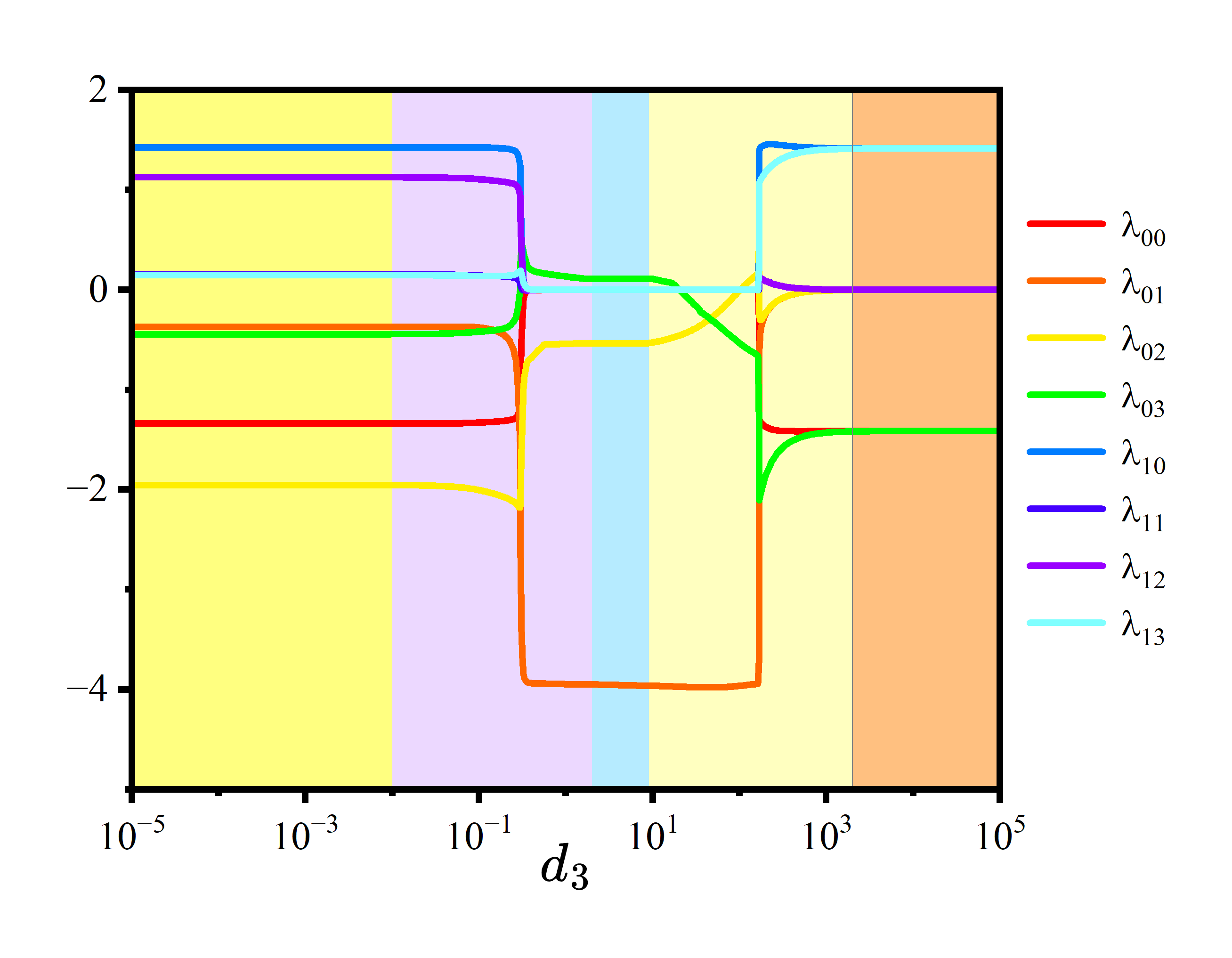}\\
\vspace{-0.5cm}
\caption{(Color online) Critical values of nondegenerate fermion-fermion interactions at $l=l_c$
with variation of the structural parameter $d_3$ at fixed $d_1=0.15$ and $d_2=200$.
The right (orange), middle (light cyan), and left (yellow) stable regions correspond to $\mathrm{FP}_1$, $\mathrm{FP}_3$, and $\mathrm{FP}_6$ in Case $\mathrm{II}$, respectively.}
\label{figure:d2=200d1=0.15}
\end{figure}

Then, paralleling this strategy, we systematically examine the parameter space in Fig.~\ref{Fig_lambda03_d123} and
identify five distinct subregions within the parameter $(d_1, d_2, d_3)$ space, which exhibit sufficient stability and
therefore give rise to a FP that coincides with
the $\mathrm{FP}_1$~(\ref{Eq_FP1}) in Case $\mathrm{I}$ and four new FPs, namely
\begin{eqnarray}
\mathrm{FP}_3&\equiv&(\lambda_{00}, \lambda_{01}, \lambda_{02}, \lambda_{03},
\lambda_{10}, \lambda_{11}, \lambda_{12},\lambda_{13})|_{l=l_c}\nonumber\\
&\approx&(-0.005, -3.958, -0.564, 0.036,  0,   0,  0,   0),\label{Eq_FP3}
\end{eqnarray}
and $\mathrm{FP}_{4,5,6}$ are defined similar to $\mathrm{FP}_3$, which take the form of
\begin{eqnarray}
\mathrm{FP}_4&\approx&(-0.002, 0.049, -0.573, -3.957, 0,  0,   0,   0),\label{Eq_FP4}\\
\mathrm{FP}_5&\approx&(-1.394, -0.107, -1.488, -0.113,  1.444,  \nonumber\\
&&0.073,  1.354,  0.077),\label{Eq_FP5}\\
\mathrm{FP}_6&\approx&(-1.367, -0.255, -1.857, -0.256,  1.450,  \nonumber\\
&&0.102, 1.179,  0.102).\label{Eq_FP6}
\end{eqnarray}
Fig.~\ref{Fig_FP3} illustrates the stable parameter subregion for $\mathrm{FP}_3$.
For brevity, the stable subregions of the remaining four FPs are provided in
Fig.~\ref{Fig_FP1-4-II} and Fig.~\ref{Fig_FP5-6-II} of Appendix~\ref{Sec_Appendix_BC_flows}.
To make the FPs explicit, let us fix $d_1$ further to $0.15$ for Fig.~\ref{figure:d2=200}
and obtain Fig.~\ref{figure:d2=200d1=0.15}, which clearly displays three FPs, i.e., $\mathrm{FP}_1$,
$\mathrm{FP}_3$, and $\mathrm{FP}_6$. FPs $\mathrm{FP}_4$ and $\mathrm{FP}_5$ can be obtained in a similar way
and hence are not shown here.

\begin{figure}[H]
\centering
\includegraphics[width=2.5in]{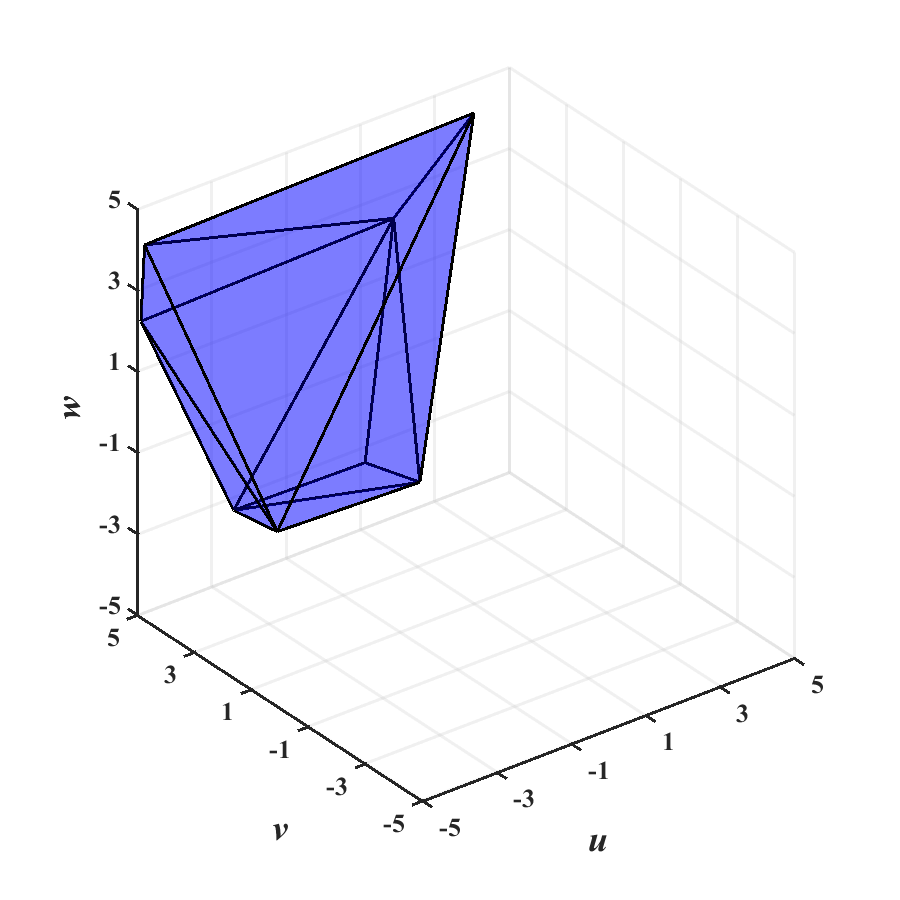}
\\
\vspace{-0.5cm}
\caption{(Color online) Approximate plane-fitting region for $\mathrm{FP}_3$
in Case $\mathrm{II}$ with $u\equiv\log{d_1}$, $v\equiv\log{d_2}$, and $w\equiv\log{d_3}$.}
\label{Fig_FP_fitting}
\end{figure}

To proceed, we are going to establish the boundary conditions for these FPs.
Compared to the symmetric case, the analysis is significantly more complex due to the three-dimensional
structural parameter space.  Let us consider $\mathrm{FP}_3$ shown in Fig.~\ref{Fig_FP3} as a representative case.
Its precise boundary in the $d_{1}$-$d_2$-$d_{3}$ space depicted in Fig.~\ref{Fig_FP3}(a)
exhibits intricate topology and is very challenging to derive the strict analytical description.
To characterize this boundary, we utilize the plane fitting to construct a polyhedra approximation displayed in
Fig.~\ref{Fig_FP_fitting}, which completely encloses the stable region.
Although such fitting is not strictly exact, this polyhedron sufficiently captures the boundary of $\mathrm{FP}_3$.
Through combined analysis of Fig.~\ref{Fig_FP3}(a) and Fig.~\ref{Fig_FP_fitting}, we obtain the approximate boundary conditions
for $\mathrm{FP}_3$ in Case II ($\mathrm{BC}$-$\mathrm{FP}_3$-$\mathrm{case}$ $\mathrm{II}$), which are approximately
bounded by eight planes
\begin{small}
\begin{numcases}{}
0.005378u - 0.999761v + 0.021197w + 4.887992 = 0,& \nonumber \\
0.000000u + 1.000000v + 0.000000w - 4.950000 = 0,& \nonumber \\
0.025832u + 0.059839v + 0.997874w + 0.945173 = 0,& \nonumber \\
0.009543u + 0.016010v + 0.999826w + 0.940853 = 0,& \label{Eq_condition_4} \\
0.015761u + 0.698293v - 0.715638w - 1.768178 = 0,& \nonumber \\
0.003611u + 0.728126v - 0.685434w - 0.776433 = 0,& \nonumber \\
0.623011u + 0.159034v - 0.765876w - 0.101688 = 0,& \nonumber \\
0.996904u + 0.057250v - 0.053894w + 4.772559 = 0, \nonumber
\end{numcases}
\end{small}
%%\begin{small}
%%\begin{numcases}{}
%%\!\!\!\mathrm{PL1}:0.005378u - 0.999761v + 0.021197w + 4.887992 = 0,& \nonumber \\
%%\!\!\!\mathrm{PL2}:0.000000u + 1.000000v + 0.000000w - 4.950000 = 0,& \nonumber\\
%%\!\!\!\mathrm{PL3}:0.025832u + 0.059839v + 0.997874w + 0.945173 = 0,& \nonumber \\
%%\!\!\!\mathrm{PL4}:0.009543u + 0.016010v + 0.999826w + 0.940853 = 0,& \label{Eq_condition_4} \\
%%\!\!\!\mathrm{PL5}:0.015761u + 0.698293v - 0.715638w - 1.768178 = 0,& \nonumber \\
%%\!\!\!\mathrm{PL6}:0.003611u + 0.728126v - 0.685434w - 0.776433 = 0,& \nonumber \\
%%\!\!\!\mathrm{PL7}:0.623011u + 0.159034v - 0.765876w - 0.101688 = 0,& \nonumber \\
%%\!\!\!\mathrm{PL8}:0.996904u + 0.057250v - 0.053894w + 4.772559 = 0,& \nonumber
%%\end{numcases}
%%\end{small}
\!\!\!with $u\equiv\log{d_1}$, $v\equiv\log{d_2}$, and $w\equiv\log{d_3}$ and each line denoting a
plane in the $d_{1}$-$d_2$-$d_{3}$ space. The BCs for $\mathrm{FP}_{1,4,5,6}$ can be derived
by carrying out the analogous procedures, which
are presented in Eqs.~(\ref{Eq_condition_3})-(\ref{Eq_condition_7}) of Appendix~\ref{Sec_Appendix_BC_flows}.

\begin{table*}[htbp]
\centering
\caption{Collections of all FPs and their boundary conditions (BC).}
\label{table:fixpoint}
\vspace{0.2cm}
\renewcommand{\arraystretch}{1.5}
\begin{tabular}{>{\centering\arraybackslash}p{1cm}|>{\centering\arraybackslash}p{1.05cm}|>{\centering\arraybackslash}p{1.05cm}|>{\centering\arraybackslash}p{1.05cm}|>{\centering\arraybackslash}p{1.05cm}|>{\centering\arraybackslash}p{1.05cm}|>{\centering\arraybackslash}p{1.05cm}|>{\centering\arraybackslash}p{1.05cm}|>{\centering\arraybackslash}p{1.05cm}|>{\centering\arraybackslash}p{3.9cm}}
\hline\hline
$\mathrm{FPs}$ & $\lambda_{00}$ & $\lambda_{01}$ & $\lambda_{02}$ & $\lambda_{03}$
& $\lambda_{10}$ & $\lambda_{11}$ & $\lambda_{12}$ & $\lambda_{13}$ & $\mathrm{BCs}$\\
\hline
$\mathrm{FP}_1$ & -1.414 & 0 & 0 & -1.414 & 1.414 & 0 & 0 & 1.414 & $\mathrm{BC}$-$\mathrm{FP}_1$-$\mathrm{Case}$ $\mathrm{I}$~(\ref{Eq_condition_1}) or $\mathrm{BC}$-$\mathrm{FP}_1$-$\mathrm{Case}$ $\mathrm{II}$~(\ref{Eq_condition_3}) or $\mathrm{BC}$-$\mathrm{FP}_1$-$\mathrm{Case}$ $\mathrm{III}$~(\ref{Eq_condition_8})\\
\hline
$\mathrm{FP}_2$ & 0 & -0.288 & -0.288 & -3.979 & 0 & 0 & 0 & 0 & $\mathrm{BC}$-$\mathrm{FP}_2$-$\mathrm{Case}$ $\mathrm{I}$~(\ref{Eq_condition_2}) \\
\hline
$\mathrm{FP}_3$ & -0.005 & -3.958 & -0.564 & 0.036 & 0 & 0 & 0 & 0 & $\mathrm{BC}$-$\mathrm{FP}_3$-$\mathrm{Case}$ $\mathrm{II}$~(\ref{Eq_condition_4}) \\
\hline
$\mathrm{FP}_4$ & -0.002 & 0.049 & -0.573 & -3.957 & 0 & 0 & 0 & 0 & $\mathrm{BC}$-$\mathrm{FP}_4$-$\mathrm{Case}$ $\mathrm{II}$~(\ref{Eq_condition_5}) \\
\hline
$\mathrm{FP}_5$ & -1.394 & -0.107 & -1.488 & -0.113 & 1.444 & 0.073 & 1.354 & 0.077 & $\mathrm{BC}$-$\mathrm{FP}_5$-$\mathrm{Case}$ $\mathrm{II}$~(\ref{Eq_condition_6}) \\
\hline
$\mathrm{FP}_6$ & -1.367 & -0.255 & -1.857 & -0.256 & 1.450 & 0.102 & 1.179 & 0.102 & $\mathrm{BC}$-$\mathrm{FP}_6$-$\mathrm{Case}$ $\mathrm{II}$~~(\ref{Eq_condition_7}) \\
\hline
$\mathrm{FP}_7 $& -0.003 & -0.558 & -3.962 & 0.029 & 0 & 0 & 0 & 0 & $\mathrm{BC}$-$\mathrm{FP}_7$-$\mathrm{Case}$ $\mathrm{III}$~~(\ref{Eq_condition_9}) \\
\hline
$\mathrm{FP}_8$ & -0.001 & -0.525 & -0.010 & -3.962 & 0 & 0 & 0 & 0 & $\mathrm{BC}$-$\mathrm{FP}_8$-$\mathrm{Case}$ $\mathrm{III}$~~(\ref{Eq_condition_10}) \\
\hline
$\mathrm{FP}_9$ & -1.366 & -1.867 & -0.252 & -0.248 & 1.450 & 1.175 & 0.098 & 0.098 & $\mathrm{BC}$-$\mathrm{FP}_9$-$\mathrm{Case}$ $\mathrm{III}$~~(\ref{Eq_condition_11}) \\
\hline
$\mathrm{FP}_{10}$ & -1.384 & -1.384 & 0.001 & -0.001 & 1.424 & 1.424 & -0.001 & 0.001 & $\mathrm{BC}$-$\mathrm{FP}_{10}$-$\mathrm{Case}$ $\mathrm{III}$~~(\ref{Eq_condition_12}) \\
\hline\hline
\end{tabular}
\label{table:fixpoint}
\end{table*}

\subsubsection{Case $\mathrm{III}$ and discussions}

Furthermore, we consider the Case $\mathrm{III}$ characterized by $\frac{1}{4}d_2^2 < d_1^2$ and $d_3^2+\frac{1}{4}d_2^2>d_0^2$.
Analogous to the Case-$\mathrm{II}$, it involves all four structural parameters $d_{0,1,2,3}$.
Following the procedure in Sec.~\ref{Sec_Case-I}, we fix $d_0=0.1$, which does not qualitatively alter the fermionic couplings, and
vary the parameters $d_{1,2,3}$. A systematic numerical analysis
reveals that there exist five distinct stable regions for the critical values of fermion-fermion
interactions as illustrated in Figs.~\ref{Fig_FP1-7-III}-\ref{Fig_FP10-III} of Appendix~\ref{Sec_Appendix_BC_flows}.
After comparing with the FPs found in both Case $\mathrm{I}$ and Case $\mathrm{II}$, we identify
Figs.~\ref{Fig_FP1-7-III}(a)-(h) as the $\mathrm{FP}_1$~(\ref{Eq_FP1}), whereas the remaining regions in
Figs.~\ref{Fig_FP1-7-III}-\ref{Fig_FP10-III} give rise to four new FPs, which are denominated and
expressed as follows,
\begin{eqnarray}
\mathrm{FP}_7&\equiv&(\lambda_{00}, \lambda_{01}, \lambda_{02}, \lambda_{03},
\lambda_{10}, \lambda_{11}, \lambda_{12},\lambda_{13})|_{l=l_c}\nonumber\\
&=&(-0.003, -0.558, -3.962,  0.029,    0,    0,     0,     0),\label{Eq_FP7}
\end{eqnarray}
and $\mathrm{FP}_{8,9,10}$ are defined similar to $\mathrm{FP}_7$, approximately taking the form of
\begin{eqnarray}
\mathrm{FP}_8&\approx&(-0.001, -0.525, -0.010, -3.962,   0,    \nonumber\\
&&   0,     0,     0),\label{Eq_FP8}\\
\mathrm{FP}_9&\approx&(-1.366, -1.867, -0.252, -0.248,  1.450, \nonumber\\
&&1.175,  0.098,  0.098),\label{Eq_FP9}\\
\mathrm{FP}_{10}&\approx&(-1.384, -1.384, 0.001, -0.001, 1.424, \nonumber\\
&& 1.424, -0.001,  0.001).\label{Eq_FP10}
\end{eqnarray}
Paralleling the same approach employed in Case $\mathrm{III}$, we carry out the plane fitting analysis
to delineate the related regions associated with these FPs.  The resulting boundary conditions for Case III are presented in Eqs.~(\ref{Eq_condition_8})-(\ref{Eq_condition_12}) of Appendix~\ref{Sec_Appendix_BC_flows}.

To recapitulate, we uncover a richer fixed-point distribution for
both symmetric and asymmetric situations in the structural parameter space, with 10
distinct FPs emerging from three cases.  Table~\ref{table:fixpoint} catalogues
every fixed point together with its boundary conditions. Among them, $\mathrm{FP}_1$ emerges
as a unique common fixed point in both symmetric and asymmetric regimes. In addition, $\mathrm{FP}_{5,6}$
and $\mathrm{FP}_{9,10}$ share identical dominant interaction sets $(\lambda_{00},\lambda_{02},\lambda_{10},\lambda_{12})$ and
$(\lambda_{00},\lambda_{01},\lambda_{10},\lambda_{11})$, respectively, whose large magnitudes firmly control
the fixed-point structure and hence dominate the low-energy physics. The
remaining five FPs ($\mathrm{FP}_{2,3,4,7,8}$) are each governed by a single overwhelmingly strong coupling,
while their subdominant interactions vary widely.
In principle, these FPs, arising from fermion-fermion interactions, dictate the system's
low-energy behavior~\cite{Vafek2014PRB,Vafek2010PRB,Yang2010PRB,Roy2018PRX,Chubukov2016PRX,Zhang-Wang2025PRB}.
We are going to examine the potential phase transitions accompanied by the FPs
in the forthcoming section (Sec.~\ref{Sec_phase_transition}).

\begin{figure}[htbp]
\centering
\includegraphics[width=3.25in]{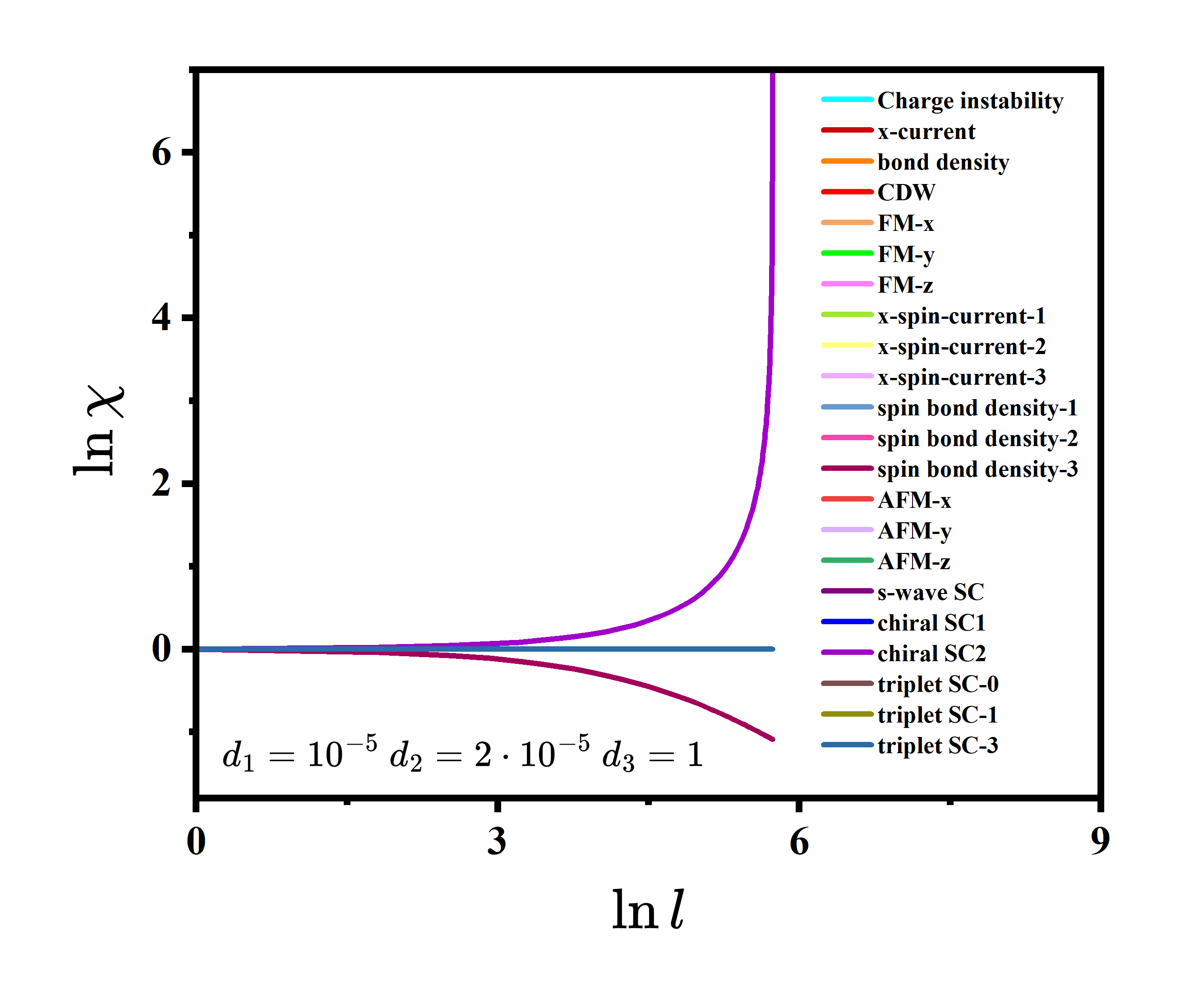}
\vspace{-0.5cm}
\caption{(Color online) Schematic illustration of the transition from
the QBCP semimetallic state to the $X$ phase as the system approaches
a given FP at the critical energy scale $l_c$. Hereby, the $X$ phase denotes the symmetry-broken
phase associated with the leading instability of a given FP.}
\label{figure:shiyitu2}
\end{figure}

\section{Phase transitions driven by four-fermion interactions}\label{Sec_phase_transition}

\begin{table}[htbp]
\centering
\caption{Twelve sorts of candidate instabilities induced by fermion-fermion interactions~\cite{Vafek2010PRB,Vafek2014PRB}.}\label{table:xiangbian}
\vspace{0.2cm}
\renewcommand{\arraystretch}{1.5} % Ôö¼ÓÐиßΪĬÈϵÄ1.5 ±¶
\begin{tabular}{>{\centering\arraybackslash}p{1.7cm}|>{\centering\arraybackslash}p{3.2cm}|>{\centering\arraybackslash}p{2.6cm}}  % ÉèÖÃÁÐ¿í²¢¾ÓÖÐ
\hline\hline
\makecell{Order \\ parameters} & \makecell{Vertex matrixes \\ of fermionic bilinears} & \makecell{Potential phases \\ } \\
\hline
$\Delta_1^c$ & $\mathcal{M}_1^c=\tau_0\otimes\sigma_0$ & Charge instability\\
\hline
$\Delta_2^c$ & $\mathcal{M}_2^c=\tau_0\otimes\sigma_1$ & $x$-current\\
\hline
$\Delta_3^c$ & $\mathcal{M}_3^c=\tau_0\otimes\sigma_2$ & bond density\\
\hline
$\Delta_4^c$ & $\mathcal{M}_4^c=\tau_0\otimes\sigma_3$ & charge density wave\\
\hline
$\vec{\Delta}_1^s$ & $\vec{\mathcal{M}}_1^s=\vec{\tau}\otimes\sigma_0$ & Ferromagnet\\
\hline
$\vec{\Delta}_2^s$ & $\vec{\mathcal{M}}_2^s=\vec{\tau}\otimes\sigma_1$ & $x$-spin-current\\
\hline
$\vec{\Delta}_3^s$ & $\vec{\mathcal{M}}_3^s=\vec{\tau}\otimes\sigma_2$ & spin bond density\\
\hline
$\vec{\Delta}_4^s$ & $\vec{\mathcal{M}}_4^s=\vec{\tau}\otimes\sigma_3$ & AFM\\
\hline
$\Delta_1^{\mathrm{pp}}$ & $\mathcal{M}_1^{\mathrm{pp}}=\tau_2\otimes\sigma_3$ & $s$-wave SC\\
\hline
$\Delta_2^{\mathrm{pp}}$ & $\mathcal{M}_2^{\mathrm{pp}}=\tau_2\otimes\sigma_1$ & chiral SC1\\
\hline
$\Delta_3^{\mathrm{pp}}$ & $\mathcal{M}_3^{\mathrm{pp}}=\tau_2\otimes\sigma_0$ & chiral SC2\\
\hline
$\Delta_4^{\mathrm{pp}}$ & $\mathcal{M}_4^{\mathrm{pp}}=\tau_{0,1,3}\otimes\sigma_2$ & triplet SC\\
\hline\hline
\end{tabular}
\end{table}

As established in Sec.~\ref{Sec_fate_ff}, the interplay between microscopic structural parameters and
fermion-fermion interactions yields 10 stable FPs whose locations in parameter space are catalogued
in Table~\ref{table:fixpoint}. Figure~\ref{Fig_evolution} reveals a critical divergence of the fermionic couplings in
the vicinity of each FP, signaling a distinct instability and the onset of a symmetry-broken phase~\cite{Vafek2012PRB,Vafek2014PRB,Wang2017PRB,Maiti2010PRB,Altland2006Book,Vojta2003RPP,
Metzner2000PRL,Halboth2000RPB,Eberlein2014PRB,Chubukov2012ARCMP,Nandkishore2012NP,Chubukov2016PRX,Roy2018PRX,Wang2020NPB}.
Figure~\ref{figure:shiyitu2} schematically illustrates the phase transition. Starting from the
QBCP semimetallic phase at $l=0$, the system evolves under RG flow and, and upon approaching a FP,
transitions into an ``$X$ phase." In this sense, the associated phase transitions are dictated by
the FPs catalogued in Table~\ref{table:fixpoint}.

This therefore motivates us to investigate which symmetry-breaking patterns
and concomitant phase transitions dominate near these FPs. Resolving this is instructive to determine the emergent quantum phases
governing the low-energy behavior of 2D kagom\'{e} QBCP materials.

\begin{figure}[htbp]
\centering
\includegraphics[width=3.0in]{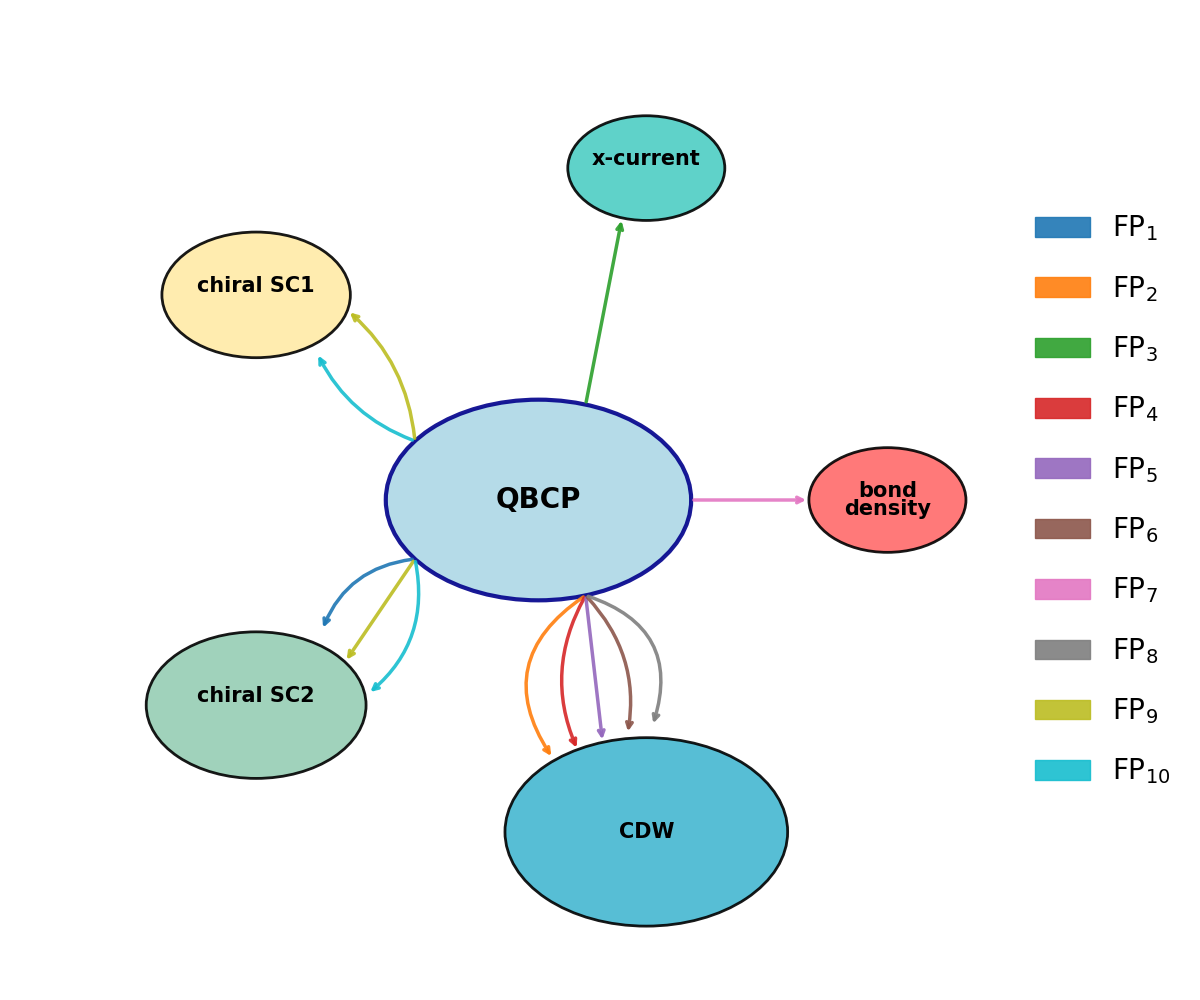}
\vspace{-0.5cm}
\caption{(Color online) Schematic depiction of the dominant instabilities associated with each distinct class of FPs
(further details are provided in Table~\ref{table:xiangbian}).}
\label{figure:abcdefghij}
\end{figure}

\subsection{Source terms}\label{Subsection_sourc_terms}

To this end, we adopt the approach of Refs.~\cite{Vafek2010PRB,Vafek2014PRB,Roy2018PRX,Roy2009.05055}
and introduce external symmetry-breaking perturbations encoded in the source terms,
\begin{eqnarray}
S_{\mathrm{sou}}&=&\int\mathrm{d}\tau\int\mathrm{d^2}\mathbf{x}
\sum_{i=1}^4(\Delta_{i}^{c}\Psi^{\dag}\mathcal{M}_{i}^{c}\Psi
+\vec{\Delta}_{i}^{s}\cdot\Psi^{\dag}\vec{\mathcal{M}}_{i}^{s}\Psi)\nonumber\\
&&+\int\mathrm{d}\tau\int\mathrm{d^2}\mathbf{x}
\sum_{i=1}^4(\Delta_{i}^{\mathrm{pp}}\Psi^{\dag}\mathcal{M}_{i}^{\mathrm{pp}}\Psi+\mathrm{h.c}),\label{Eq_source_term}
\end{eqnarray}
where the vertex matrices $\mathcal{M}_i^{c/s}$ with $i=1,2,3,4$ represent charge/spin fermion bilinears in
the particle-hole channel, while $\mathcal{M}_i^{\mathrm{pp}}$ correspond to the particle-particle bilinears.
In addition, the couplings $\Delta_i^{c/s}$ and $\Delta_i^{\mathrm{pp}}$ serve as the strengths of associated
fermion-source terms, which are relevant to the order parameters accompanied by corresponding symmetry breakings.
Table~\ref{table:xiangbian} catalogs the potential symmetry-allowed fermion bilinears and their associated phase transitions~\cite{Vafek2010PRB,Vafek2014PRB}.

As illustrated in Fig.~\ref{figure:shiyitu2} and discussed above,
FPs are inherently linked to specific instabilities that signify symmetry breakings.
To judge the concrete state of the ``$X$ phase" in Fig.~\ref{figure:shiyitu2},
it is necessary to evaluate and compare the susceptibilities $\chi_i$ corresponding to the potential
instabilities in the vicinity of each FP.
These are defined as the second derivatives of the free-energy density $f$ with respect to the source fields $\Delta_i$~\cite{Vafek2014PRB,Vafek2010PRB,Yang2010PRB,Roy2018PRX,Chubukov2016PRX,Zhang-Wang2025PRB},
\begin{equation}
\chi^{c,s,\mathrm{pp}}_i \equiv -\frac{\partial^2 f}{\partial \Delta^{c,s,\mathrm{pp}}_i
\partial \Delta^{c,s,\mathrm{pp}}_i} \Big|_{\Delta^{c,s,\mathrm{pp}}_i=0},\label{Eq_chi_i}
\end{equation}
with $i=1,2,3,4$. In principle, divergent susceptibilities herald the onset of ordered phases~\cite{Metzner2000PRL,Halboth2000RPB,Maiti2010PRB,Vafek2012PRB,Vafek2014PRB,Chubukov2016PRX,Roy2018PRX}.
To track their energy-scale evolution toward each FP, we add the source terms~(\ref{Eq_source_term})
into the effective action~(\ref{Eq_S_eff}) and then derive the related
RG equations for $\Delta^{c,s,\mathrm{pp}}_i$ by paralleling the analogous procedure in Sec.~\ref{Sec_RG_analysis},
which are explicitly provided in Appendix~\ref{Appendix_1L-Delta_i}.

To proceed, we compute energy-dependent susceptibilities at each FP by combining Eq.~(\ref{Eq_chi_i})
and the coupled RG equations for fermionic couplings (\ref{Eq_RG_flows}) and
source terms (\ref{Eq_source_1})-(\ref{Eq_source_2}). All the dominant instabilities that are identified as the channels with the most strongly divergent susceptibilities~\cite{Vafek2012PRB,Vafek2014PRB,Wang2017PRB,Maiti2010PRB,Altland2006Book,Vojta2003RPP,
Metzner2000PRL,Halboth2000RPB,Eberlein2014PRB,Chubukov2012ARCMP,Nandkishore2012NP,Chubukov2016PRX,
Roy2018PRX,Nandkishore2008.05485} are determined from the candidate phases in Table~\ref{table:xiangbian}. Results are presented sequentially below.

\begin{figure}[htbp]
\centering
\includegraphics[width=3in]{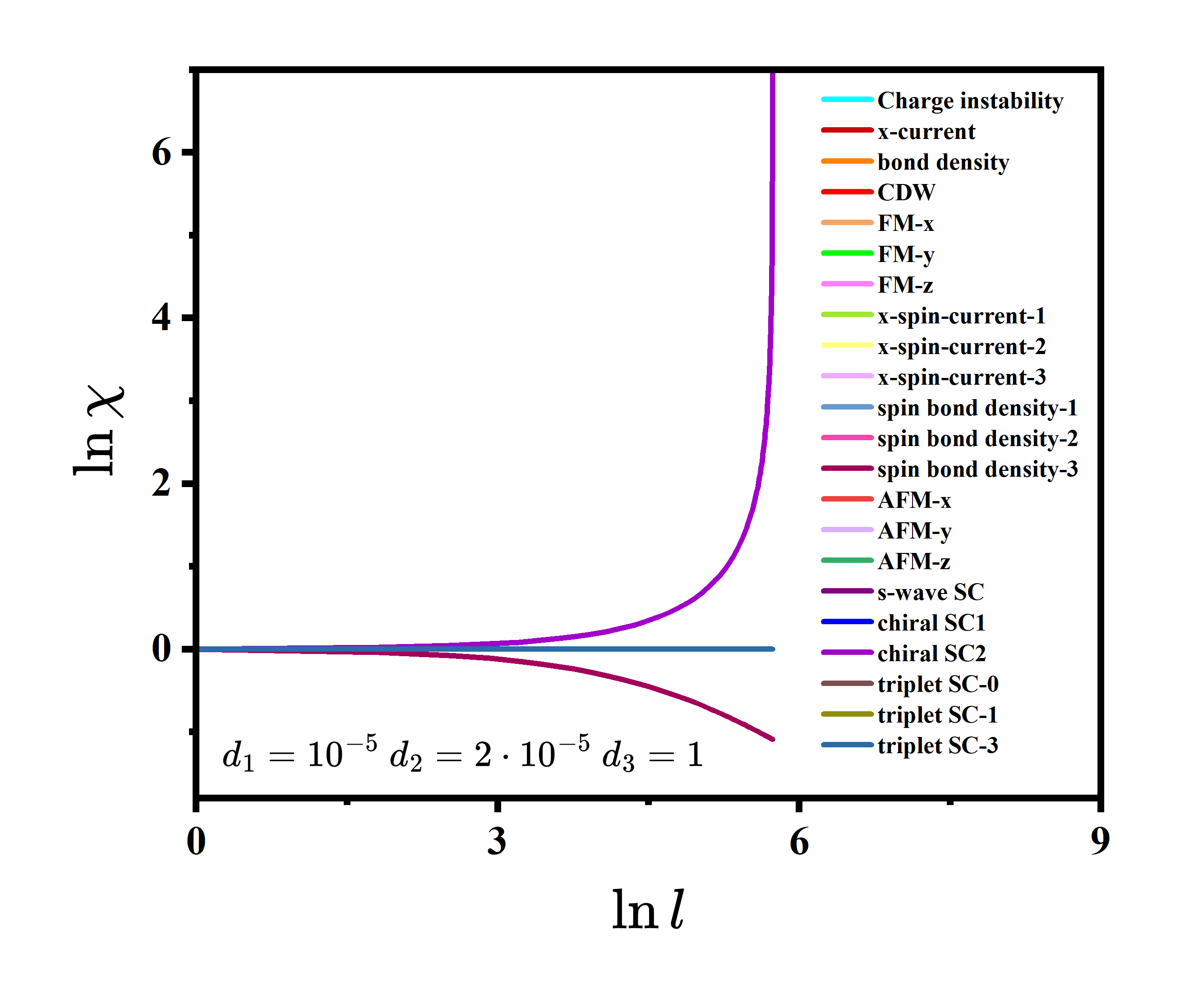}
\vspace{-0.5cm}
\caption{(Color online) Energy-dependent susceptibilities of candidate instabilities near $\mathrm{FP}_1$ in Case $\mathrm{I}$
(the tendencies are similar for $\mathrm{FP}_1$ in both Case $\mathrm{II}$ and Case $\mathrm{III}$).}
\label{Fig_chi_1}
\end{figure}

\begin{figure*}[htbp]
\centering
\subfigure[]{
\includegraphics[width=2.3in]{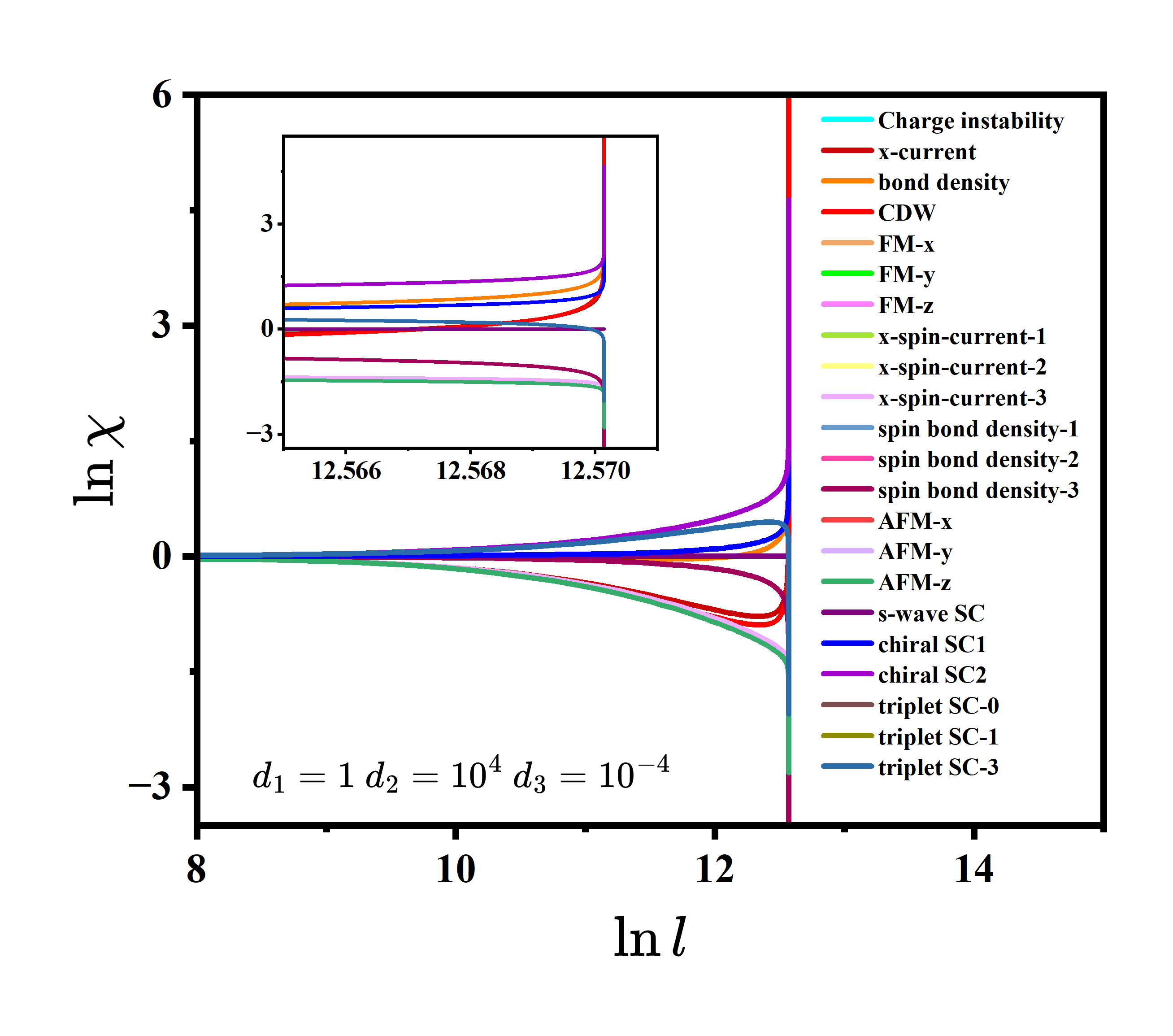}
}\hspace{-0.5cm}
\subfigure[]{
\includegraphics[width=2.3in]{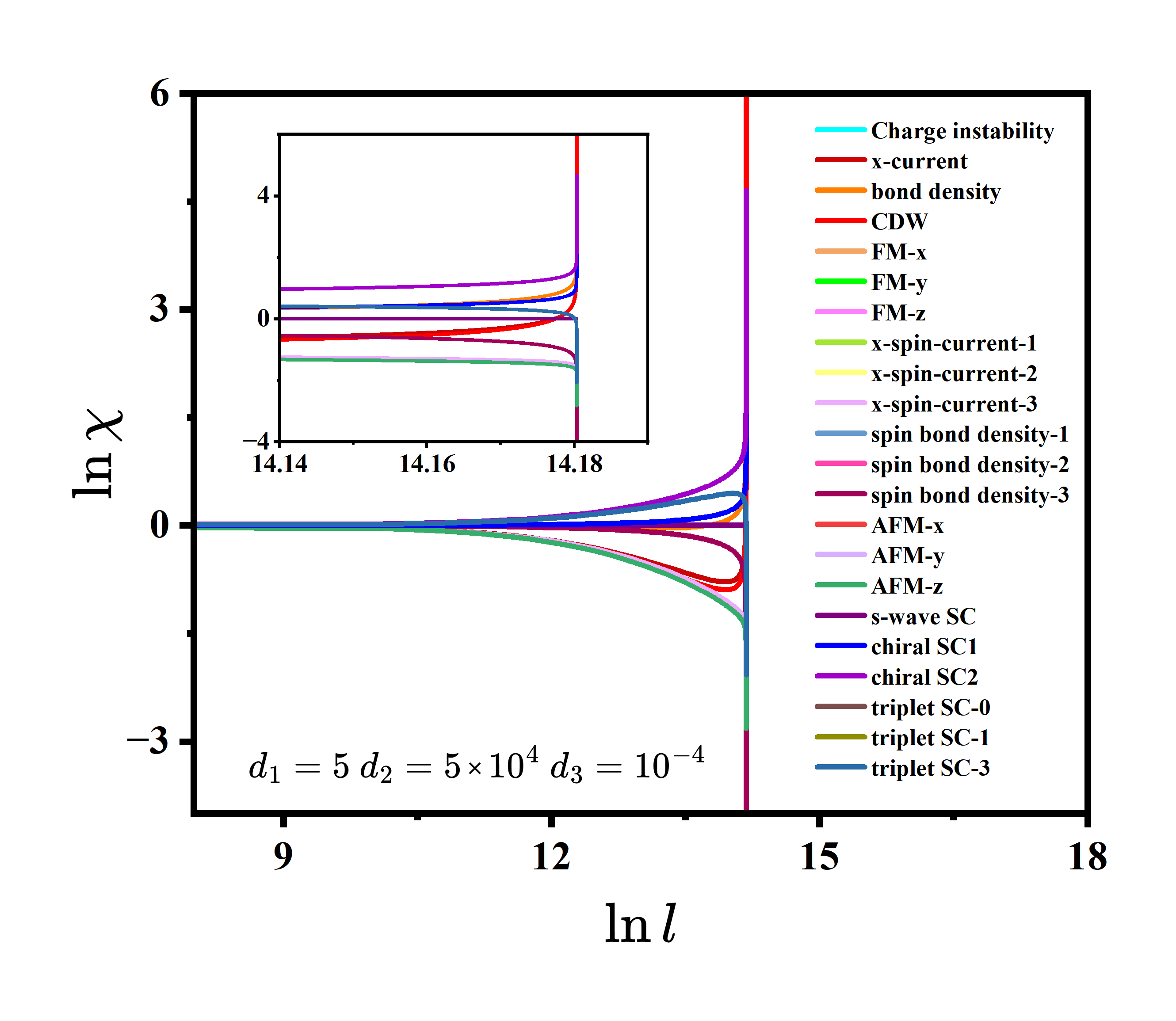}
}\hspace{-0.5cm}
\subfigure[]{
\includegraphics[width=2.3in]{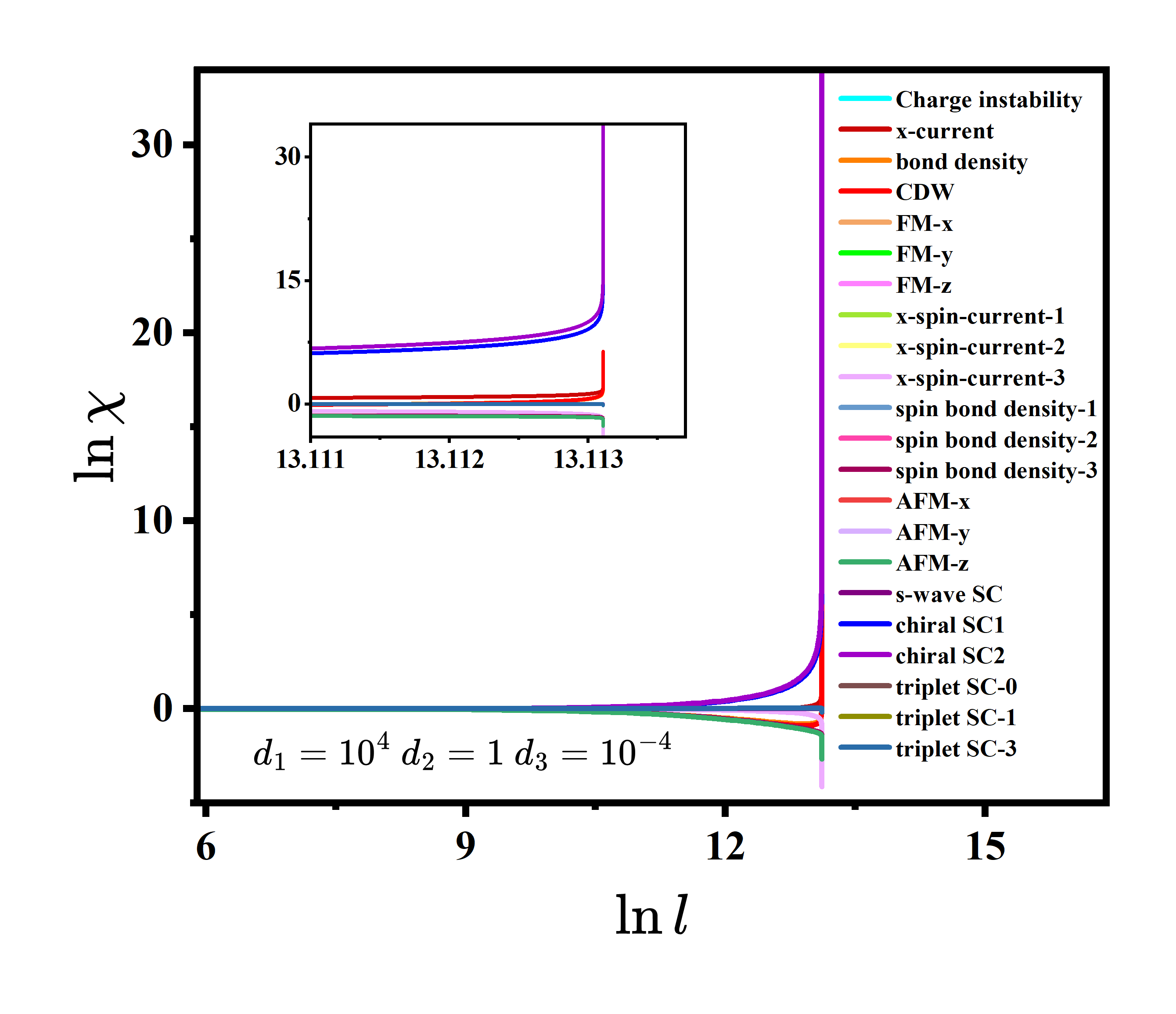}
}
\\ \vspace{-0.3cm}
\subfigure[]{
\includegraphics[width=2.3in]{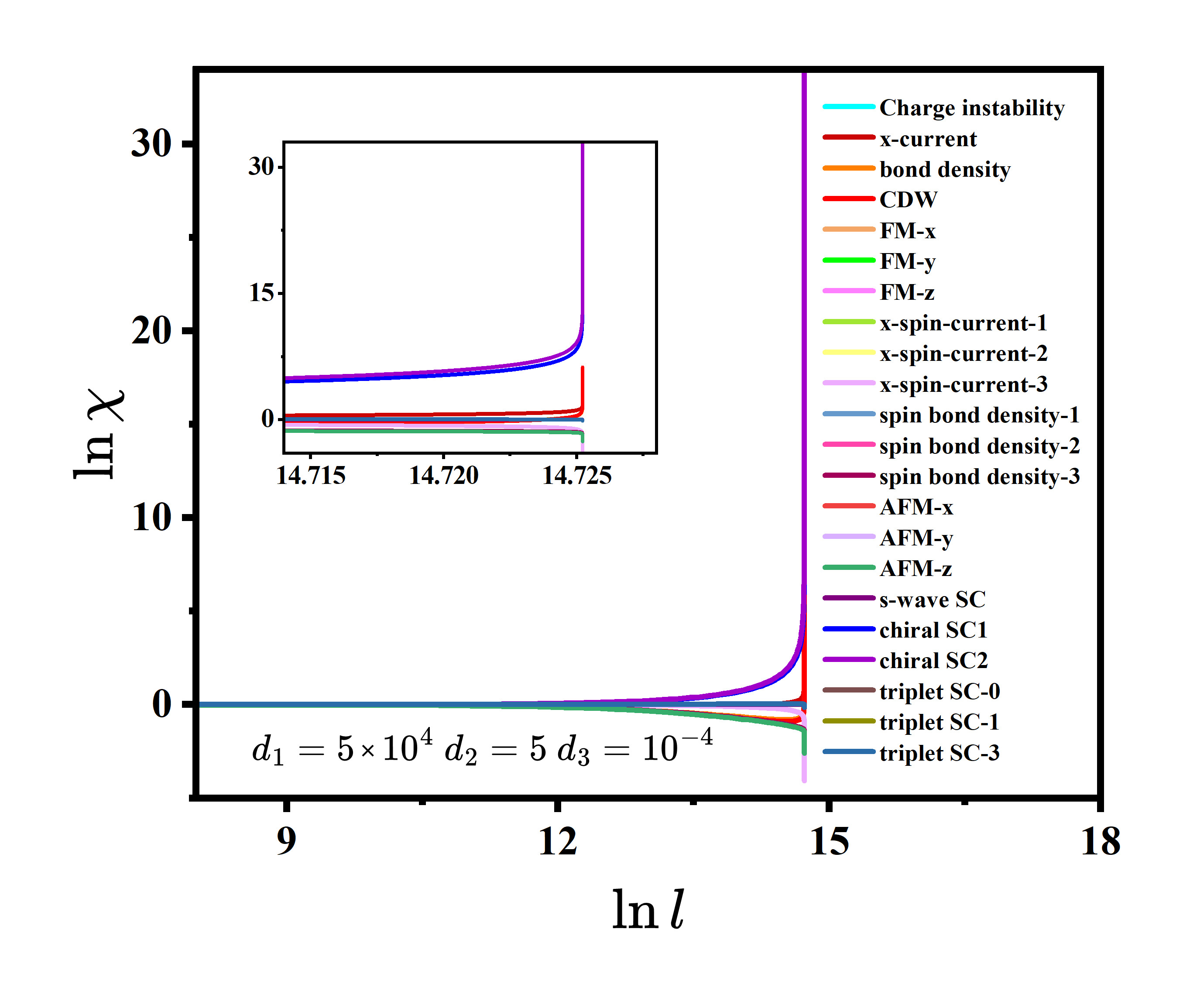}
}\hspace{-0.5cm}
\subfigure[]{
\includegraphics[width=2.3in]{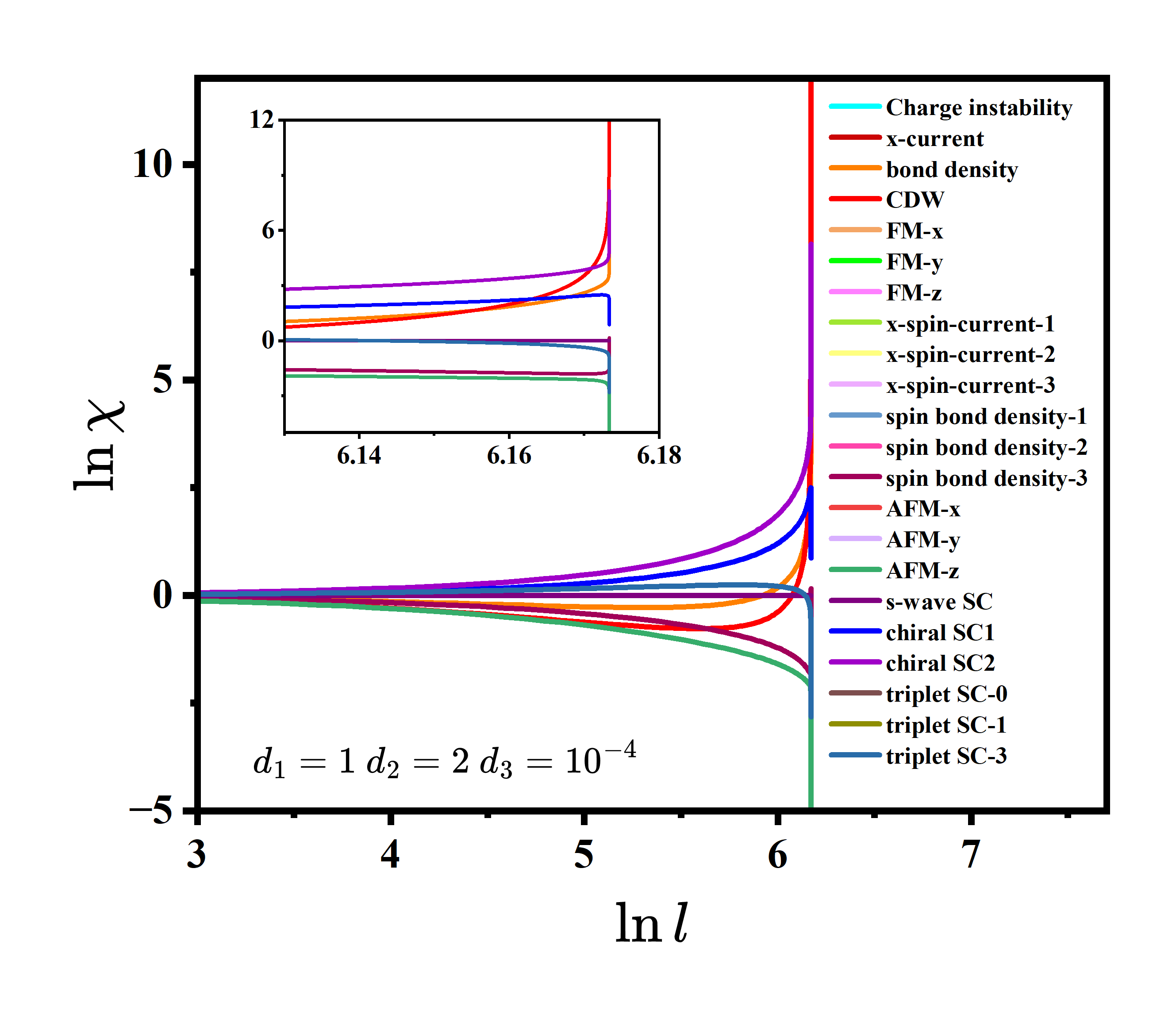}
}\hspace{-0.5cm}
\subfigure[]{
\includegraphics[width=2.3in]{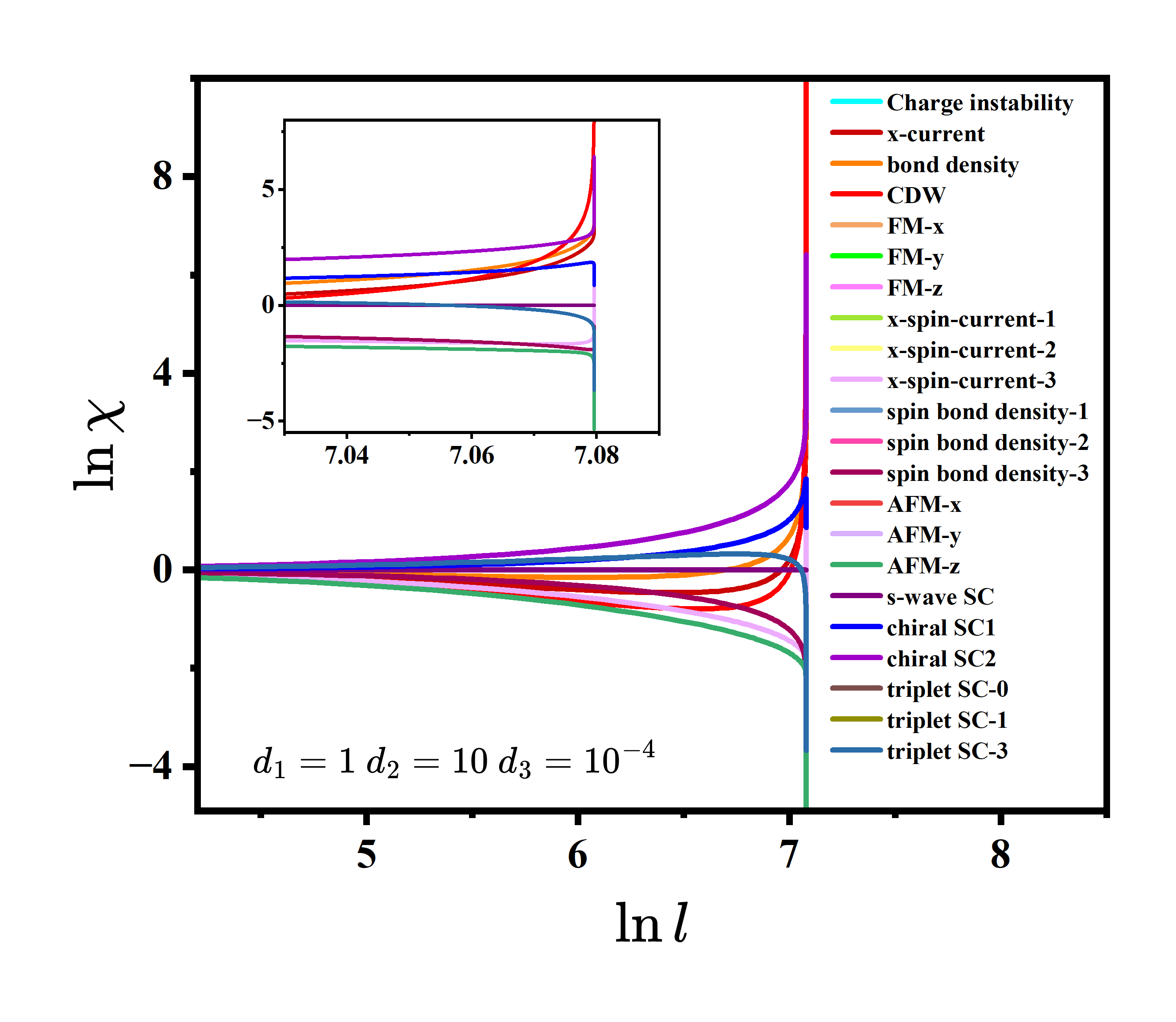}
}
\\ \vspace{-0.3cm}
\subfigure[]{
\includegraphics[width=2.3in]{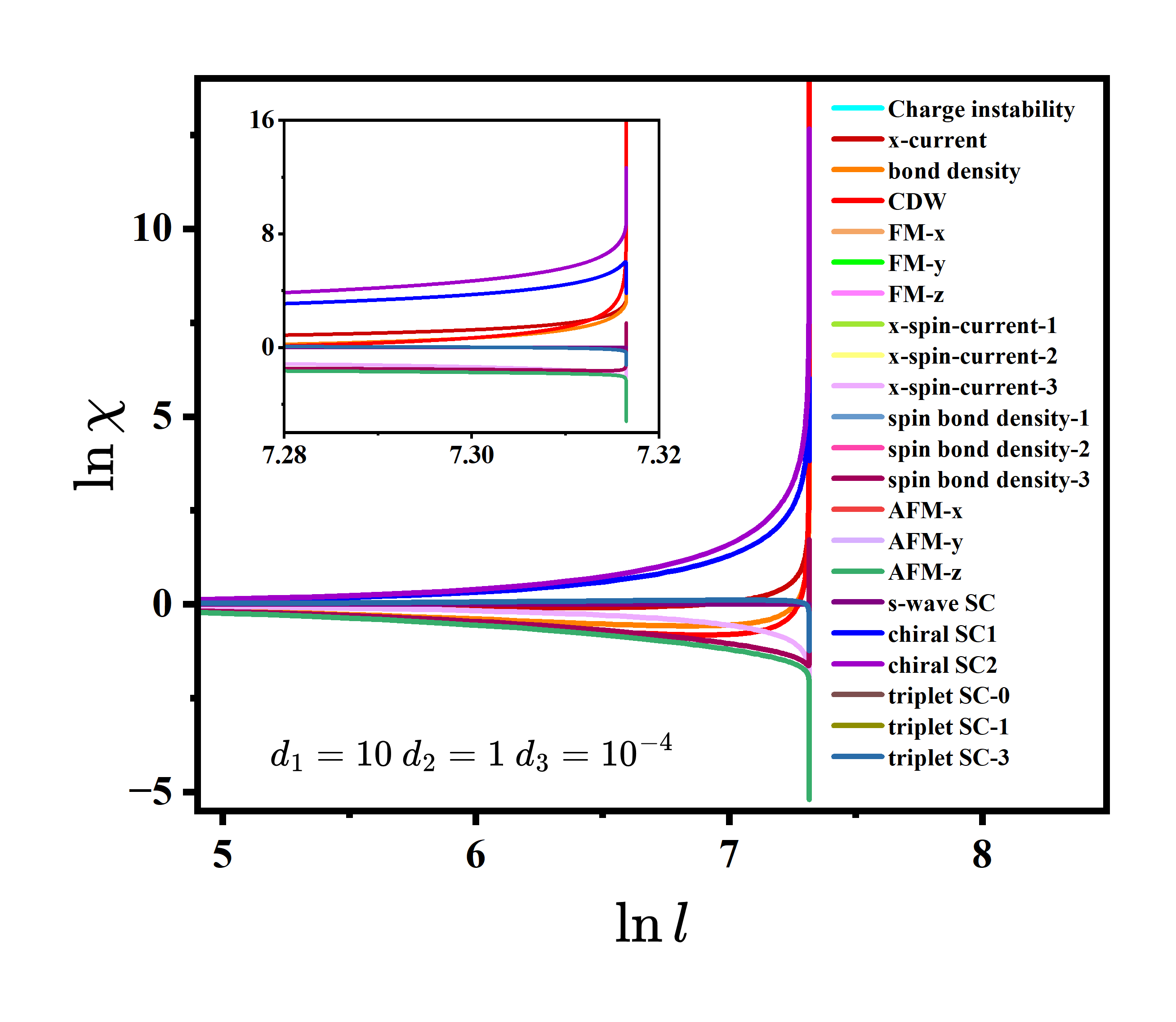}
}\hspace{-0.5cm}
\subfigure[]{
\includegraphics[width=2.3in]{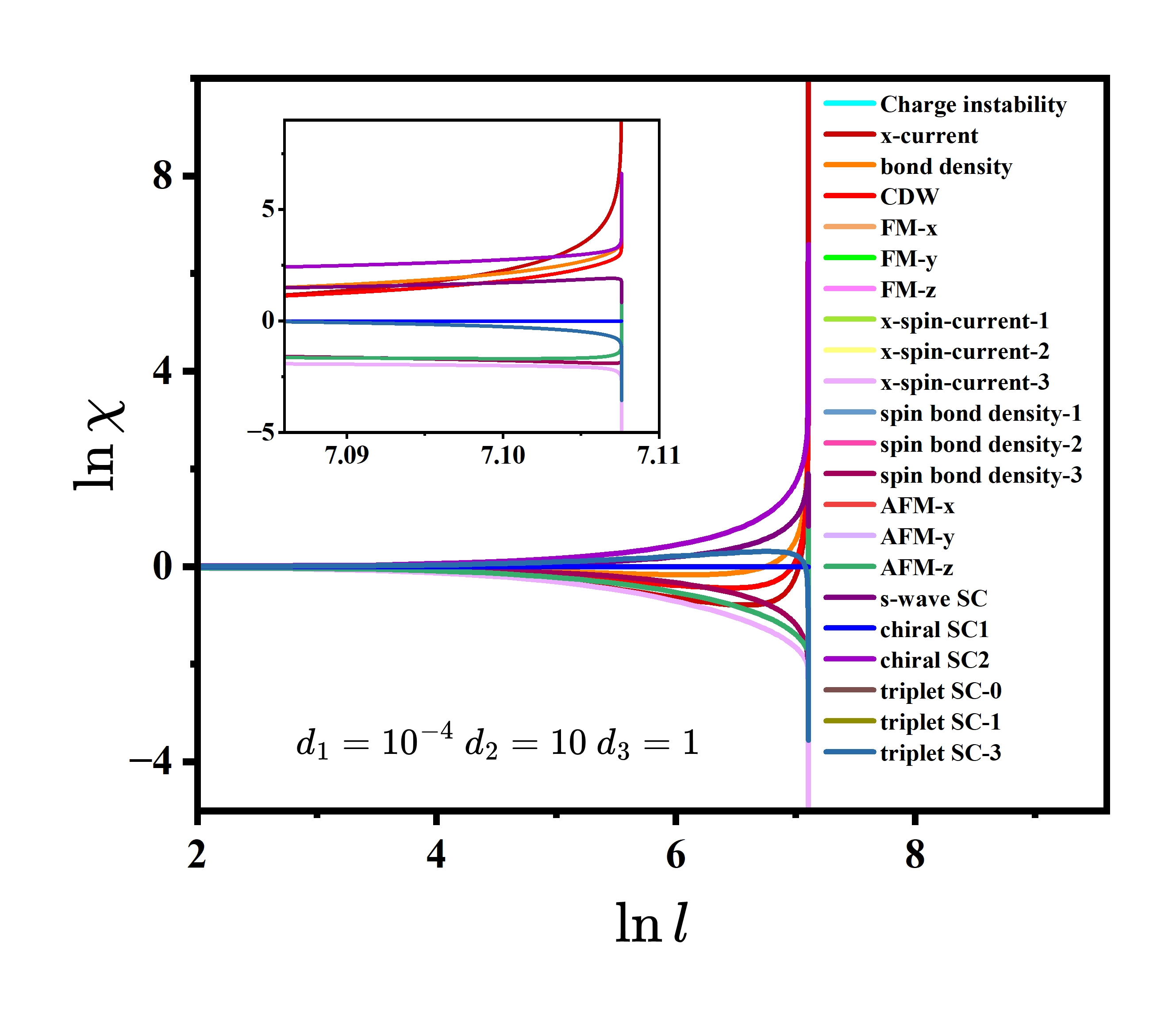}
}\hspace{-0.5cm}
\subfigure[]{
\includegraphics[width=2.3in]{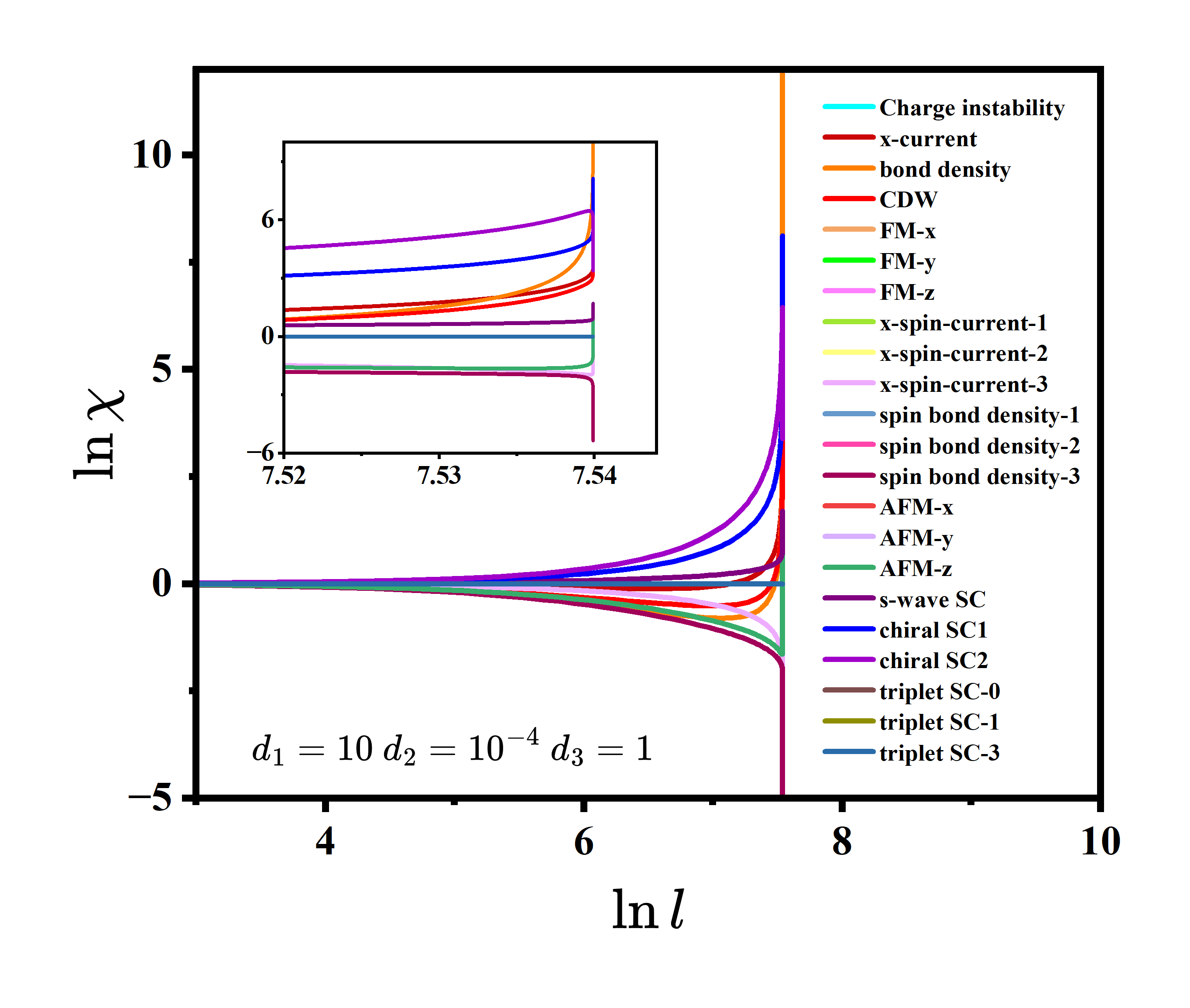}
}
\\ \vspace{-0.3cm}
\caption{(Color online) Energy-dependent susceptibilities of candidate instabilities for Group-$\mathrm{B}$ near
(a) $\mathrm{FP}_5$, (b) $\mathrm{FP}_6$, (c) $\mathrm{FP}_9$, and (d) $\mathrm{FP}_{10}$,
for Group-$\mathrm{C}$ near (e) $\mathrm{FP}_2$, (f) $\mathrm{FP}_4$, and (g) $\mathrm{FP}_8$,
and for Group-$\mathrm{D}$ near (h) $\mathrm{FP}_3$ and (i) $\mathrm{FP}_7$.}
\label{Fig_chi_2}
\end{figure*}

\subsection{Instability analysis: An illustrative example}

We now proceed to identify the phase transitions favored in the vicinity of each FP by evaluating the susceptibilities of the candidate states listed in Table~\ref{table:xiangbian}. In accordance with standard criteria, the instability associated with the most rapidly diverging susceptibility is considered as the leading instability and determines the ordered phase.
Consequently, different fixed points yield distinct leading instabilities, giving rise to different classes of phase transitions, namely, the ``$X$ phases" illustrated in Fig.~\ref{figure:shiyitu2}. The FPs are obtained from the RG equations~(\ref{Eq_RG_flows}), which depend on fermion¨Cfermion interactions and the initial conditions of the QBCP system listed in Table~\ref{table:fixpoint}. By tuning these initial conditions, one can alter the RG flow trajectory and direct the system toward different FPs, thereby realizing different leading instabilities and their associated ordered phases.

As a concrete example, let us consider the $\mathrm{FP}_3$.
At first, supposing the microscopic structural parameters $d_1$, $d_2$, and $d_3$ that are dependent upon the materials point are restricted to the boundary condtion $\mathrm{BC}$-$\mathrm{FP}_3$-$\mathrm{case}$ $\mathrm{II}$~(\ref{Eq_condition_4}).
Solving the RG equations~(\ref{Eq_RG_flows}) with these initial conditions reveals that the system flows inevitably
toward fixed point $\mathrm{FP}_3$~(\ref{Eq_FP3}). Accordingly, as shown schematically in Fig.~\ref{figure:shiyitu2},
the system would undergo a phase transition from the QBCP semimetallic phase into the ordered $X$
phase dictated by $\mathrm{FP}_3$.

Next, we turn to identifying the $X$ phase in Fig.~\ref{figure:shiyitu2}. Following the the strategy
outlined in Sec.~\ref{Subsection_sourc_terms}, we compute the susceptibilities for all candidate states listed in Table~\ref{table:xiangbian} within the RG framework. After a detailed numerical analysis, we obtain the energy
scale-dependent susceptibilities illustrated in Fig.~\ref{Fig_chi_2}(h). The results indicate that the $x$-current
phase exhibits the most rapidly diverging susceptibility, establishing it as the leading instability.
Accordingly, for $\mathrm{FP}_3$, the ``$X$ phase" in Fig.~\ref{figure:shiyitu2} corresponds to the $x$-current
ordered state. In addition, three subleading instabilities including chiral SC2, bond density, and CDW
emerge as competing orders. These subordinate states may become leading instabilities in the presence of
impurities or other perturbations acting on the QBCP system.

Paralleling the analysis for $\mathrm{FP}_3$, we can examine the related phase transitions for all the other FPs.
To present these results more intuitively, Fig.~\ref{figure:abcdefghij} offers a schematic illustration of the main results.
In what follows, we are going to address each case in detail.

\subsection{Results and discussions}\label{Sec_results_chi}

To simplify the analysis, we cluster the 10 FPs collected in Table~\ref{table:fixpoint} into
four distinct groups. Specifically, Group-$\mathrm{A}$ consists of $\mathrm{FP}_1$ which appears in all three cases,
Group-$\mathrm{B}$ includes  $\mathrm{FP}_{5,6}$ and $\mathrm{FP}_{9,10}$ where each pair has an identical interaction structure, Group-$\mathrm{C}$ contains $\mathrm{FP}_2$, $\mathrm{FP}_4$, and $\mathrm{FP}_8$ characterized by equivalent
dominant interaction strengths, and Group-$\mathrm{D}$ corresponds to the remaining $\mathrm{FP}_3$ and $\mathrm{FP}_7$.
Then, we are going to judge the favorable phase transitions around these FPs by evaluating the susceptibilities of
candidate states in Table~\ref{table:xiangbian}. While only the leading instability (strongest divergence) condenses into a phase, subleading instabilities remain physically significant as they may dominate under varying external conditions. As a consequence,
we report both leading and subleading instabilities.

The leading instability that exhibits the strongest divergent susceptibility
is generally relevant to the most probable phase near a given fixed point~\cite{Metzner2000PRL,Halboth2000RPB,Maiti2010PRB,Vafek2012PRB,Vafek2014PRB,Chubukov2016PRX,Roy2018PRX}.
Subleading instabilities, while secondary, remain essential as they may be the optimal
substitutes under appropriate external perturbations.
Since the fixed point itself dictates the basic physics, the $\mathrm{FP}_1$
displays the same doubly degenerate leading instabilities (i.e., $s$-wave SC and chiral SC2) in both Case $\mathrm{II}$
and Case $\mathrm{III}$ although the initial conditions shown in Table~\ref{table:fixpoint} are different (see Fig.~\ref{Fig_chi_1}).
%%In all three cases the sub-leading instability is CDW, whose susceptibility remains markedly smaller.
The subleading instability for all three cases corresponds to the CDW, with susceptibility significantly
smaller than those of $s$-wave SC and chiral SC2.
As to Group-$\mathrm{B}$, the numerical analysis shown in Fig.~\ref{Fig_chi_2}(a)-(b) indicates that
$\mathrm{FP}_5$ and $\mathrm{FP}_6$ share the CDW as their common leading instability. Likewise,
Fig.~\ref{Fig_chi_2}(c)-(d) suggest that $\mathrm{FP}_9$ and $\mathrm{FP}_{10}$ both display a
degenerate leading instability between chiral SC1 and SC2. This implies that the identical interaction
structures is prone to yielding identical instabilities. Turning to the subleading instabilities,
the $x$-current state is subordinate to the CDW near $\mathrm{FP}_5$ and $\mathrm{FP}_6$,
while $\mathrm{FP}_9$ and $\mathrm{FP}_{10}$ select the bond density and CDW as their subleading
instabilities, respectively.

With respect to Group-$\mathrm{C}$, Fig.~\ref{Fig_chi_2}(e)-(g) exhibit that the leading
instability again goes to the CDW at all three FPs, i.e., $\mathrm{FP}_2$, $\mathrm{FP}_4$, and $\mathrm{FP}_8$.
This confirms an essential role of the dominant interactions in determining the low-energy physics.
In this circumstance, the chiral SC2 is less strong to the CDW at all three FPs.
Additionally, the $x$-current and bond density are also present as the subleading instabilities
near $\mathrm{FP}_4$. As for Group-$\mathrm{D}$, we find from Fig.~\ref{Fig_chi_2}(h)-(i) that two FPs
exhibit distinct evolutions of susceptibilities. Specifically, $\mathrm{FP}_3$ and $\mathrm{FP}_7$
are equipped with the $x$-current and bond density as the dominant instabilities, respectively.
In addition, the chiral SC2, bond density, and CDW serve as three nearly degenerate subleading instabilities
near $\mathrm{FP}_3$, while the chiral SC1 is only a subordinate state compared to the bond
density around $\mathrm{FP}_7$.

\begin{table}[htbp]
\centering
\caption{The leading and subleading instabilities near distinct kinds of FPs~\cite{Roy2018PRX}.}
\vspace{0.2cm}
\renewcommand{\arraystretch}{1.5} % Ôö¼ÓÐиßΪĬÈϵÄ1.5 ±¶
\begin{tabular}{>{\centering\arraybackslash}p{1.0cm}|>{\centering\arraybackslash}p{3.3cm}|>{\centering\arraybackslash}p{3.2cm}}  % ÉèÖÃÁÐ¿í²¢¾ÓÖÐ
\hline\hline
\makecell{FPs} & \makecell{Leading Instability} & \makecell{Subleading Instability} \\
\hline
$\mathrm{FP}_1$ & $s$-wave SC, chiral SC2 & CDW\\
\hline
$\mathrm{FP}_2$ & CDW & chiral SC2\\
\hline
$\mathrm{FP}_3$ & $x$-current & chiral SC2, bond density, CDW\\
\hline
$\mathrm{FP}_4$ & CDW & chiral SC2, bond density, $x$-current\\
\hline
$\mathrm{FP}_5$ & CDW & $x$-current\\
\hline
$\mathrm{FP}_6$ & CDW & $x$-current\\
\hline
$\mathrm{FP}_7$ & bond density & chiral SC1\\
\hline
$\mathrm{FP}_8$ & CDW & chiral SC2\\
\hline
$\mathrm{FP}_9$ & chiral SC2, chiral SC1 & bond density, CDW\\
\hline
$\mathrm{FP}_{10}$ & chiral SC2, chiral SC1 & bond density, CDW\\
\hline\hline
\end{tabular}
\label{table:leading}
\end{table}

To wrap up, we examine both leading and subleading instabilities from the candidates in Table~\ref{table:xiangbian}
as approaching all 10 distinct FPs cataloged in Table~\ref{table:fixpoint}. The basic results are presented
in Table~\ref{table:leading}. These would be helpful to improve our understanding of the low-energy behavior of
2D kagom\'{e} QBCP materials.

\section{Comparisons and discussions}

To properly highlight and clarify the specific contributions and key findings of the present work,
it is necessary to provide a comparative analysis with prior studies. To this end, we present in
this section a comparison from several perspectives, including theoretical strategies
and principal results. Additionally, we provide a brief overview of recent experimental progress and discuss the
potential investigation in the future.

\subsection{Theoretical strategies}

The theoretical framework for two-dimensional QBCP systems was initially developed by
Sun \emph{et al}.~\cite{Fradkin2008PRB,Fradkin2009PRL}, who studied TRS-preserving systems with $C_4$  or $C_6$
rotational symmetry. Their generic two-band Hamiltonian revealed that short-range repulsive interactions are marginally relevant, leading to spontaneous symmetry breaking into QAH or nematic phases. Murray and Vafek~\cite{Vafek2014PRB} extended this work by introducing asymmetry parameters and systematically parametrizing fermion-fermion interactions. A microscopic realization on the kagom\'{e} lattice was provided by Zhu \emph{et al}.~\cite{Zhu2016PRL}, who numerically confirmed spontaneous TRS breaking and the emergence of the QAH phase. Wang \emph{et al}.~\cite{Wang2017PRB} later incorporated disorder effects into the TRS-preserving systems, while the interplay between asymmetry parameters and disorder was further explored in subsequent work~\cite{Wang2020-arxiv}.  More recently, the effects of fermion-fermion interactions on the kagom\'{e} QBCP system, where TRS is explicitly broken while preserving rotational and particle¨Chole symmetries, has been investigated in Ref.~\cite{Zhai2021NPB}.

Compared with these earlier works, within the present work we focus on the 2D TRS-broken QBCP system that has been relatively less studied, and we consider a more general low-energy Hamiltonian~(\ref{Eq_H0_density}) that allows time-reversal, rotational, and particle-hole symmetries to be broken simultaneously. By imposing the constraints $d_0 = 0$ (restoring particle-hole symmetry) and $d_2 = 2d_1$ (restoring rotational symmetry), it reduces to a special case studied in Ref.~\cite{Zhai2021NPB}. The present framework therefore incorporates earlier kagom\'{e} QBCP models into a unified parameter space and, accordingly, yields more kinds of fermion-fermion interactions with a richer phase structure.

In addition, we go beyond the mean-field theory and employ the Wilsonian RG
approach~\cite{Shankar1994RMP,Wilson1975RMP,Polchinski9210046} which
takes into account all one-loop corrections to build the coupled RG equations of all interaction parameters.
This allows us to make a consistent treatment of interplay of all kinds of interactions, seek all
possible FPs in the parameter space, and particularly identify all underlying dominant
instabilities in the low-energy regime, thereby providing a more complete picture of the low-energy physics.

\subsection{Principal results}

As mentioned in Sec.~\ref{Sec_model}, two-dimensional QBCP systems can be broadly classified into two categories depending on whether time-reversal symmetry (TRS) is preserved (checkerboard lattice) or broken (kagom\'{e} lattice).

The majority of previous investigations have focused on
TRS-preserving systems. In the spinless case, Sun \emph{et al.}~\cite{Fradkin2009PRL}
identified the quantum anomalous Hall (QAH) phase as the dominant instability.
For the spin-$1/2$ case, in contrast, a richer competition emerges among multiple ordered states,
including QAH, quantum spin Hall (QSH), nematic, and nematic spin-nematic (NSN) phases.
Murray and Vafek~\cite{Vafek2014PRB} analyzed the RG flow of symmetry-allowed fermion couplings,
identifying stable FPs corresponding to QAH and QSH instabilities, with the final ordering
determined by coupling-constant ratios. Wang \emph{et al.}~\cite{Wang2017PRB} later incorporated
disorder effects and showed that disorder can modify the fixed-point structure and influence the
dominant phase. Further work~\cite{Wang2020-arxiv} extending this considered
the rotational and particle¨Chole asymmetry systems, uncovering the energy-scale dependence of
these asymmetry parameters and thereby arriving at a more general fixed-point structure.

As for the TRS-broken kagom\'{e} QBCP systems, they have attracted less attention.
In the TRS-breaking kagom\'{e} QBCP case, a symmetry-constrained scenario was considered
in which rotational and particle¨Chole symmetries were still preserved. This analysis
identified two stable fixed points together with their corresponding leading
instabilities~\cite{Zhai2021NPB}.

In contrast, the present work extends the model to a general asymmetric
QBCP system on the kagom\'{e} lattice, in which TRS, rotational symmetry, and
particle¨Chole symmetry are all relaxed simultaneously. By solving the RG flow
equations for all four-fermion couplings, we identify 10 stable FPs, determine
their corresponding initial-condition regions and associated leading and subleading
instabilities as detailed in Sec.~\ref{Sec_fate_ff} and  Sec.~\ref{Sec_phase_transition}.
Compared with the two FPs obtained under previously imposed high-symmetry constraints~\cite{Zhai2021NPB},
the present results reveal a markedly richer fixed-point structure and phase diagram,
indicative of a more intricate multi-fixed-point competition within the extended parameter space.
Importantly, in the appropriate high-symmetry limits, a subset of our results recovers the previously
reported kagom\'{e} QBCP behavior, whereas the additional fixed points and competing instabilities
arise exclusively in the fully generalized asymmetric parameter space considered here.

\subsection{Experimental progress}

In the nearly two decades since the theoretical description of 2D QBCPs was
introduced~\cite{Fradkin2008PRB, Fradkin2009PRL}, significant progress has been made on both
theoretical and computational fronts~\cite{Vafek2014PRB,Zhu2016PRL,Wang2017PRB,Wang2020-arxiv,Zhai2021NPB}.
Direct experimental realization of ideal QBCP systems, however, remains a considerable challenge~\cite{Yin2022Nature,Ye2018Nature,Ortiz2020PRL,Jiang2021NM,Cheng2023NL}.
Notably, to the best of our knowledge, TRS-breaking kagom\'{e} QBCP system, which is most directly
pertinent to the present work, has not yet been directly realized experimentally. Nevertheless,
broader experimental efforts relevant to 2D QBCP physics are ongoing, driven by platforms such
as checkerboard-type 2D systems, artificially tunable lattices, and the rapidly expanding family
of kagom\'{e} materials~\cite{Hu2023NL,Brown2022Science,Cai2026QF}.

Important experimental progress has already been achieved in checkerboard-type
2D platforms relevant to QBCP physics. Monolayer Cu$_2$N, for instance, has been shown
to realize a checkerboard lattice, with QBCP-related features observed near the Fermi
level~\cite{Hu2023NL}. In parallel, artificially tunable platforms have enabled more
direct experimental probes of QBCP. A representative example involves ultracold atoms
in a honeycomb optical lattice, where the QBCP was systematically investigated and particularly the winding
numbers associated with both linear and QBCP dispersions were experimentally extracted, providing unambiguous
evidence of QBCP in a highly controllable artificial lattice~\cite{Brown2022Science}.
Turning to the kagom\'{e} platform itself, although direct experimental realization of the
idealized TRS-breaking kagom\'{e} QBCP considered in the present work remains lacking, experimental
research on kagom\'{e} materials and kagom\'{e} thin films has developed rapidly in recent years.
These systems have now become versatile platforms for exploring a wide range of phenomena, including
flat bands, Dirac cones, van Hove singularities, anomalous Hall effects, charge-density-wave order,
nematicity, and TRS breaking~\cite{Cai2026QF, Yin2022Nature, Ye2018Nature, Ortiz2020PRL, Jiang2021NM,
Cheng2023NL, Ren2024NC, Zhang2023NL, Inoue2019APL}.

In this sense, although the specific kagom\'{e} QBCP corresponding exactly to our model has not
yet been experimentally realized, the broader experimental progress offers strong motivation for the
present theoretical study. Accordingly, a systematic study of interaction-driven instabilities, fixed-point
structures, and quantum critical behavior in TRS-breaking kagom\'{e} QBCP systems remains both timely and
well motivated.

\subsection{Outlook}

Looking ahead, unlike
2D kagom\'{e} QBCP systems, where the finite DOS allows reliable treatment within the one-loop RG framework,
systems with a Van Hove singularity (VHS) exhibit a DOS that renders conventional perturbative methods
inadequate~\cite{Chubukov2024NPJ}. It would therefore be interesting to explore the physics of VHS systems
within an extended theoretical framework in future work.

In addition, we confine our attention to short-range fermion-fermion interactions. This restriction is physically reasonable as the long-range interactions are largely screened in realistic settings, which is also consistent with most previous weak-coupling RG studies on QBCP systems \cite{Fradkin2009PRL, Vafek2014PRB, Zhu2016PRL, Wang2017PRB, Wang2020-arxiv, Zhai2021NPB}. Nevertheless, if the long-range interaction remains insufficiently screened, it may qualitatively modify the RG flows, FPs, and instabilities. A comprehensive treatment of this scenario lies beyond the scope of the present study and is reserved for future work.

\section{Summary}\label{Sec_summary}

In summary, we systematically investigate the effects of 16 marginally relevant fermion-fermion interactions on the low-energy critical behavior of 2D kagom\'{e} systems hosting a QBCP~\cite{Fradkin2009PRL,Vafek2014PRB,Venderbos2016PRB,Wu2016PRL,
Zhu2016PRL}. Employing a momentum-shell RG approach~\cite{Shankar1994RMP,Wilson1975RMP,Polchinski9210046}
to unbiasedly treat the complex interplay of these interactions, we derive energy-dependent flow equations
for all coupling strengths. A comprehensive analysis of these RG flows allows us to identify the governing
FPs and their corresponding instabilities at the lowest-energy limit.

We begin by analyzing the energy-dependent evolutions of fermion-fermion interactions.
Our numerical calculations reveal that the couplings diverge at a critical low-energy scale,
signaling the onset of quantum criticality. This critical behavior is governed by a certain FP ~\cite{Vojta2003RPP,Sachdev2011Book,Metzner2000PRL,Maiti2010PRB,Vafek2014PRB,Chubukov2016PRX,
Roy2018PRX,Nandkishore2012NP,Vafek2012PRB}, whose character is highly sensitive to the structural
parameters $d_{0,1,2,3}$ in the effective action~(\ref{Eq_S_eff}). The requirement of a stable
quadratic band crossing at $\mathbf{k}=0$ imposes a key constraint on
these parameters, which leads to a natural classification of the system into three distinct
cases (Case I, II, and III), as detailed in Sec.~\ref{Sec_classification_FPs}.
In the rotationally symmetric Case I, we identify two stable FPs: $\mathrm{FP}_1$~(\ref{Eq_FP1})
and $\mathrm{FP}_2$~(\ref{Eq_FP2}). In contrast, the asymmetric Case II and Case III exhibit a richer fixed-point structure,
including the common FP$_1$ and nine additional FPs as presented in Table~\ref{table:fixpoint}.
To proceed, we employ the linear and plane fitting analysis to appropriately determine the initial conditions of all these
FPs, which are characterized by the boundary conditions presented in Eqs.~(\ref{Eq_condition_1}),
(\ref{Eq_condition_2}), (\ref{Eq_condition_4}), and Eqs.~(\ref{Eq_condition_3})-(\ref{Eq_condition_12}).

In principle, symmetry breaking near these FPs can induce a variety of
instabilities~\cite{Metzner2000PRL,Halboth2000RPB,Maiti2010PRB,Vafek2014PRB,Chubukov2016PRX,Roy2018PRX}.
To identify the leading orders, we incorporate external source terms that are constituted
by fermionic bilinears to the effective theory, derive their RG evolutions~(\ref{Eq_source_1})-(\ref{Eq_source_2}),
and then compute the susceptibilities of all candidate phases as the system approaches each FP.
For clarity, we classify the 10 FPs into four groups based on their interaction structures
detailed in Sec.~\ref{Sec_results_chi}, i.e., Group-A (FP$_1$), Group-B (FP$_5$, FP$_6$, FP$_9$, and FP$_{10}$),
Group-C (FP$_2$, FP$_4$, and FP$_8$), and Group-D (FP$_3$ and FP$_7$). Our analysis reveals the leading and
subleading instabilities near all FPs, which are summarized in Table~\ref{table:leading}. At first,
CDW and chiral SC2 become two primary competing orders. CDW appears as the leading instability for
Group-C (FP$_2$, FP$_4$, and FP$_8$) and for FP$_{5,6}$ in Group-B, and as a subleading instability
in Group-A (FP$_1$), Group-D (FP$_3$), and for FP$_{9,10}$ in Group-B. Conversely, chiral SC2 dominates
as the leading instability for Group-A (FP$_1$) and for FP$_{9,10}$ in Group-B, while serving as a
subleading instability in Group-C (FP$_2$, FP$_4$, FP$_8$) and Group-D (FP$_3$).
In addition, the bond density and $x$-current state have an opportunity to be leading states at FP$_7$ of Group-D
and FP$_3$ of Group-D, while the chiral SC1 and $s$-wave SC are two degenerate dominant states at FP$_9$ and FP$_{10}$ of
Group-B and FP$_1$ of Group-A, respectively. Moreover, both bond-density and $x$-current are prone to the primary subleading states. Within Group-B, the $x$-current appears as a subleading instability at FP$_5$ and FP$_6$, whereas bond density is subleading at FP$_9$ and FP$_{10}$. In contrast, bond density appears as a subleading state at both FP$_3$ of Group-D and FP$_4$ of Group-C,
but instead $x$-current state only acts as a subleading state at FP$_4$ of Group-C.

These results elucidate rich low-energy phenomena induced by marginally relevant fermion-fermion interactions,
providing useful insights into interaction-driven quantum criticality and phase transitions in 2D
kagom\'{e} QBCP materials. We hope it will be helpful for the future exploration of structural parameter tuning
in real materials and related experimental studies.

\section*{ACKNOWLEDGEMENTS}

We thank Tian-Sen Ye, Wen Liu, and Wen-Hao Bian for the helpful discussions.
J.W. is supported  by Tianjin Natural Science Foundation Project (25JCYBJC01640) and
was partially supported by the National Natural
Science Foundation of China under Grant No. 11504360.

\appendix

\section{One-loop corrections}\label{Appendix_1L_corrections}

\begin{figure*}[htbp]
\centering
\subfigure[]{
\includegraphics[width=0.8in]{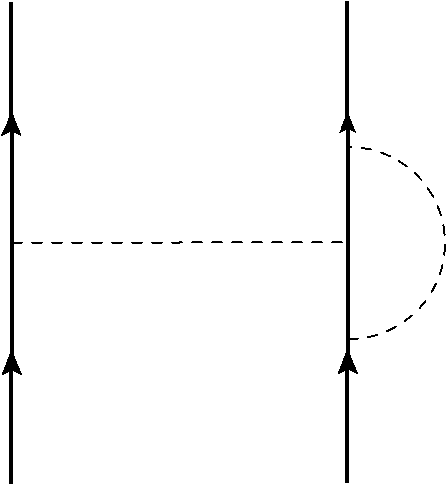}
}\hspace{0.5cm}
\subfigure[]{
\includegraphics[width=0.8in]{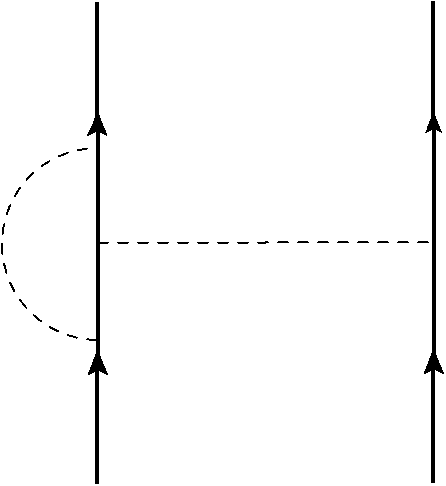}
}\hspace{0.5cm}
\subfigure[]{
\includegraphics[width=0.8in]{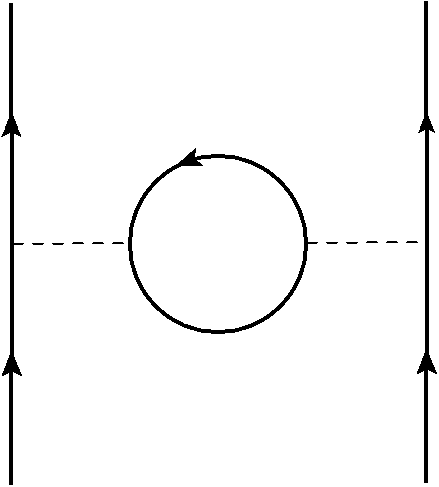}
}\hspace{0.5cm}
\subfigure[]{
\includegraphics[width=0.8in]{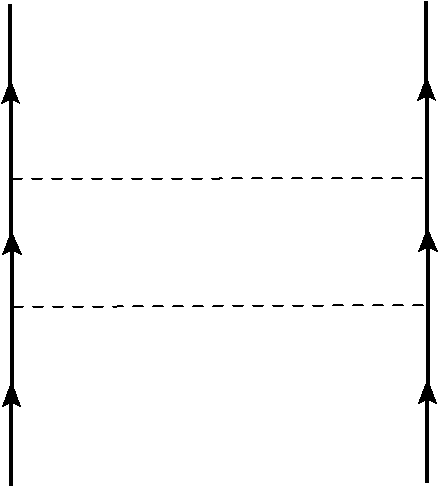}
}\hspace{0.5cm}
\subfigure[]{
\includegraphics[width=0.8in]{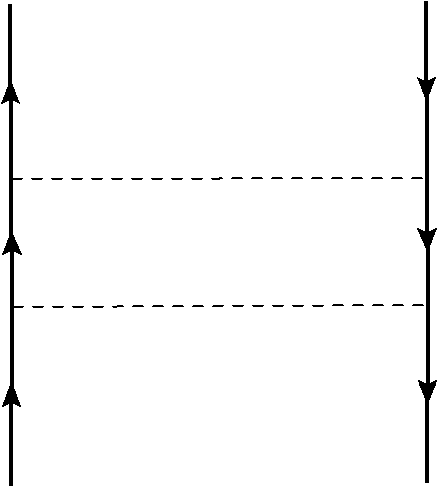}
}
\\
\vspace{-0.1cm}
\caption{One-loop Feynamn diagrams for four-fermion interaction renormalization (a)-(e) due to the
fermion-fermion interactions. The solid and dashed lines specify the fermionic propagator and fermion-fermion
interaction, respectively.}
\label{Fig12}
\end{figure*}

The one-loop quantum corrections to the fermion-fermion interactions are diagrammatically represented in Fig.~\ref{Fig12}.
For convenience, we employ dimensionless variables scaled by the lattice cutoff, i.e., $k\rightarrow k/\Lambda_0$
and $\omega\rightarrow\omega/\Lambda_0$, to simplify computations and express results compactly
~\cite{Vafek2014PRB,Stauber2005PRB, Wang2017PRB, Wang2011PRB,She2010PRB,Huh2008PRB,Wang2019JPCM}.
After long but straightforward calculations~\cite{Vafek2014PRB,Wang2017PRB}, all one-loop corrections to
the fermion-fermion interactions are obtained as follows, where $\mathcal{F}_{\mu\nu}$ denote the
corrections to $\lambda_{\mu\nu}$ with $\mu,\nu=0,1,2,3$.
\begin{figure*}[htbp]
\subfigure[]{
\includegraphics[width=1.5in]{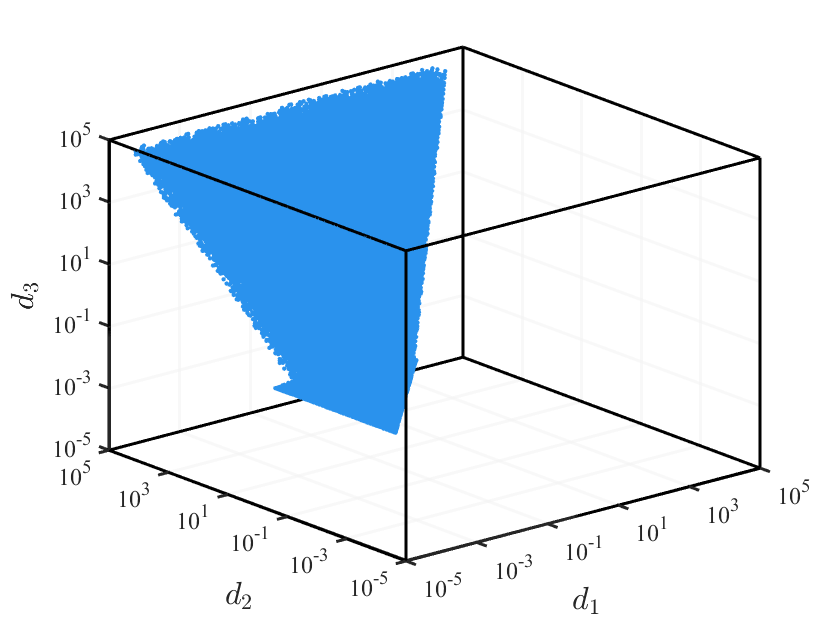}
}
\subfigure[]{
\includegraphics[width=1.5in]{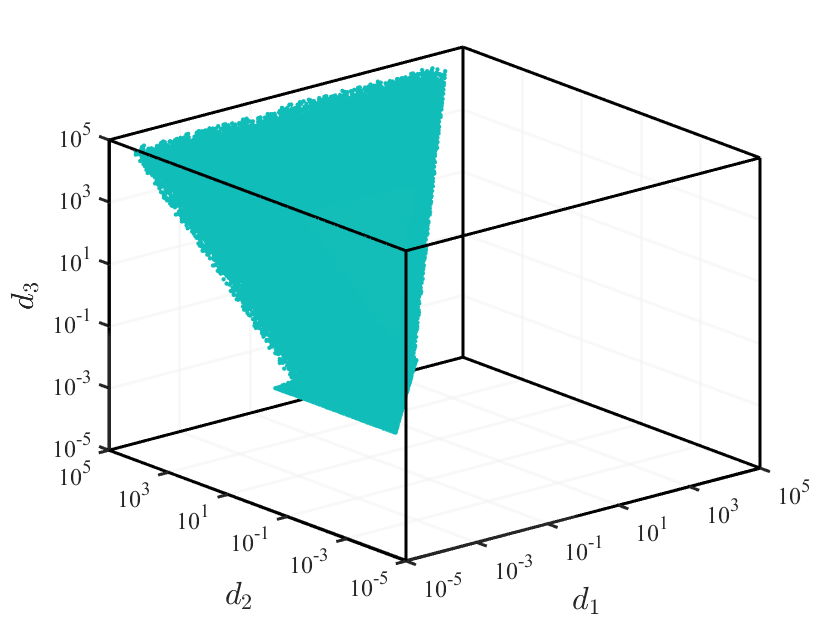}
}
\subfigure[]{
\includegraphics[width=1.5in]{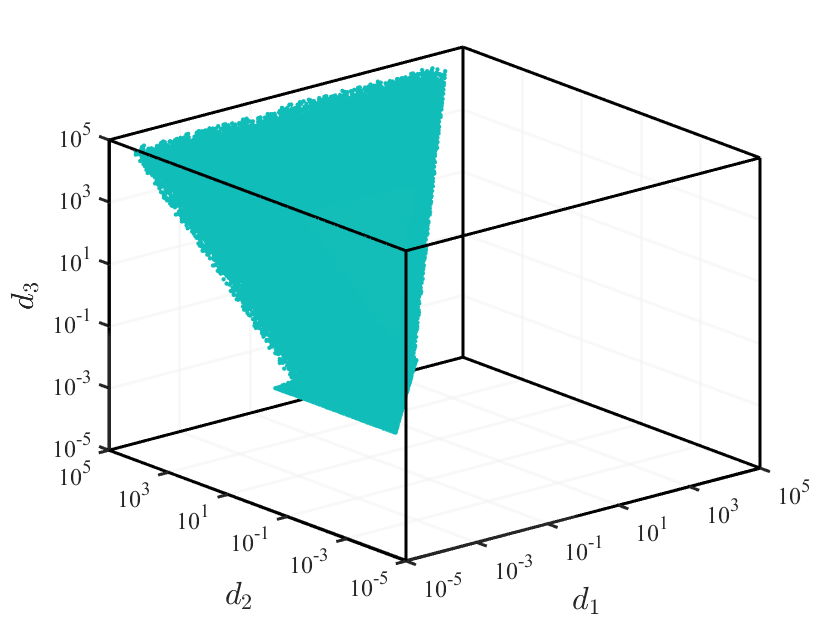}
}
\subfigure[]{
\includegraphics[width=1.5in]{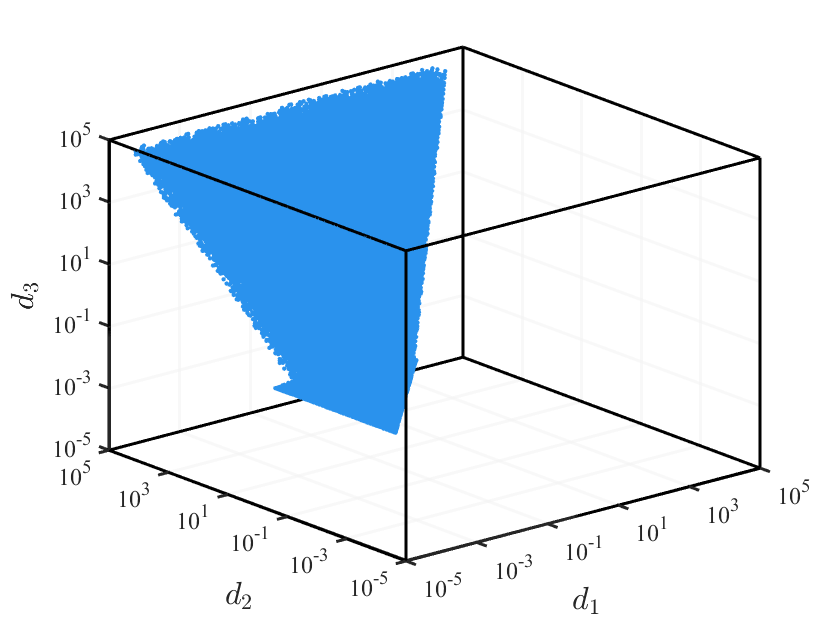}
}
\\
\vspace{-0.3cm}
\subfigure[]{
\includegraphics[width=1.5in]{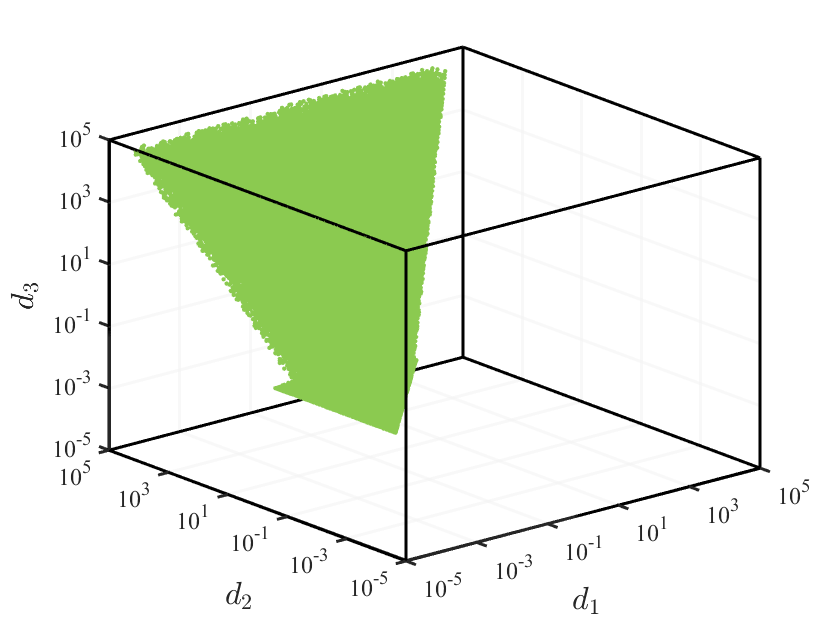}
}
\subfigure[]{
\includegraphics[width=1.5in]{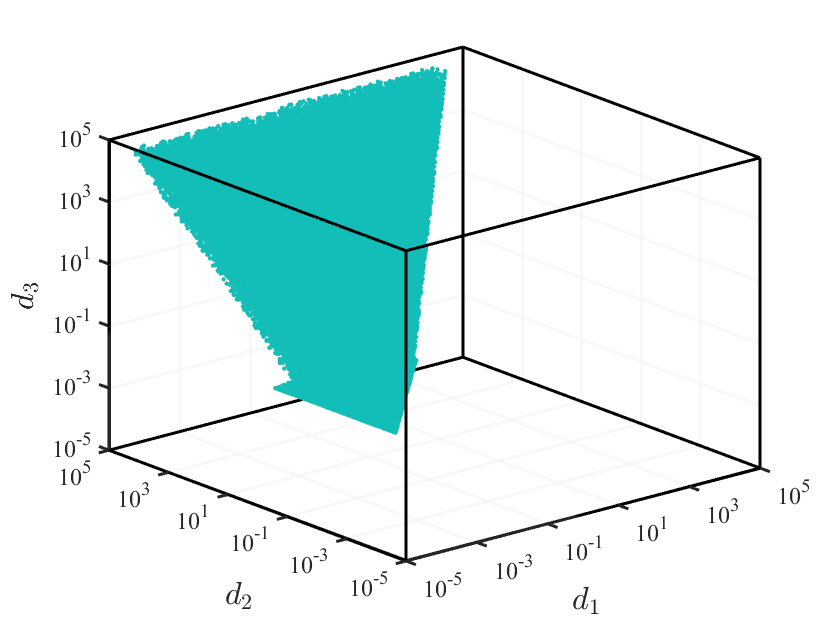}
}
\subfigure[]{
\includegraphics[width=1.5in]{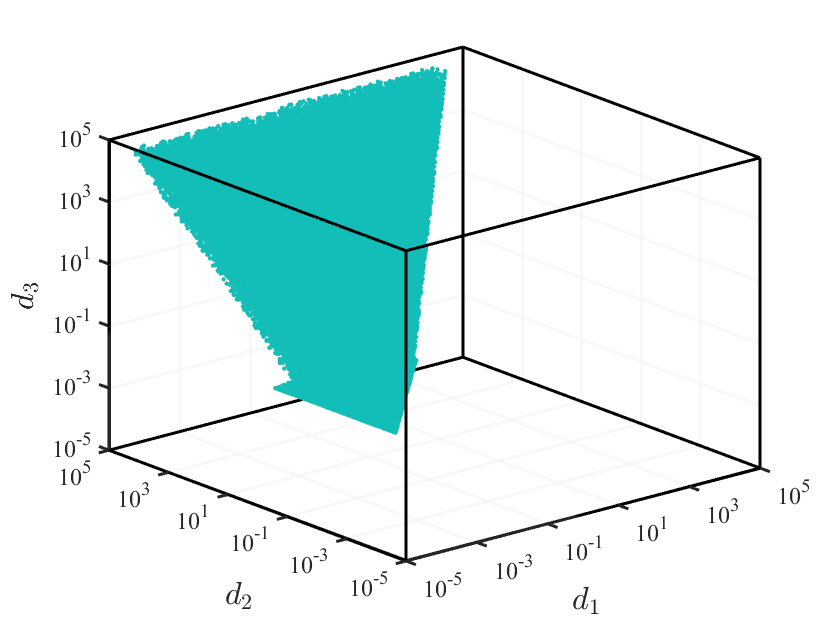}
}
\subfigure[]{
\includegraphics[width=1.5in]{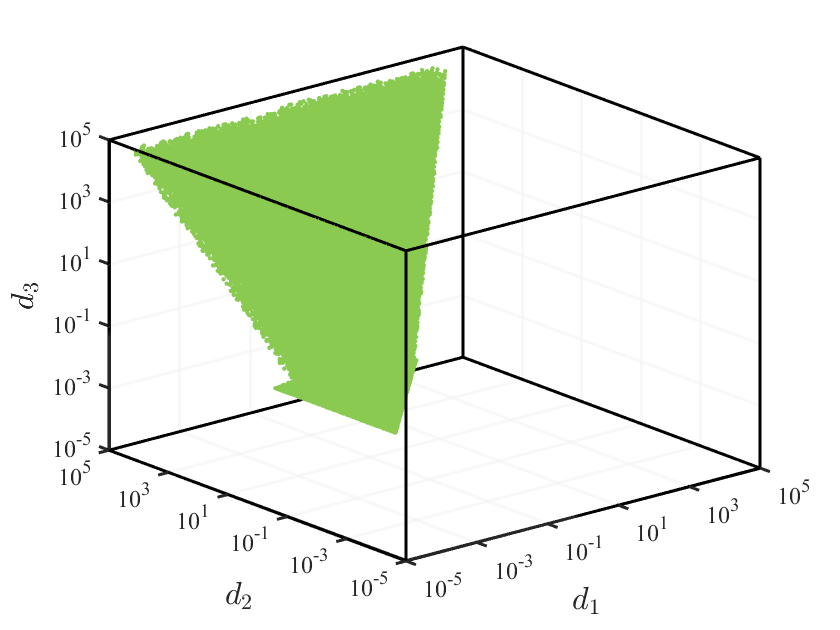}
}
\\
\vspace{-0.3cm}
\subfigure[]{
\includegraphics[width=1.5in]{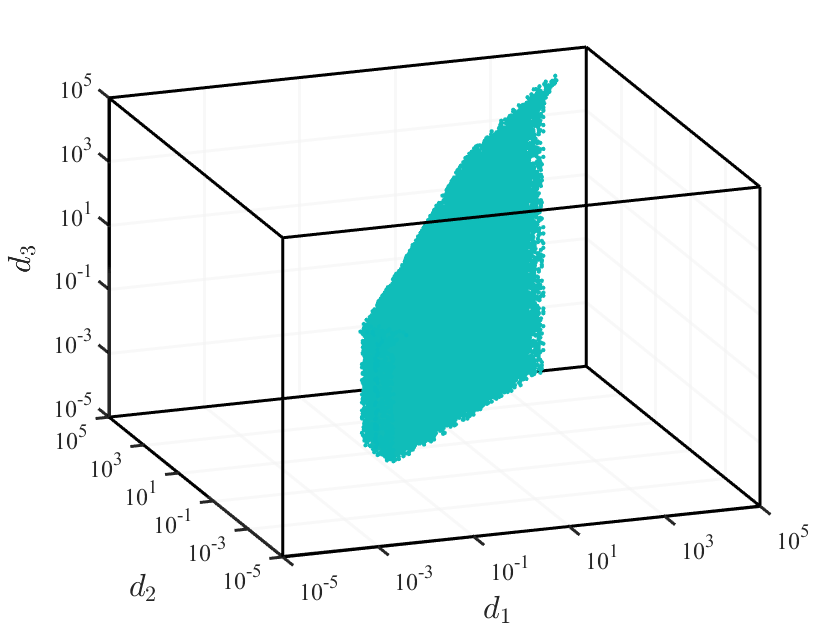}
}
\subfigure[]{
\includegraphics[width=1.5in]{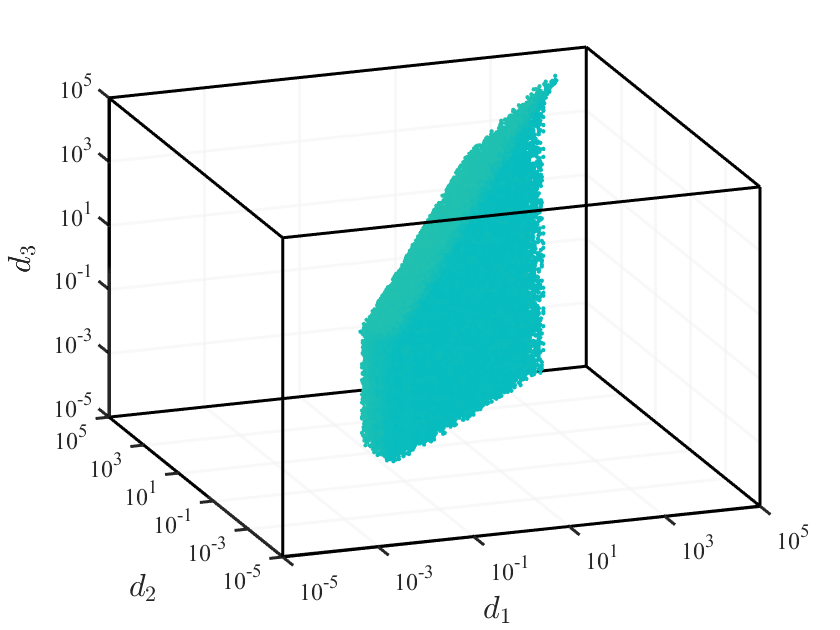}
}
\subfigure[]{
\includegraphics[width=1.5in]{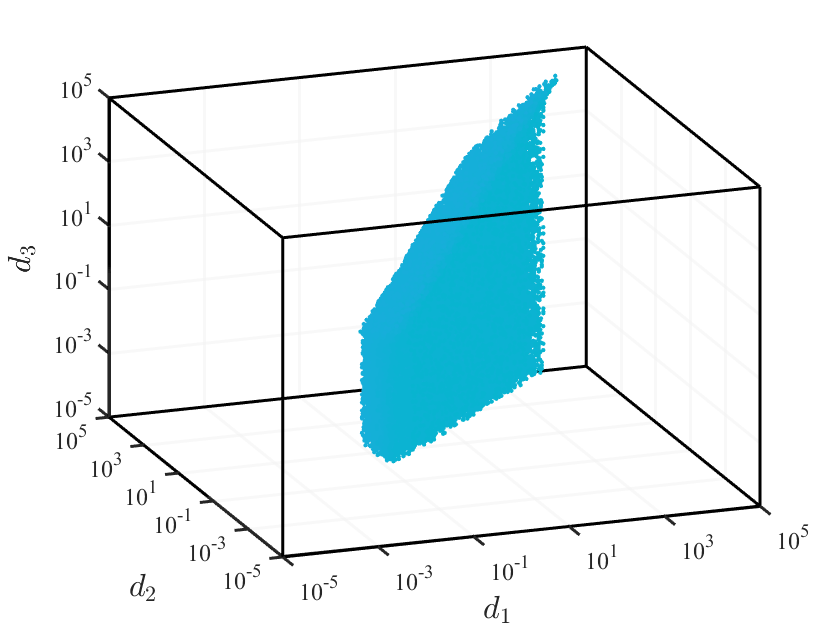}
}
\subfigure[]{
\includegraphics[width=1.5in]{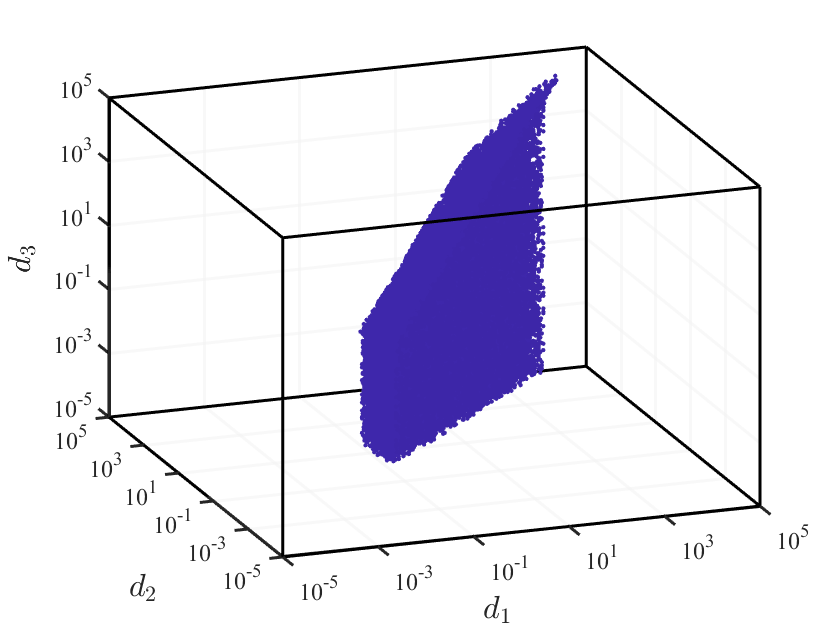}
}
\\
\vspace{-0.3cm}
\subfigure[]{
\includegraphics[width=1.5in]{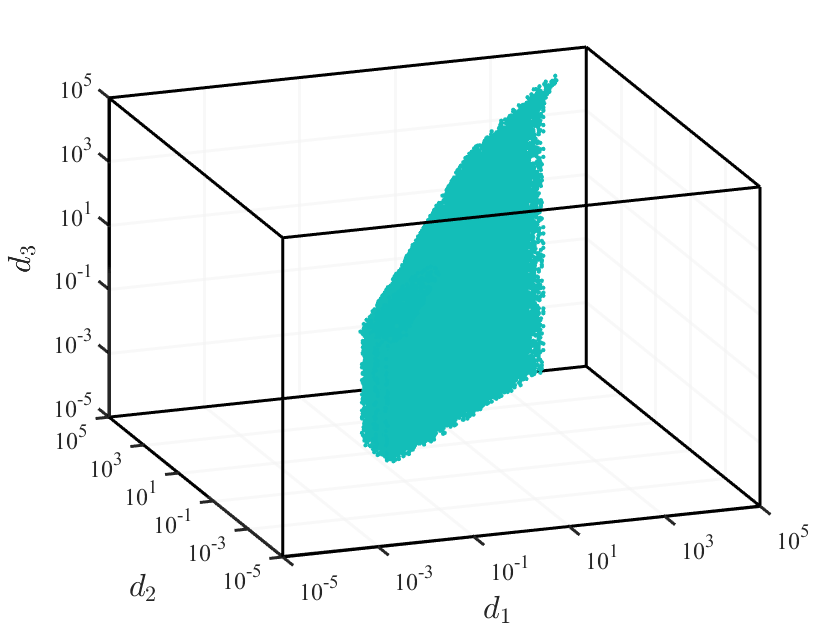}
}
\subfigure[]{
\includegraphics[width=1.5in]{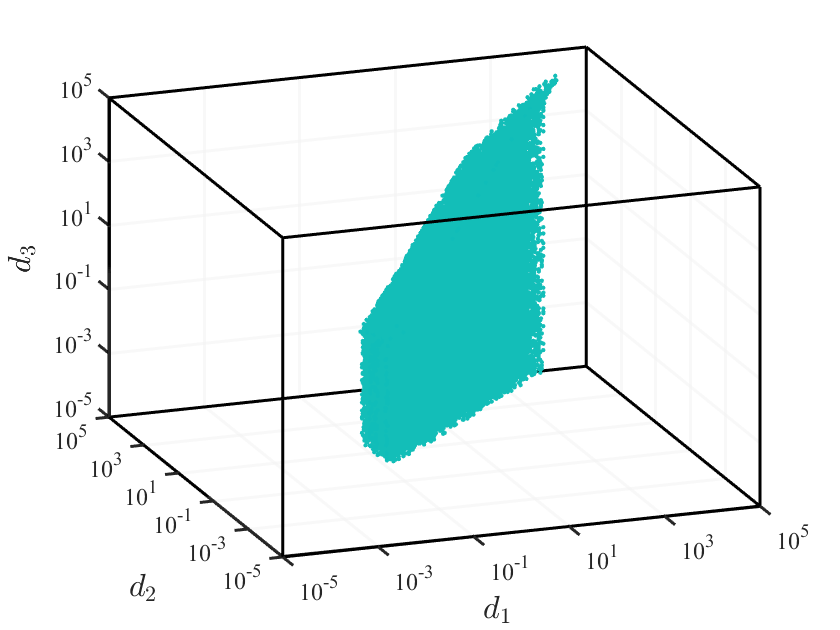}
}
\subfigure[]{
\includegraphics[width=1.5in]{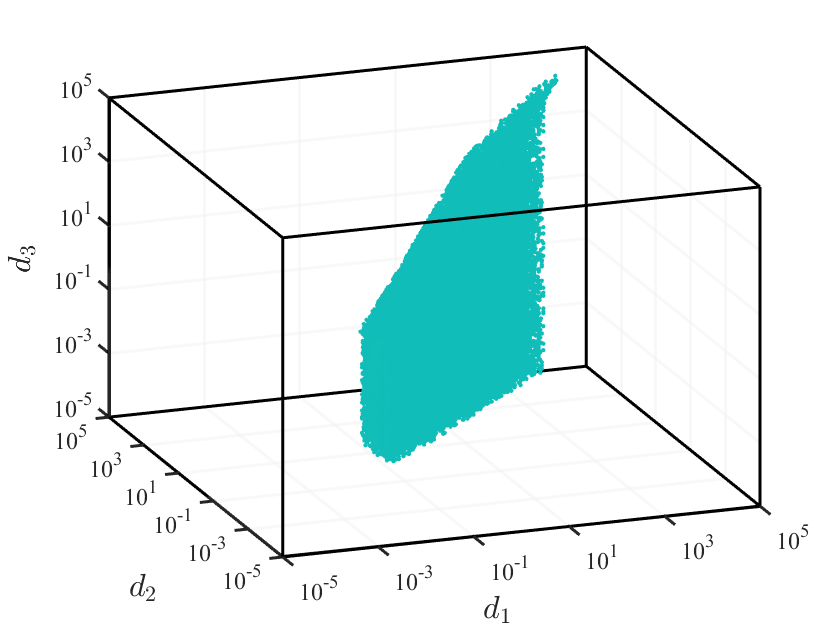}
}
\subfigure[]{
\includegraphics[width=1.5in]{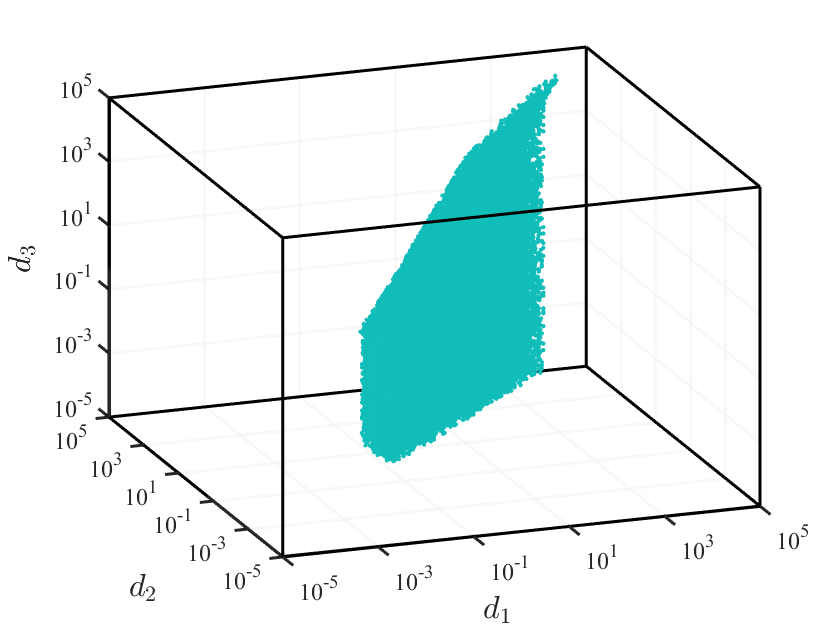}
}
\\
\vspace{-0.3cm}
\caption{(Color online) Stable subregions in the $d_1$-$d_2$-$d_3$ structural parameter
space for critical fermion-fermion interaction values at $l=l_c$
(color mapping follows the same scale as Fig.~\ref{Fig_lambda03_d123}(a)):
(a) $\lambda_{00} = -1.414$, (b) $\lambda_{01} = 0$, (c) $\lambda_{02} = 0$, (d) $\lambda_{03} = -1.414$,
(e) $\lambda_{10} = 1.414$, (f) $\lambda_{11} = 0$, (g) $\lambda_{12} = 0$, and (h) $\lambda_{13} = 1.414$,
which constitute the $\mathrm{FP}_1$ in $\mathrm{Case}$ $\mathrm{II}$, and
(i) $\lambda_{00} = -0.002$, (j) $\lambda_{01} = 0.049$, (k) $\lambda_{02} = -0.573$, (l) $\lambda_{03} = -3.957$,
(m) $\lambda_{10} = 0$, (n) $\lambda_{11} = 0$, (o) $\lambda_{12} = 0$, and (p) $\lambda_{13} = 0$,
which constitute the $\mathrm{FP}_4$ in $\mathrm{Case}$ $\mathrm{II}$~(\ref{Eq_FP4}).}
\label{Fig_FP1-4-II}
\end{figure*}
\begin{figure*}[htbp]
\subfigure[]{
\includegraphics[width=1.5in]{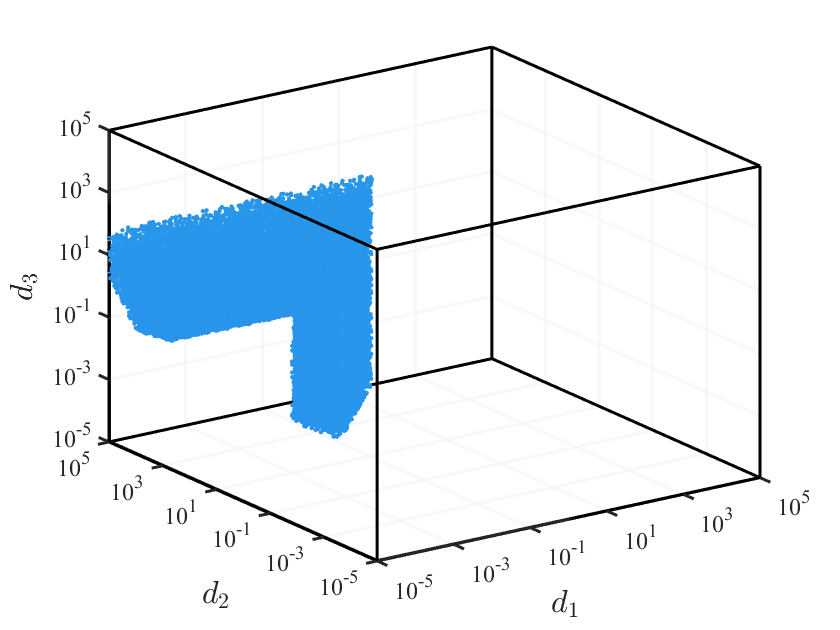}
}
\subfigure[]{
\includegraphics[width=1.5in]{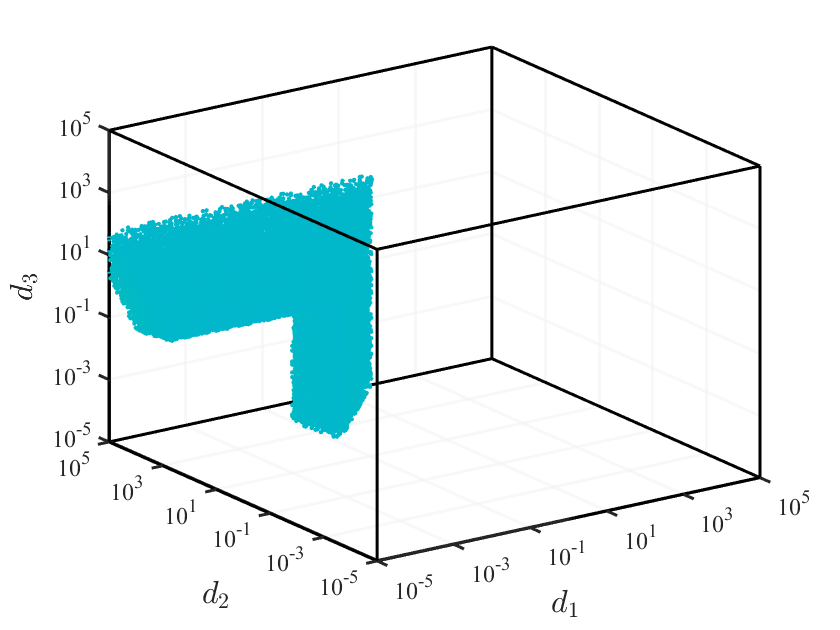}
}
\subfigure[]{
\includegraphics[width=1.5in]{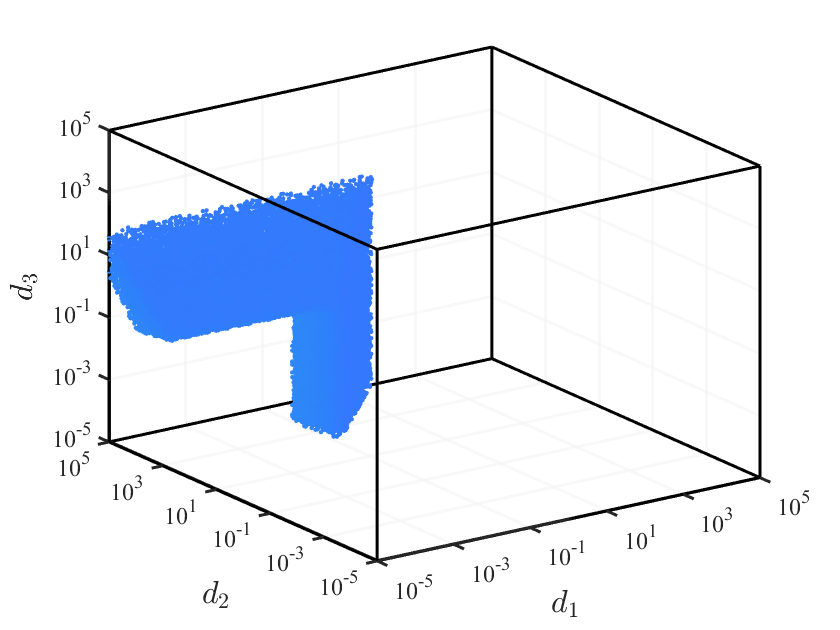}
}
\subfigure[]{
\includegraphics[width=1.5in]{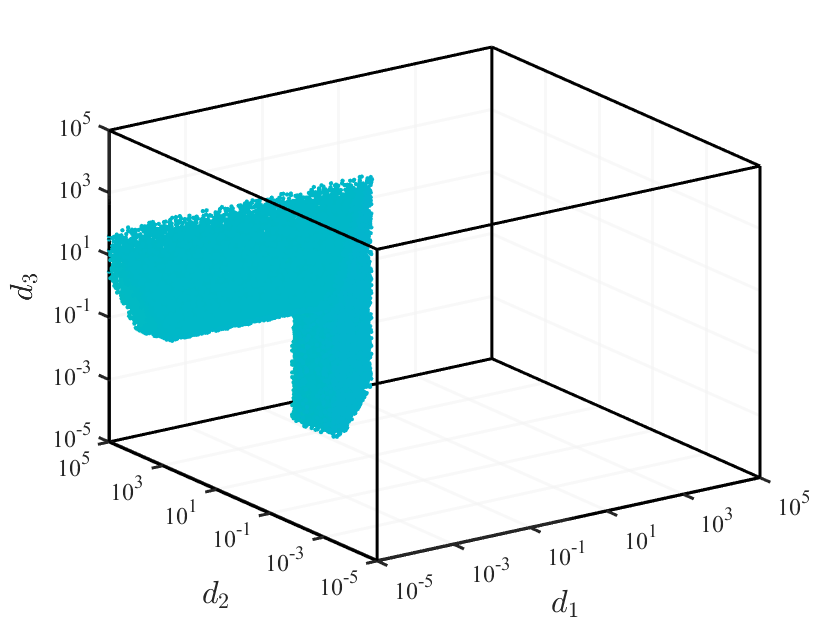}
}
\\
\vspace{-0.3cm}
\subfigure[]{
\includegraphics[width=1.5in]{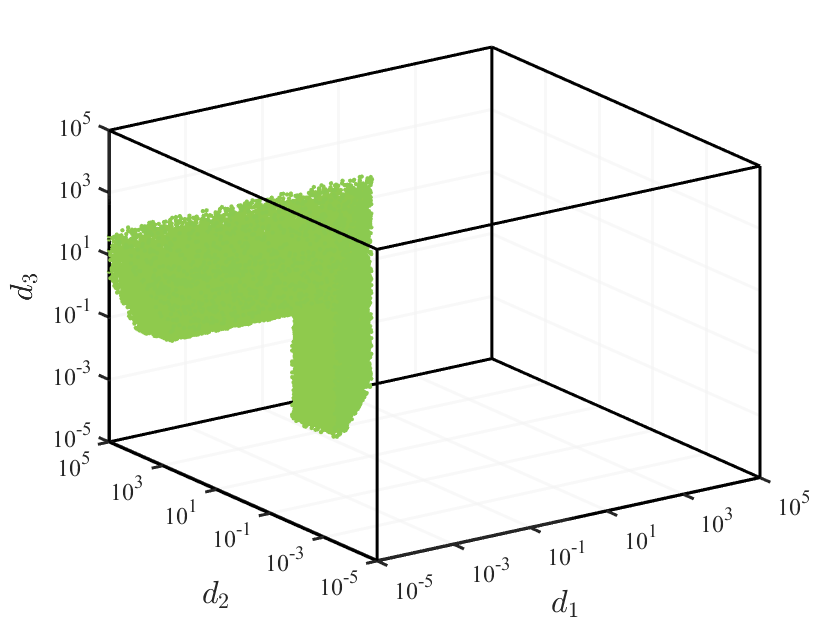}
}
\subfigure[]{
\includegraphics[width=1.5in]{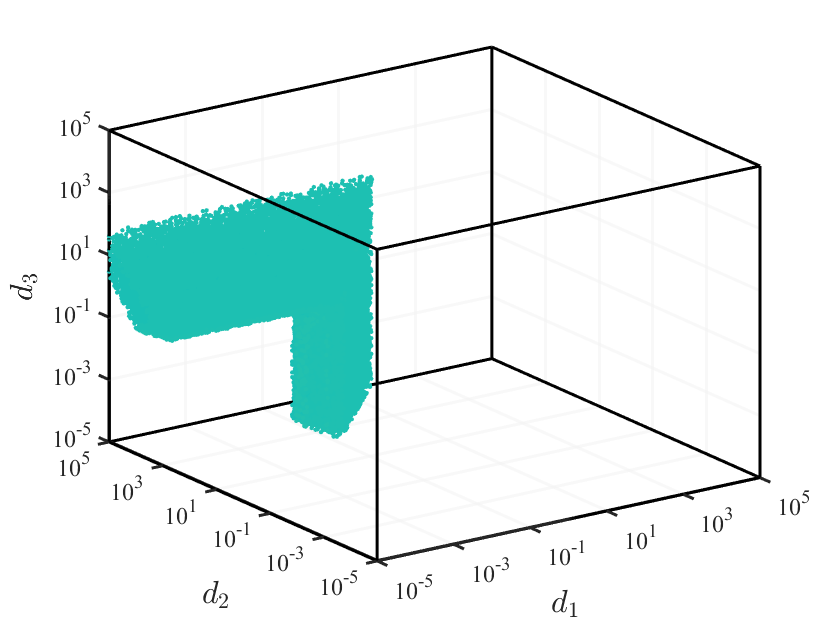}
}
\subfigure[]{
\includegraphics[width=1.5in]{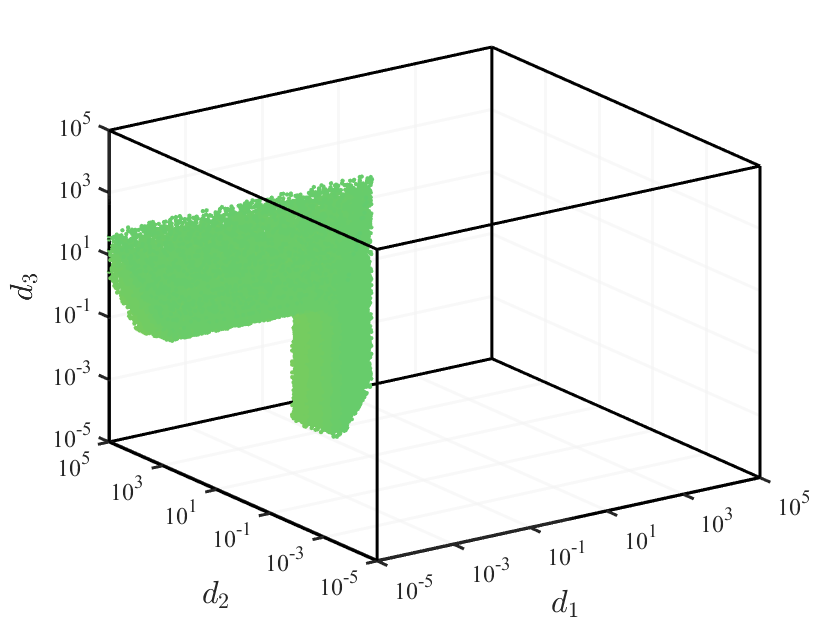}
}
\subfigure[]{
\includegraphics[width=1.5in]{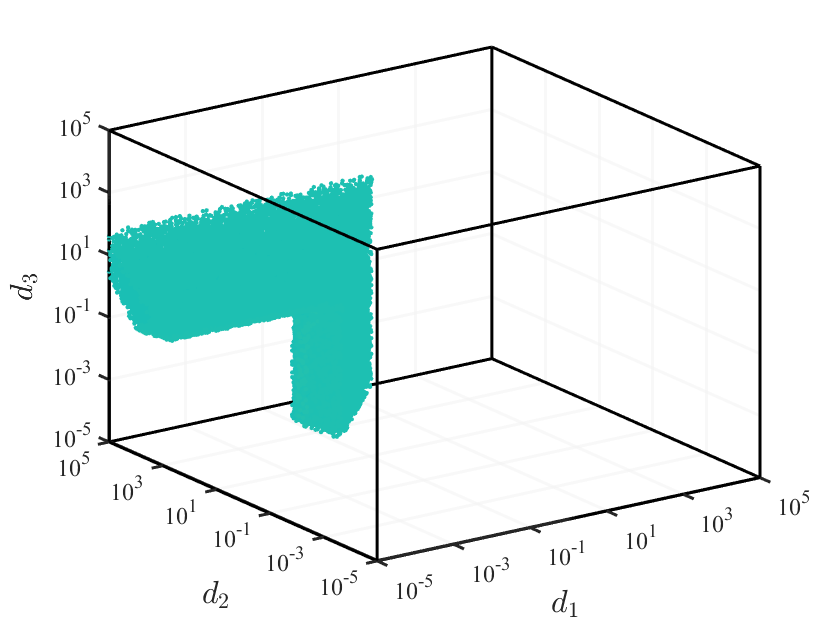}
}
\\
\vspace{-0.3cm}
\subfigure[]{
\includegraphics[width=1.5in]{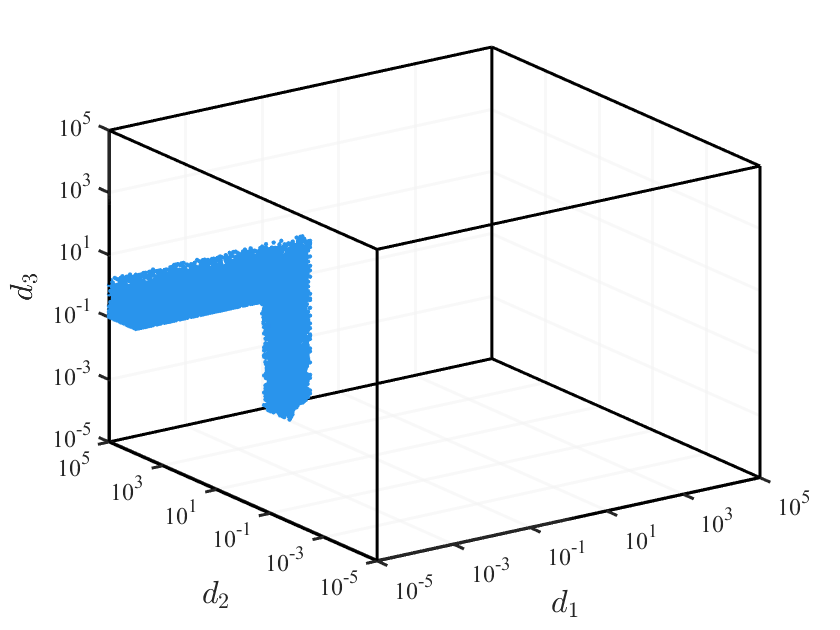}
}
\subfigure[]{
\includegraphics[width=1.5in]{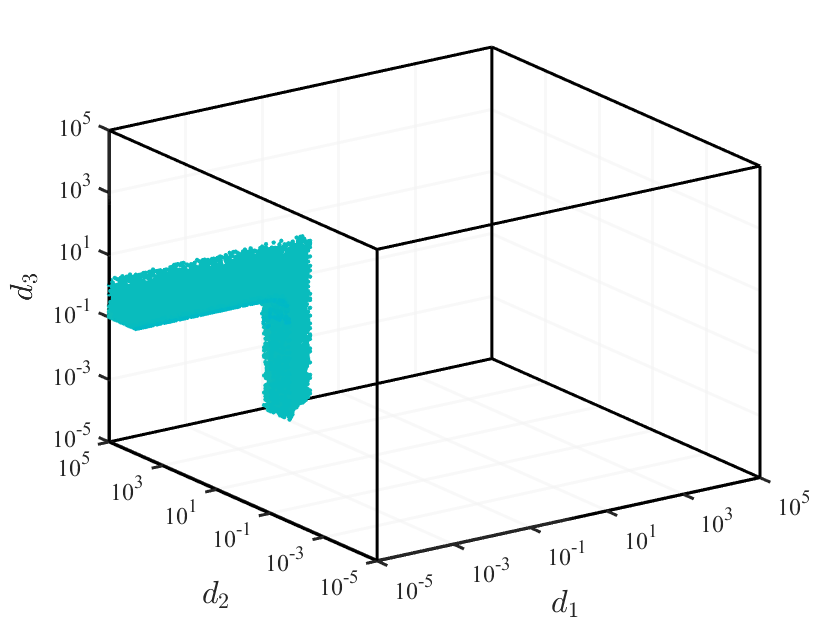}
}
\subfigure[]{
\includegraphics[width=1.5in]{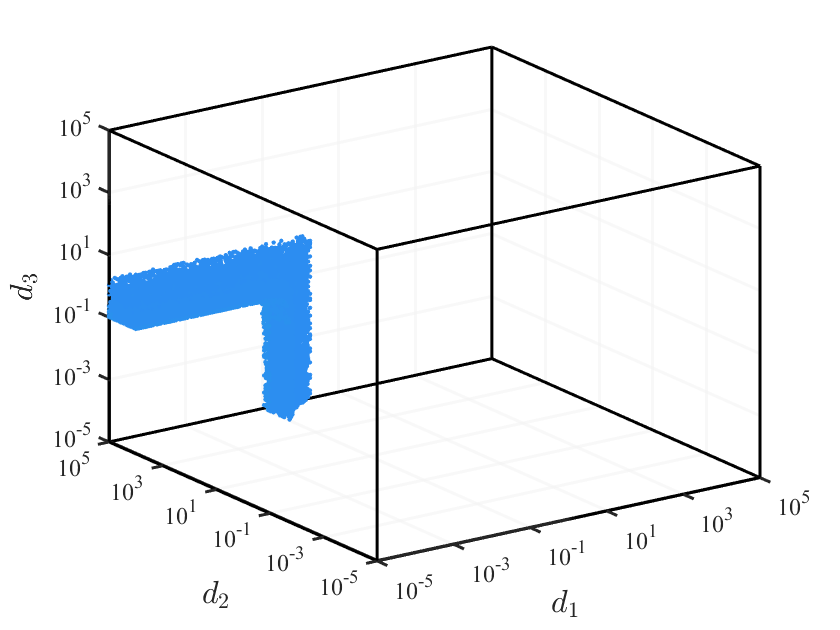}
}
\subfigure[]{
\includegraphics[width=1.5in]{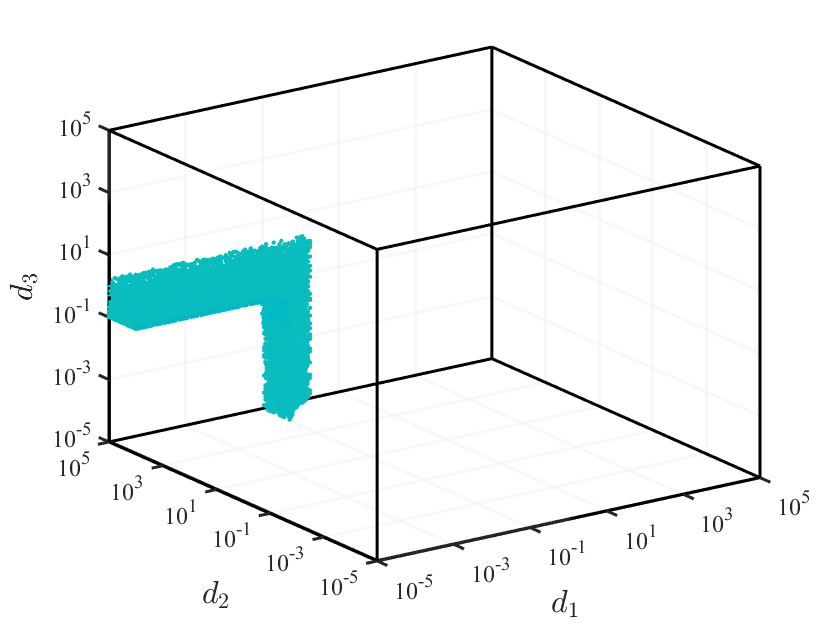}
}
\\
\vspace{-0.3cm}
\subfigure[]{
\includegraphics[width=1.5in]{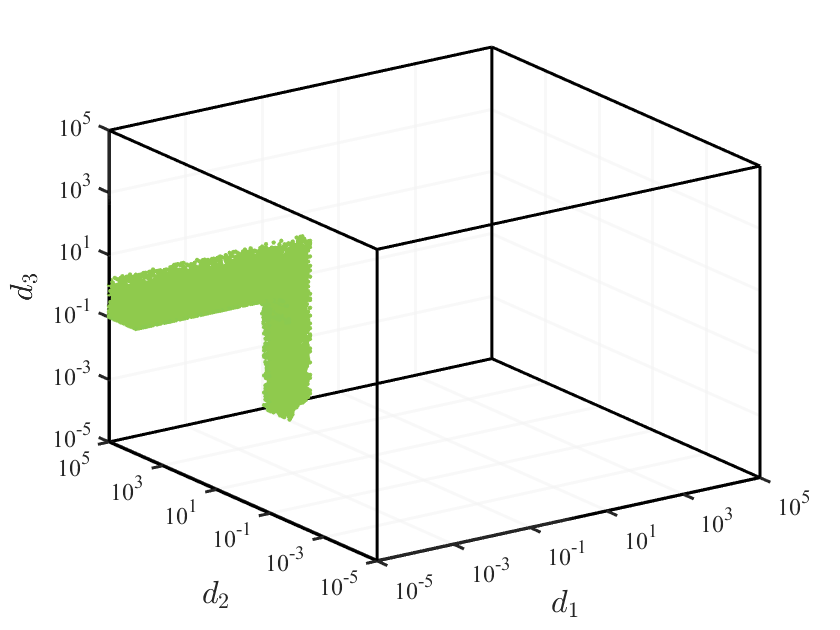}
}
\subfigure[]{
\includegraphics[width=1.5in]{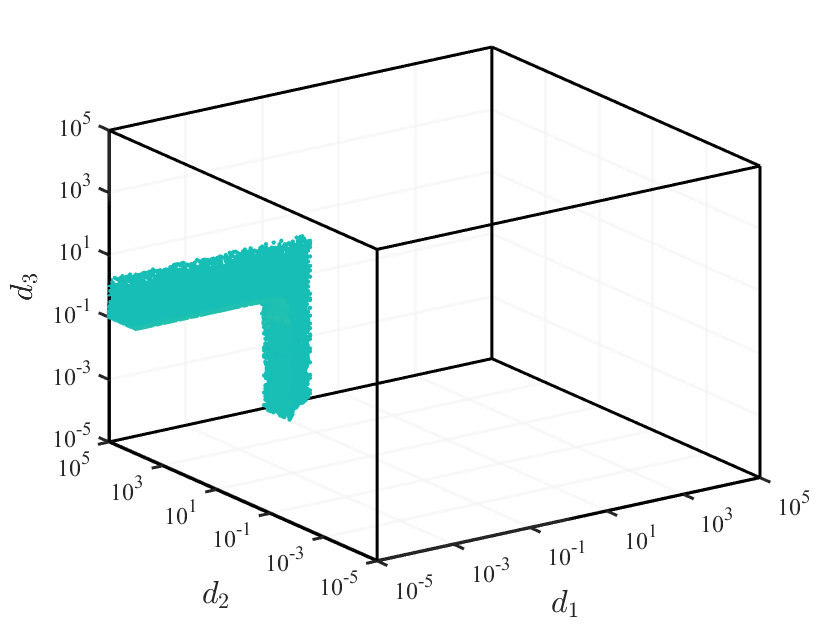}
}
\subfigure[]{
\includegraphics[width=1.5in]{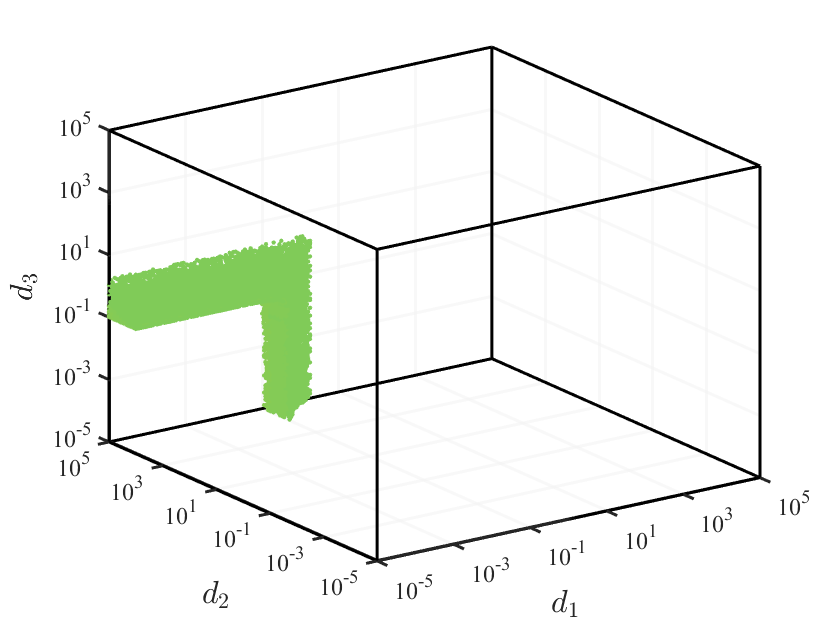}
}
\subfigure[]{
\includegraphics[width=1.5in]{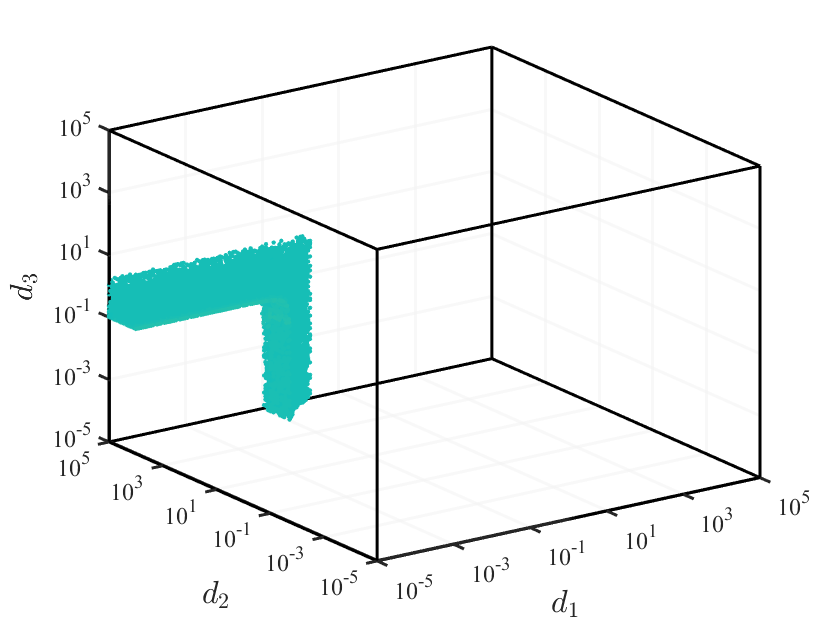}
}
\\
\vspace{-0.3cm}
\caption{(Color online) Stable subregions in the $d_1$-$d_2$-$d_3$ structural parameter
space for critical fermion-fermion interaction values at $l=l_c$
(color mapping follows the same scale as Fig.~\ref{Fig_lambda03_d123}(a)):
(a) $\lambda_{00} = -1.394$, (b) $\lambda_{01} = -0.107$, (c) $\lambda_{02} = -1.488$, (d) $\lambda_{03} = -0.113$,
(e) $\lambda_{10} = 1.444$, (f) $\lambda_{11} = 0.073$, (g) $\lambda_{12} = 1.354$, and (h) $\lambda_{13} = 0.077$,
which constitute the $\mathrm{FP}_5$ in $\mathrm{Case}$ $\mathrm{II}$~(\ref{Eq_FP5}), and
(i) $\lambda_{00} = -1.367$, (j) $\lambda_{01} = -0.255$, (k) $\lambda_{02} = -1.857$, (l) $\lambda_{03} = -0.256$,
(m) $\lambda_{10} = 1.450$, (n) $\lambda_{11} = 0.102$, (o) $\lambda_{12} = 1.179$, and (p) $\lambda_{13} = 0.102$,
which constitute the $\mathrm{FP}_6$ in $\mathrm{Case}$ $\mathrm{II}$~(\ref{Eq_FP6}).}
\label{Fig_FP5-6-II}
\end{figure*}
\begin{figure*}[htbp]
\subfigure[]{
\includegraphics[width=1.5in]{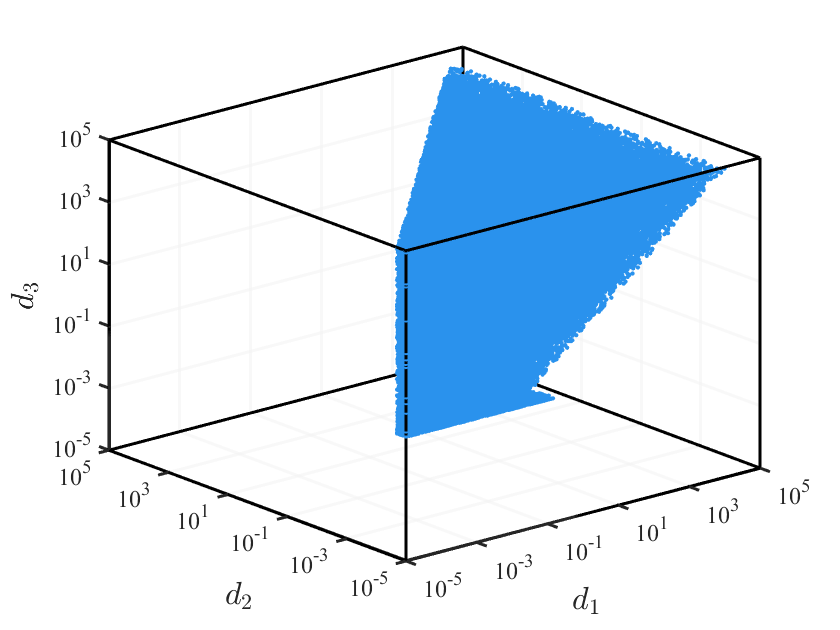}
}
\subfigure[]{
\includegraphics[width=1.5in]{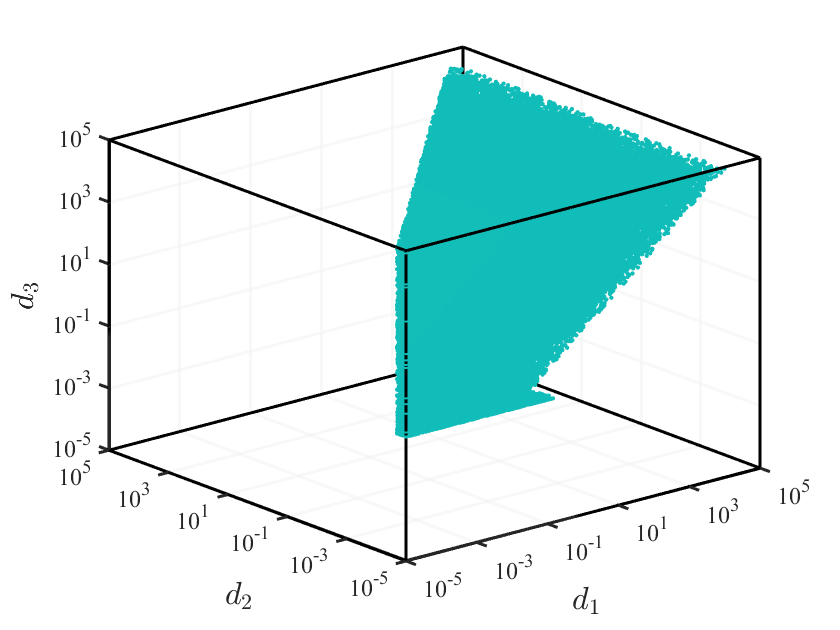}
}
\subfigure[]{
\includegraphics[width=1.5in]{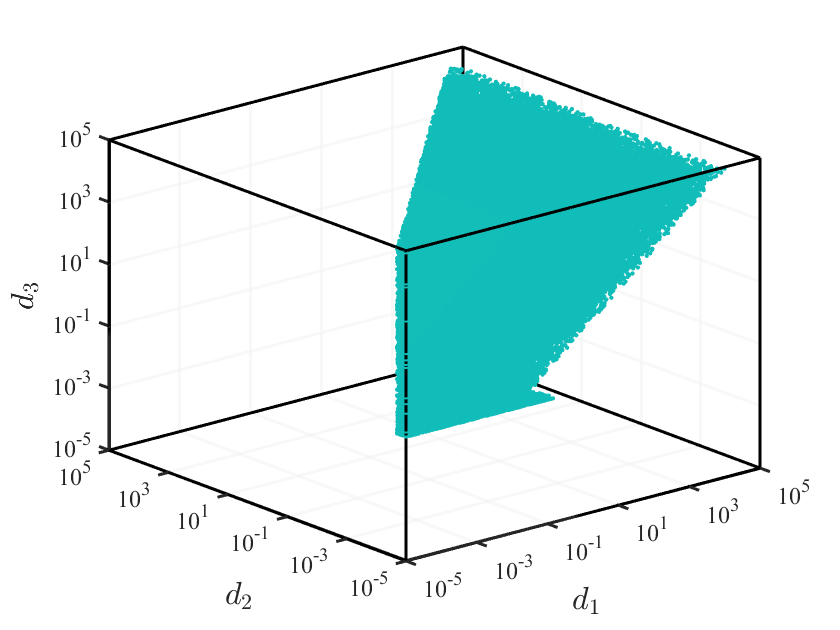}
}
\subfigure[]{
\includegraphics[width=1.5in]{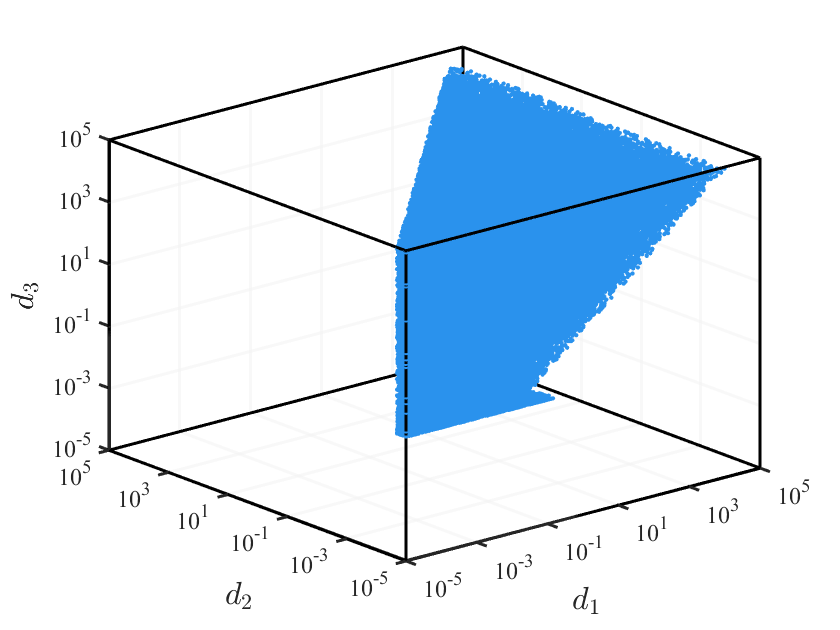}
}
\\
\vspace{-0.3cm}
\subfigure[]{
\includegraphics[width=1.5in]{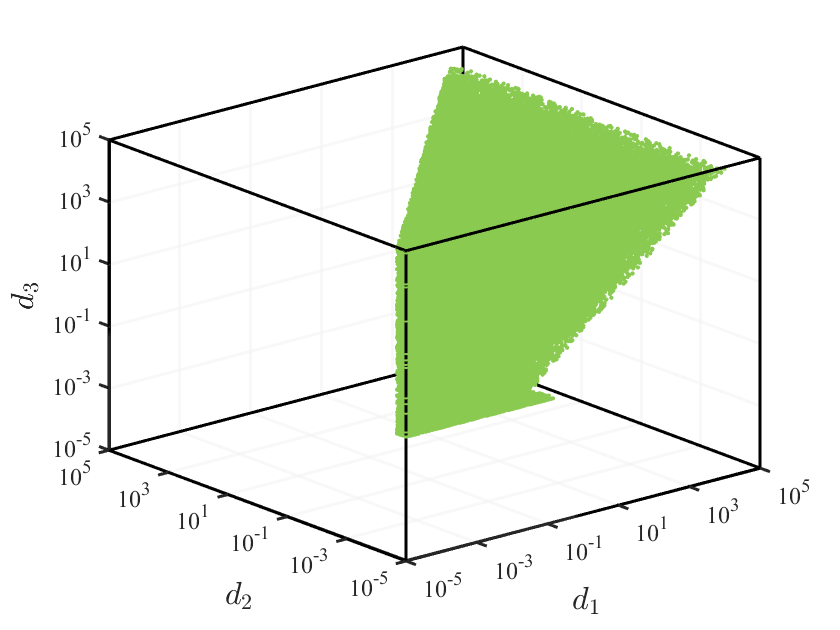}
}
\subfigure[]{
\includegraphics[width=1.5in]{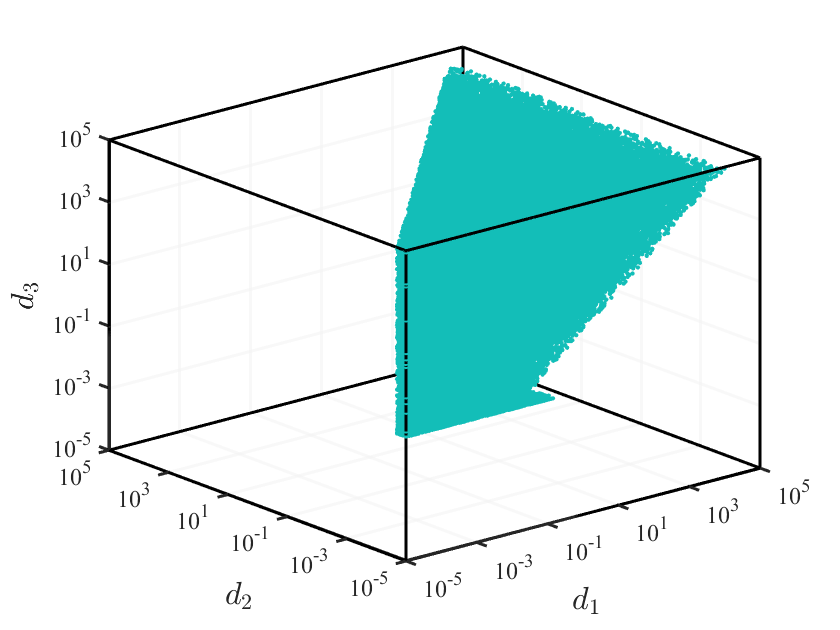}
}
\subfigure[]{
\includegraphics[width=1.5in]{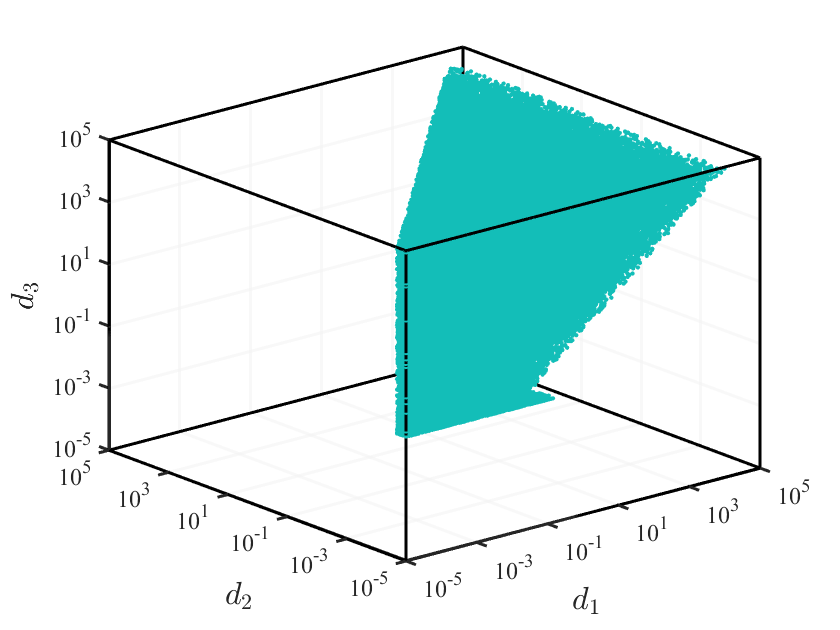}
}
\subfigure[]{
\includegraphics[width=1.5in]{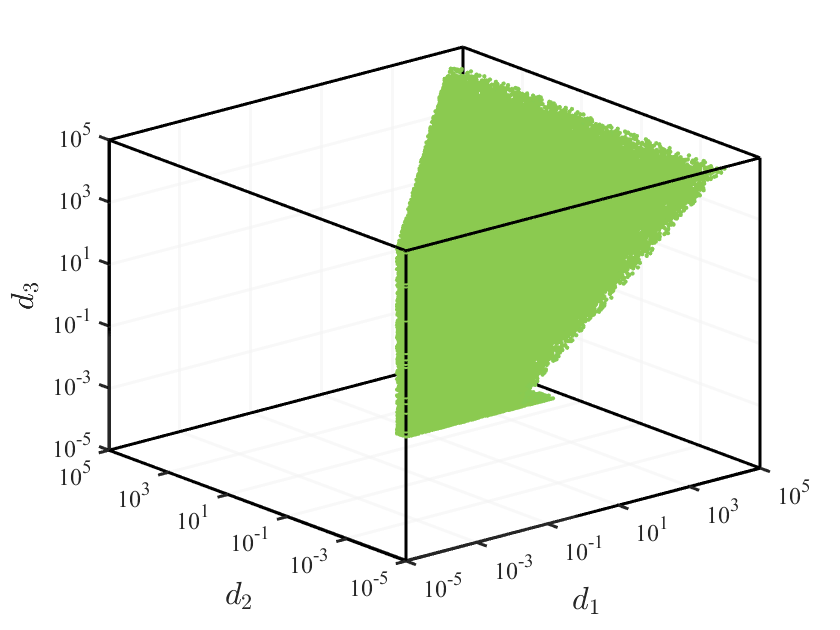}
}
\\
\vspace{-0.3cm}
\subfigure[]{
\includegraphics[width=1.5in]{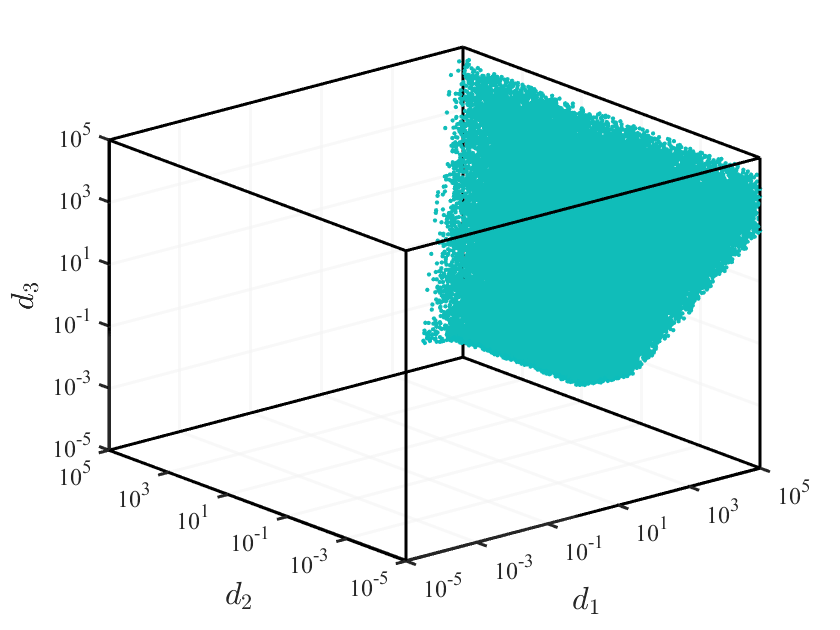}
}
\subfigure[]{
\includegraphics[width=1.5in]{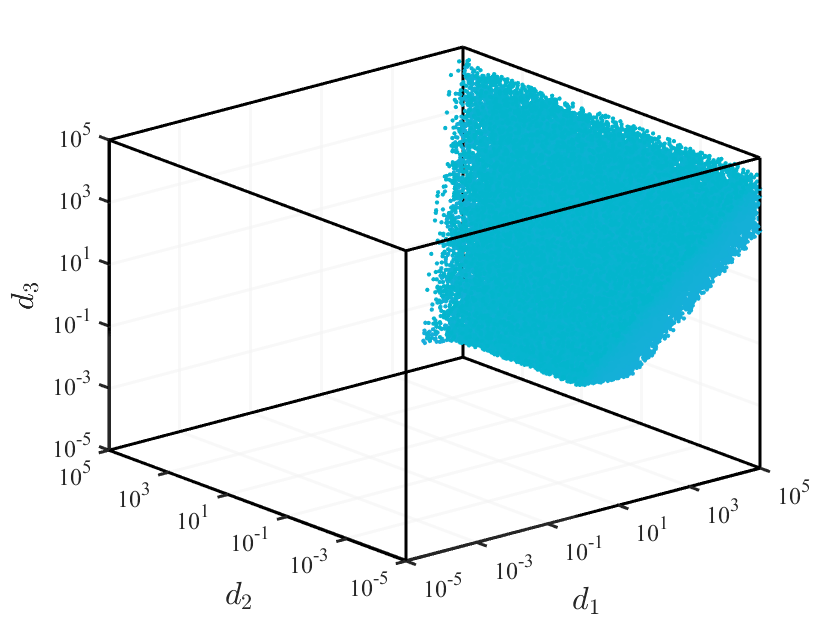}
}
\subfigure[]{
\includegraphics[width=1.5in]{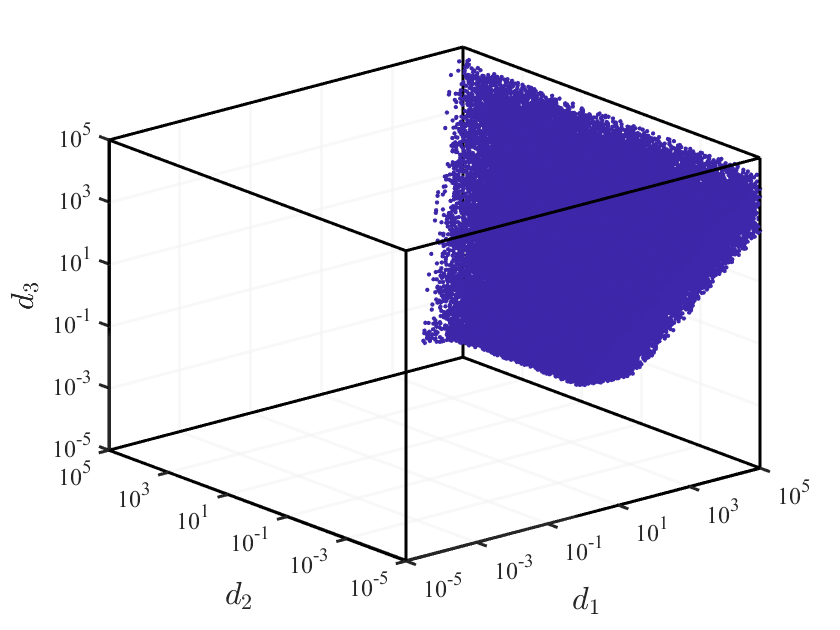}
}
\subfigure[]{
\includegraphics[width=1.5in]{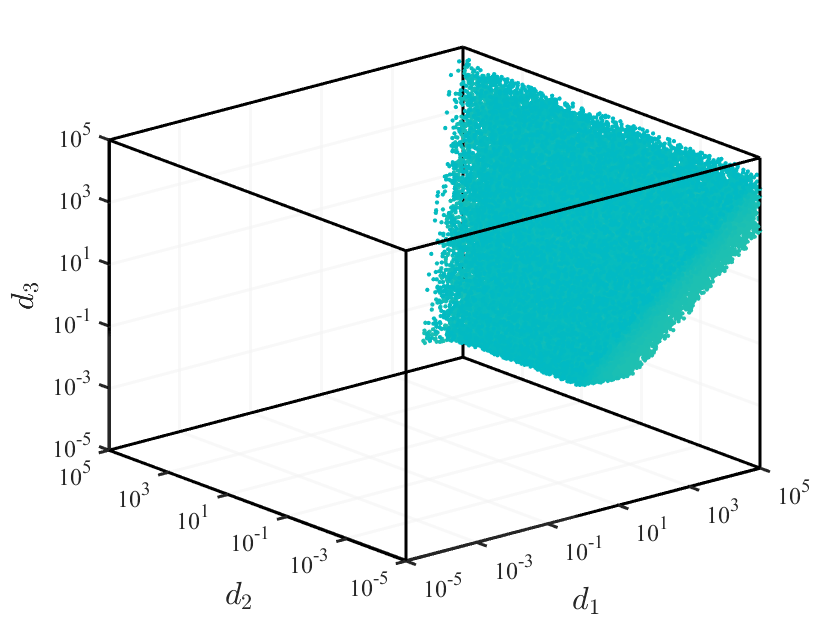}
}
\\
\vspace{-0.3cm}
\subfigure[]{
\includegraphics[width=1.5in]{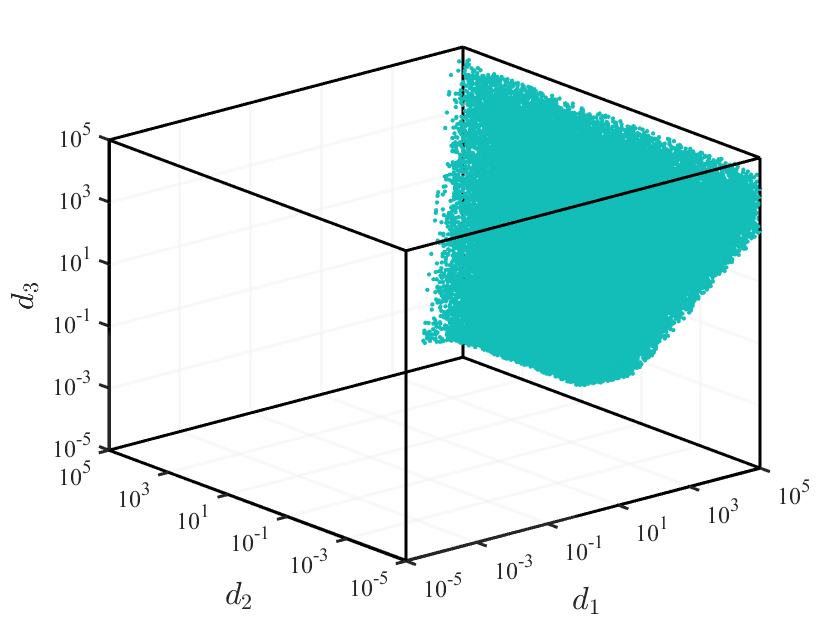}
}
\subfigure[]{
\includegraphics[width=1.5in]{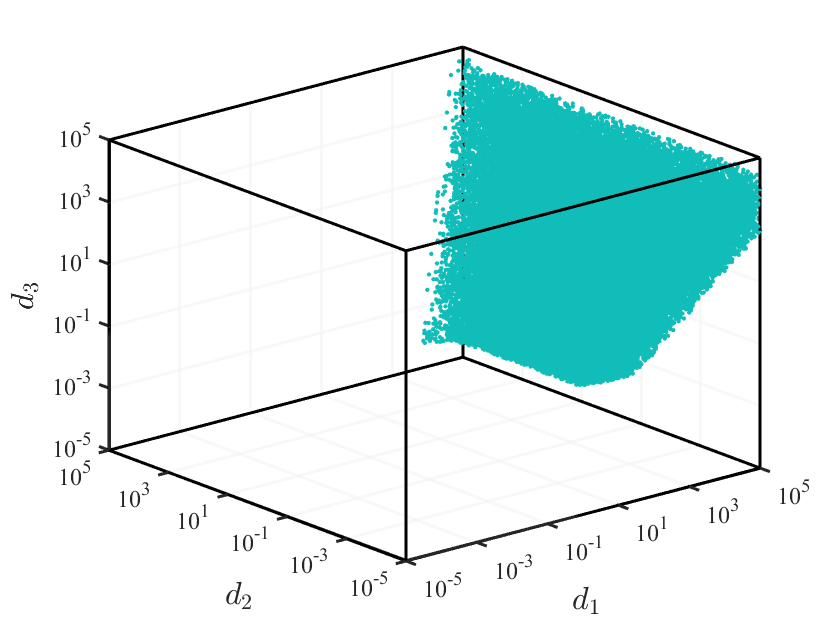}
}
\subfigure[]{
\includegraphics[width=1.5in]{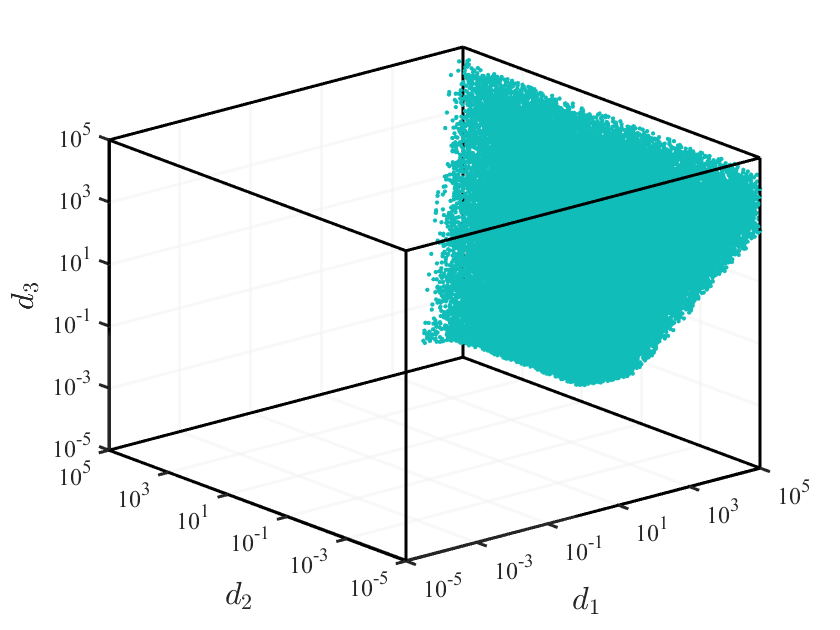}
}
\subfigure[]{
\includegraphics[width=1.5in]{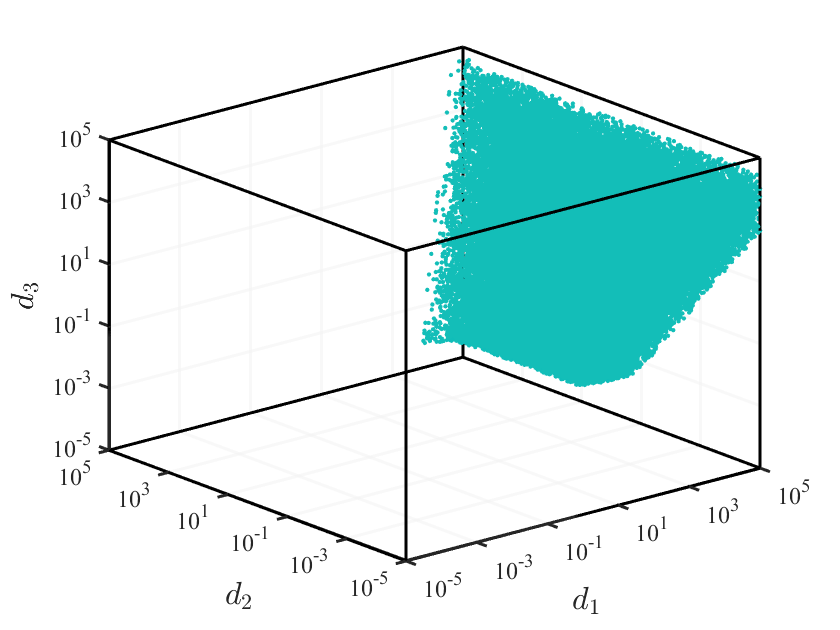}
}
\\
\vspace{-0.3cm}
\caption{(Color online) Stable subregions in the $d_1$-$d_2$-$d_3$ structural parameter
space for critical fermion-fermion interaction values at $l=l_c$
(color mapping follows the same scale as Fig.~\ref{Fig_lambda03_d123}(a)):
(a) $\lambda_{00} = -1.414$, (b) $\lambda_{01} = 0$, (c) $\lambda_{02} = 0$, (d) $\lambda_{03} = -1.414$,
(e) $\lambda_{10} = 1.414$, (f) $\lambda_{11} = 0$, (g) $\lambda_{12} = 0$, and (h) $\lambda_{13} = 1.414$,
which constitute the $\mathrm{FP}_1$ in $\mathrm{Case}$ $\mathrm{III}$, and
(i) $\lambda_{00} = -0.003$, (j) $\lambda_{01} = -0.558$, (k) $\lambda_{02} = -3.962$, (l) $\lambda_{03} = 0.029$,
(m) $\lambda_{10} = 0$, (n) $\lambda_{11} = 0$, (o) $\lambda_{12} = 0$, and (p) $\lambda_{13} = 0$,
which constitute the $\mathrm{FP}_7$ in $\mathrm{Case}$ $\mathrm{III}$~(\ref{Eq_FP7}).}
\label{Fig_FP1-7-III}
\end{figure*}
\begin{figure*}[htbp]
\subfigure[]{
\includegraphics[width=1.5in]{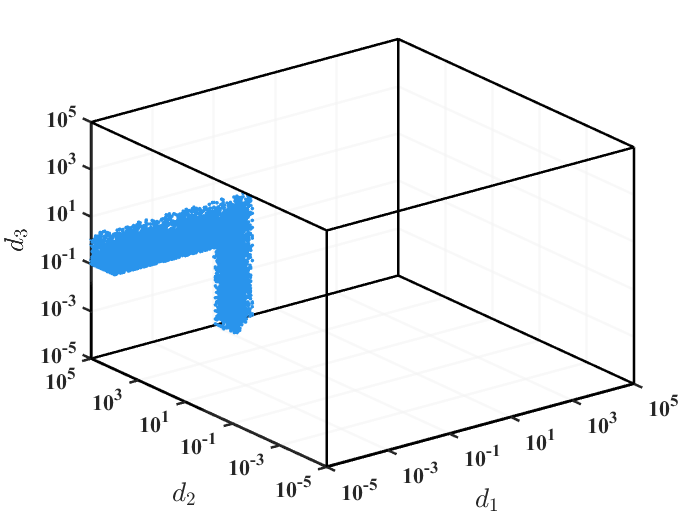}
}
\subfigure[]{
\includegraphics[width=1.5in]{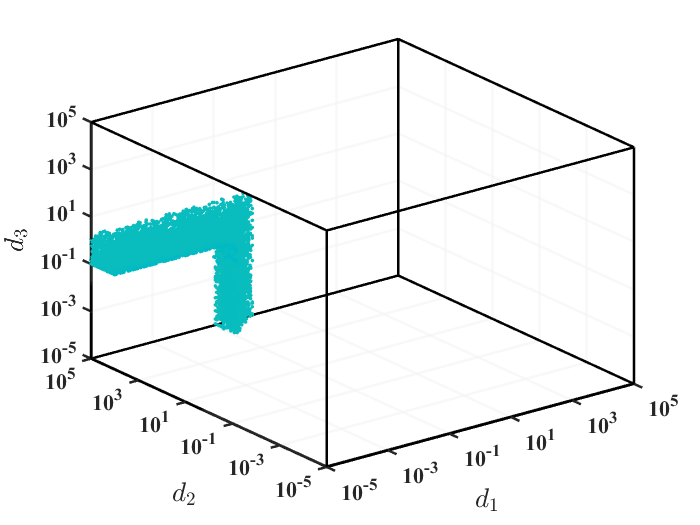}
}
\subfigure[]{
\includegraphics[width=1.5in]{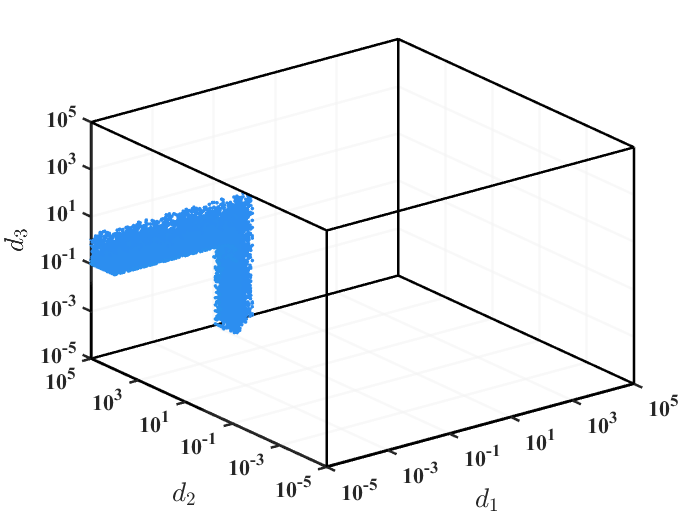}
}
\subfigure[]{
\includegraphics[width=1.5in]{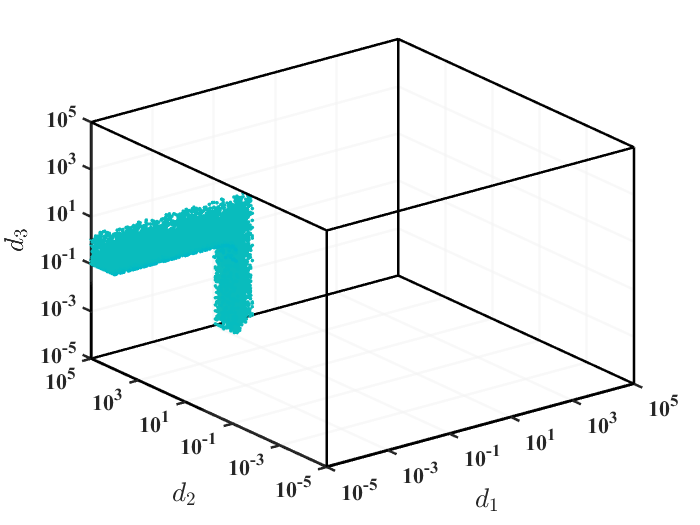}
}
\\
\vspace{-0.3cm}
\subfigure[]{
\includegraphics[width=1.5in]{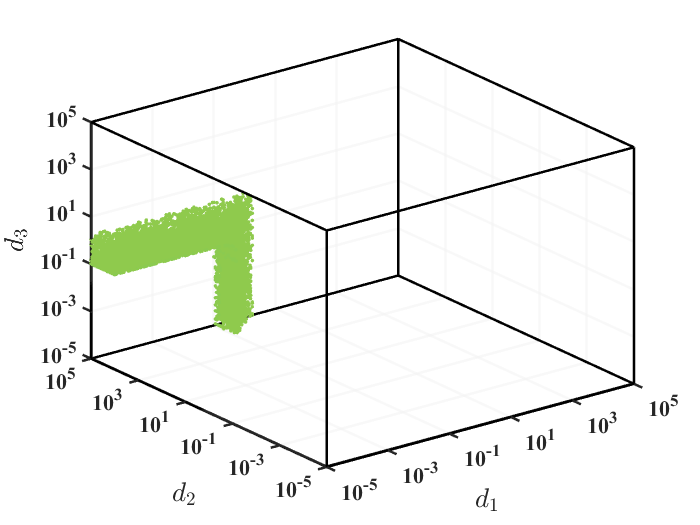}
}
\subfigure[]{
\includegraphics[width=1.5in]{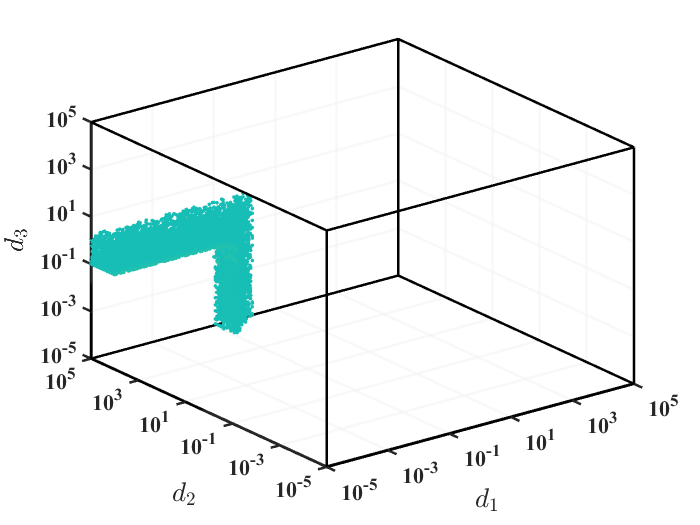}
}
\subfigure[]{
\includegraphics[width=1.5in]{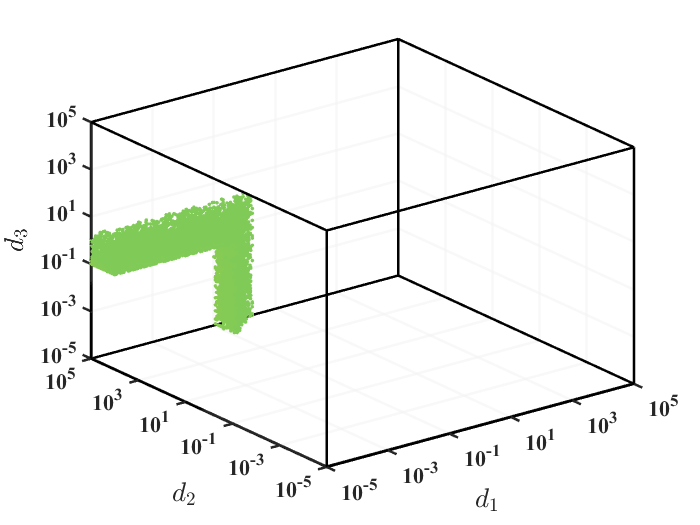}
}
\subfigure[]{
\includegraphics[width=1.5in]{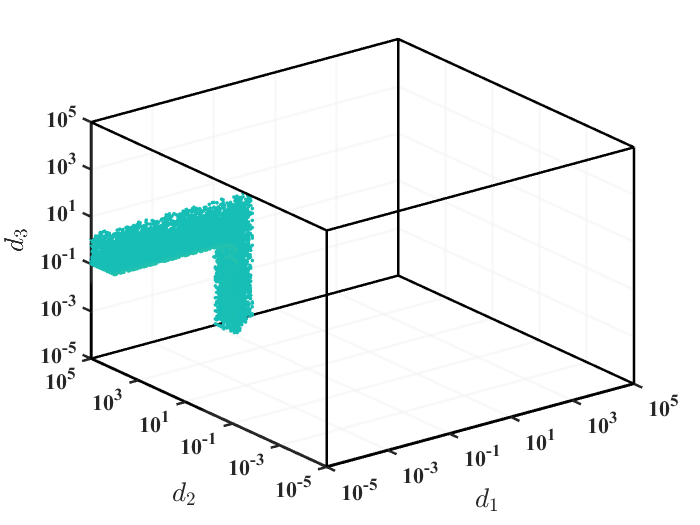}
}
\\
\vspace{-0.3cm}
\subfigure[]{
\includegraphics[width=1.5in]{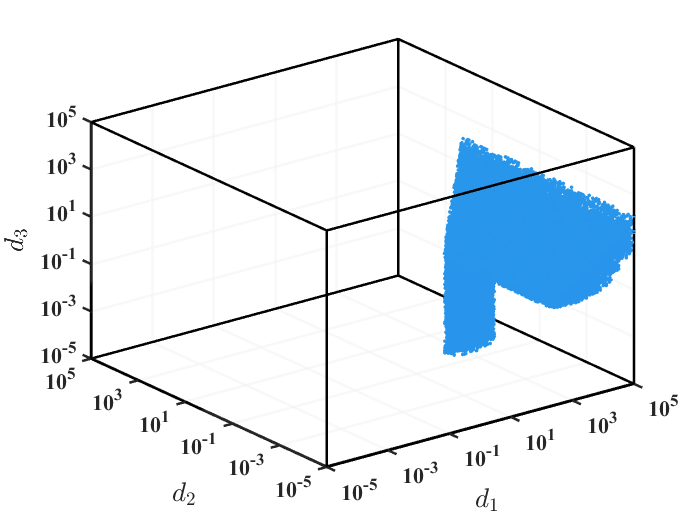}
}
\subfigure[]{
\includegraphics[width=1.5in]{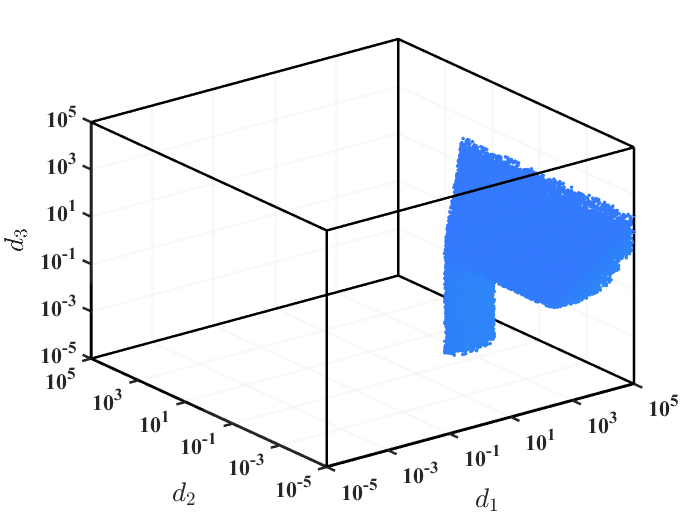}
}
\subfigure[]{
\includegraphics[width=1.5in]{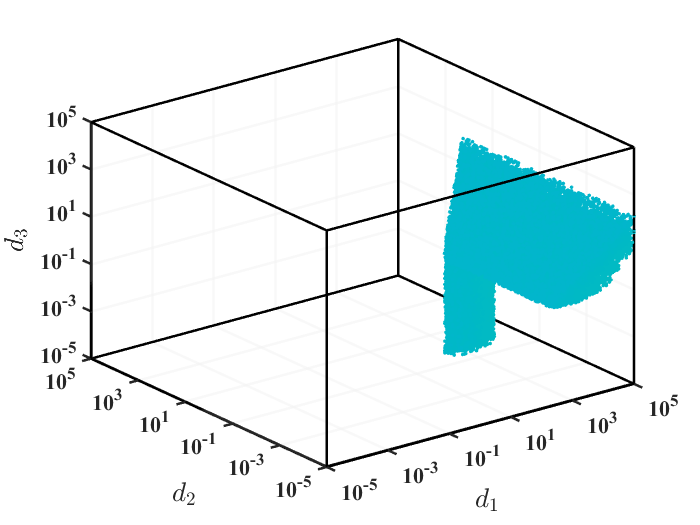}
}
\subfigure[]{
\includegraphics[width=1.5in]{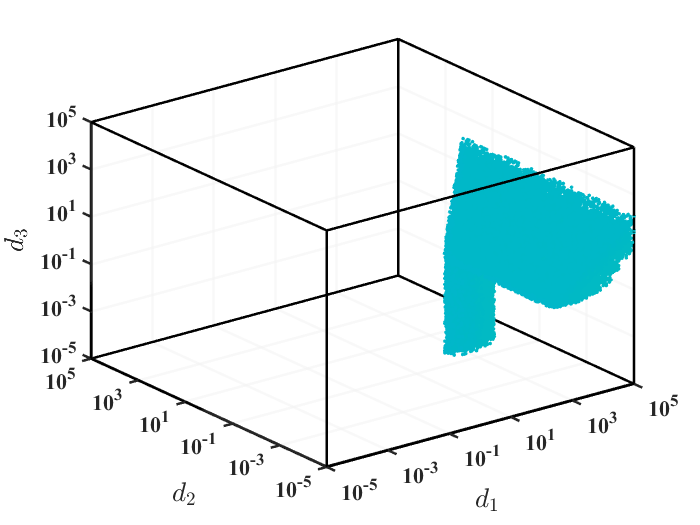}
}
\\
\vspace{-0.3cm}
\subfigure[]{
\includegraphics[width=1.5in]{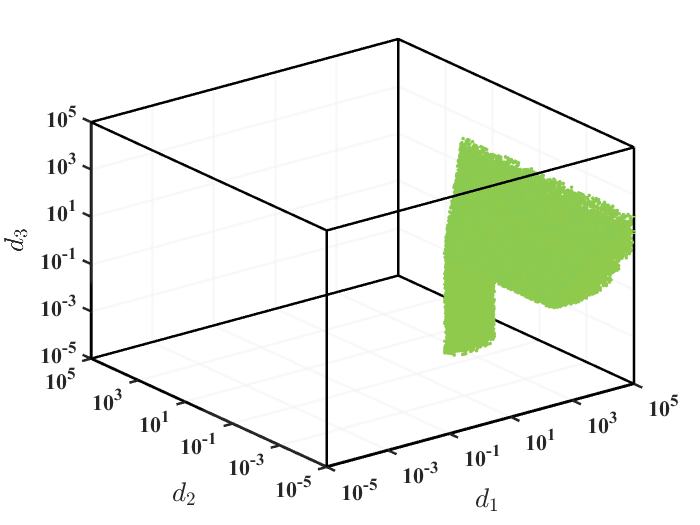}
}
\subfigure[]{
\includegraphics[width=1.5in]{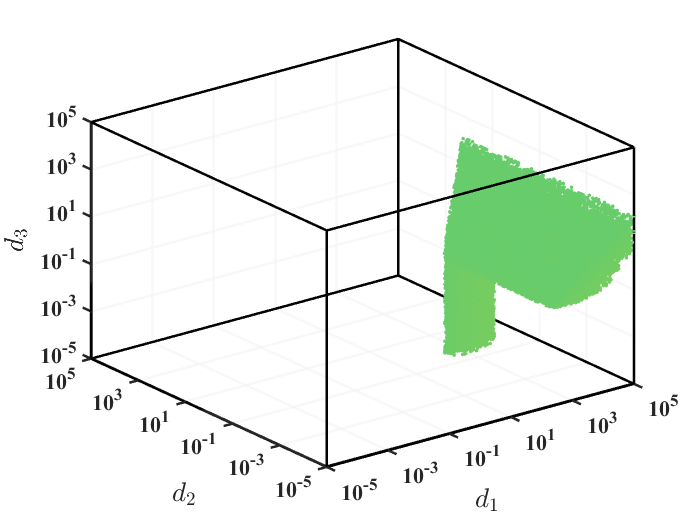}
}
\subfigure[]{
\includegraphics[width=1.5in]{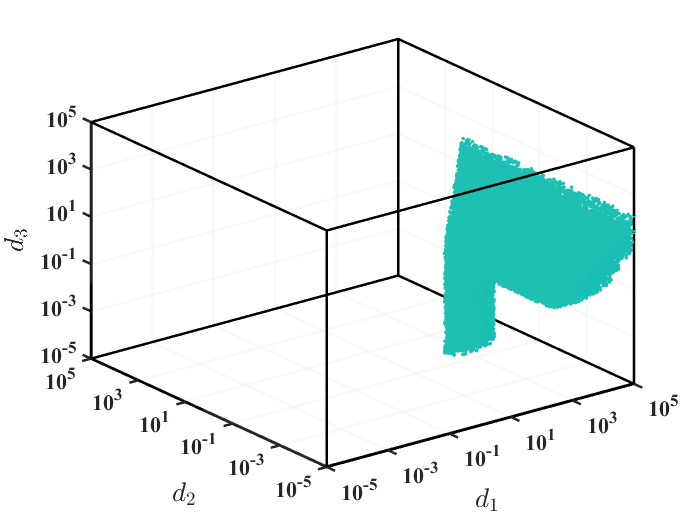}
}
\subfigure[]{
\includegraphics[width=1.5in]{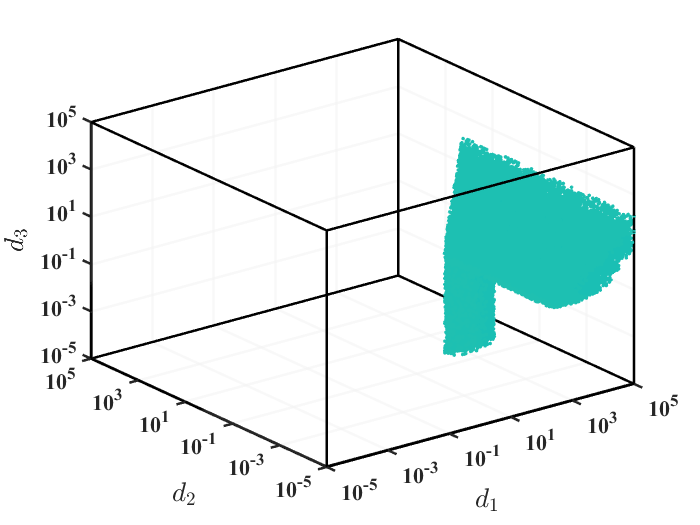}
}
\\
\vspace{-0.3cm}
\caption{(Color online) Stable subregions in the $d_1$-$d_2$-$d_3$ structural parameter
space for critical fermion-fermion interaction values at $l=l_c$
[color mapping follows the same scale as Fig.~\ref{Fig_lambda03_d123}(a)]:
(a) $\lambda_{00} = -0.001$, (b) $\lambda_{01} = -0.525$, (c) $\lambda_{02} = -0.010$, (d) $\lambda_{03} = -3.962$,
(e) $\lambda_{10} = 0$, (f) $\lambda_{11} = 0$, (g) $\lambda_{12} = 0$, and (h) $\lambda_{13} = 0$,
which constitute the $\mathrm{FP}_8$ in $\mathrm{Case}$ $\mathrm{III}$~(\ref{Eq_FP8}), and
(i) $\lambda_{00} = -1.366$, (j) $\lambda_{01} = -1.867$, (k) $\lambda_{02} = -0.252$, (l) $\lambda_{03} = -0.248$,
(m) $\lambda_{10} = 1.450$, (n) $\lambda_{11} = 1.175$, (o) $\lambda_{12} = 0.098$, and (p) $\lambda_{13} = 0.098$,
which constitute the $\mathrm{FP}_9$ in $\mathrm{Case}$ $\mathrm{III}$~(\ref{Eq_FP9}).}
\label{Fig_FP8-9-III}
\end{figure*}
\begin{figure*}[htbp]
\subfigure[]{
\includegraphics[width=1.5in]{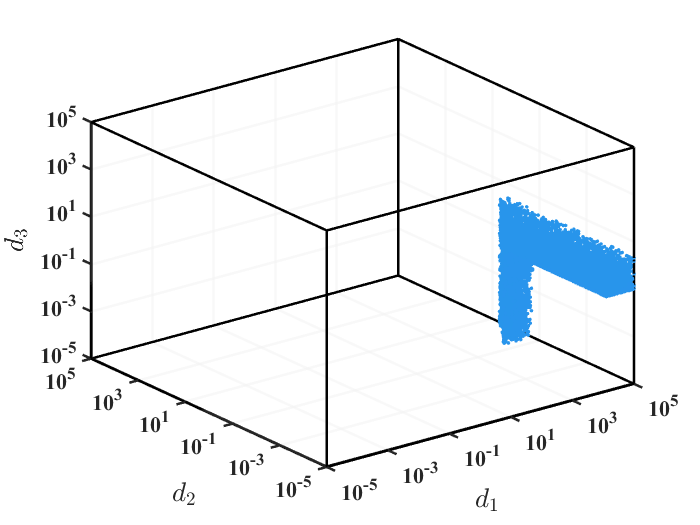}
}
\subfigure[]{
\includegraphics[width=1.5in]{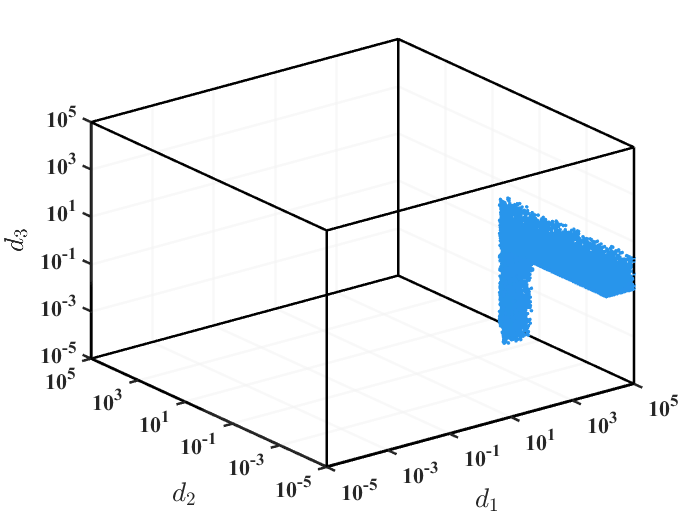}
}
\subfigure[]{
\includegraphics[width=1.5in]{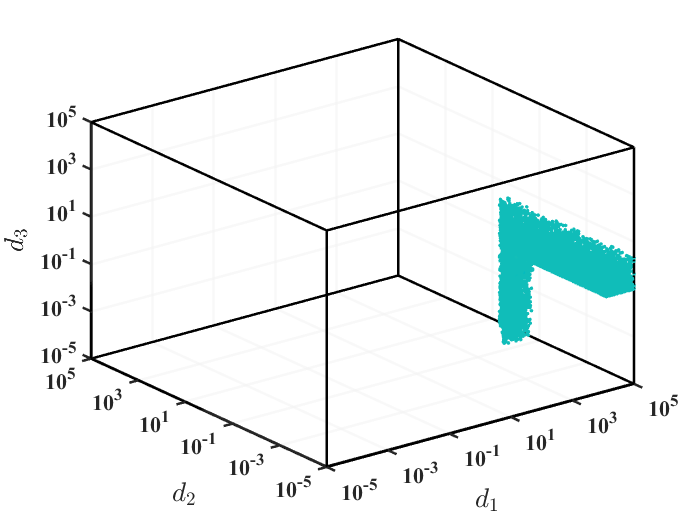}
}
\subfigure[]{
\includegraphics[width=1.5in]{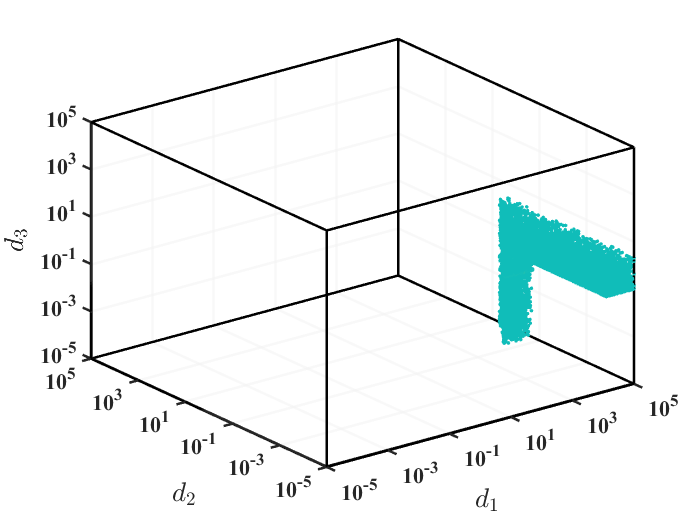}
}
\\
\vspace{-0.3cm}
\subfigure[]{
\includegraphics[width=1.5in]{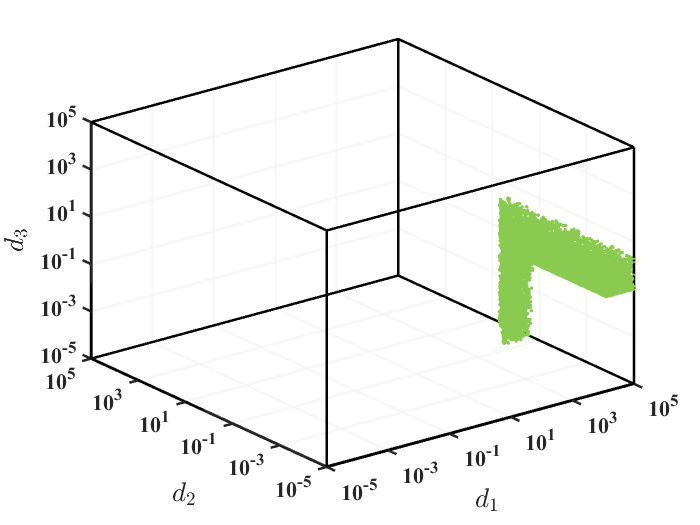}
}
\subfigure[]{
\includegraphics[width=1.5in]{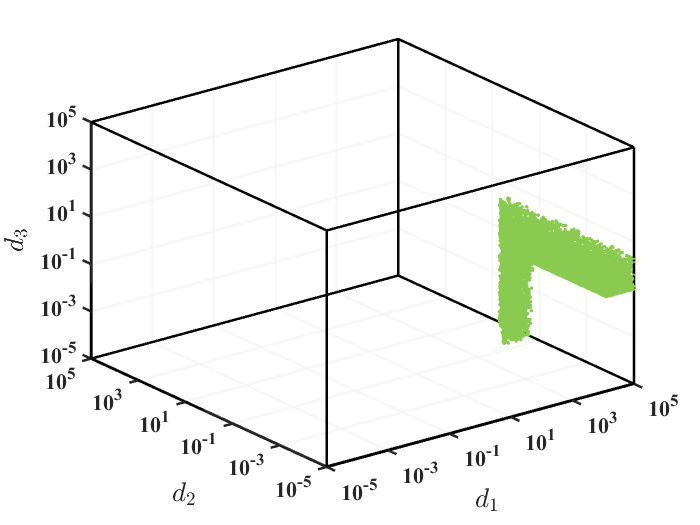}
}
\subfigure[]{
\includegraphics[width=1.5in]{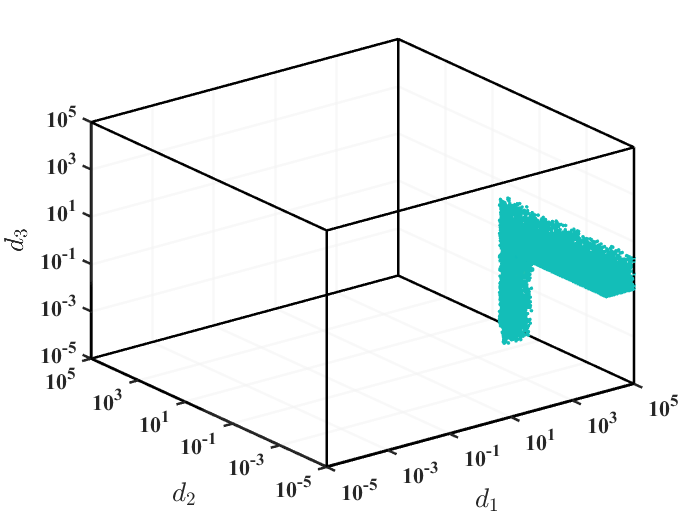}
}
\subfigure[]{
\includegraphics[width=1.5in]{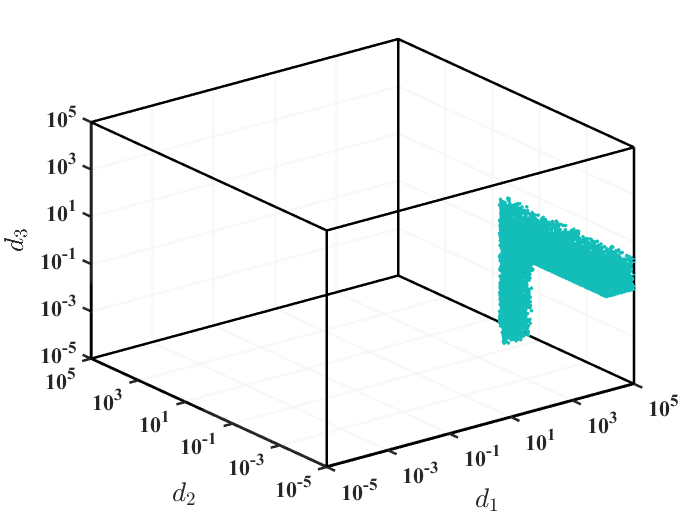}
}
\vspace{-0.3cm}
\caption{(Color online) Stable subregions in the $d_1$-$d_2$-$d_3$ structural parameter
space for critical fermion-fermion interaction values at $l=l_c$
[color mapping follows the same scale as Fig.~\ref{Fig_lambda03_d123}(a)]:
(a) $\lambda_{00} = -1.384$, (b) $\lambda_{01} = -1.384$, (c) $\lambda_{02} = 0.001$, (d) $\lambda_{03} = -0.001$,
(e) $\lambda_{10} = 1.424$, (f) $\lambda_{11} = 1.424$, (g) $\lambda_{12} = -0.001$, and
(h) $\lambda_{13} = 0.001$, which constitute $\mathrm{FP}_{10}$ in $\mathrm{Case}$ $\mathrm{III}$~(\ref{Eq_FP10}).}
\label{Fig_FP10-III}
\end{figure*}
\begin{widetext}
\begin{small}
\begin{eqnarray}
\mathcal{F}_{00}
&=&\frac{1}{16\pi^2}[(\lambda_{00}\lambda_{00}+\lambda_{01}\lambda_{01}+\lambda_{02}\lambda_{02}+\lambda_{03}\lambda_{03}+\lambda_{10}\lambda_{10}+\lambda_{11}\lambda_{11}+\lambda_{12}\lambda_{12}+\lambda_{13}\lambda_{13}+\lambda_{20}\lambda_{20}+\lambda_{21}\lambda_{21}+\lambda_{22}\lambda_{22}+\lambda_{23}\lambda_{23}\nonumber\\
&&+\lambda_{30}\lambda_{30}+\lambda_{31}\lambda_{31}+\lambda_{32}\lambda_{32}+\lambda_{33}\lambda_{33})I_0-(2\lambda_{00}\lambda_{01}+2\lambda_{10}\lambda_{11}+2\lambda_{20}\lambda_{21}+2\lambda_{30}\lambda_{31}+2\lambda_{02}\lambda_{03}+2\lambda_{12}\lambda_{13}+2\lambda_{22}\lambda_{23}\nonumber\\
&&+2\lambda_{32}\lambda_{33})I_1-(2\lambda_{00}\lambda_{02}+2\lambda_{10}\lambda_{12}+2\lambda_{20}\lambda_{22}+2\lambda_{30}\lambda_{32}+2\lambda_{01}\lambda_{03}+2\lambda_{11}\lambda_{13}+2\lambda_{21}\lambda_{23}+2\lambda_{31}\lambda_{33})I_2-(2\lambda_{00}\lambda_{03}+2\lambda_{10}\lambda_{13}\nonumber\\
&&+2\lambda_{20}\lambda_{23}+2\lambda_{30}\lambda_{33}+2\lambda_{01}\lambda_{02}+2\lambda_{11}\lambda_{12}+2\lambda_{21}\lambda_{22}+2\lambda_{31}\lambda_{32})I_3+(\lambda_{00}\lambda_{00}+\lambda_{01}\lambda_{01}+\lambda_{02}\lambda_{02}+\lambda_{03}\lambda_{03}+\lambda_{10}\lambda_{10}+\lambda_{11}\lambda_{11}\nonumber\\
&&+\lambda_{12}\lambda_{12}+\lambda_{13}\lambda_{13}+\lambda_{20}\lambda_{20}+\lambda_{21}\lambda_{21}+\lambda_{22}\lambda_{22}+\lambda_{23}\lambda_{23}+\lambda_{30}\lambda_{30}+\lambda_{31}\lambda_{31}+\lambda_{32}\lambda_{32}+\lambda_{33}\lambda_{33})J_0+(2\lambda_{00}\lambda_{01}+2\lambda_{10}\lambda_{11}\nonumber\\
&&+2\lambda_{20}\lambda_{21}+2\lambda_{30}\lambda_{31}-2\lambda_{02}\lambda_{03}-2\lambda_{12}\lambda_{13}-2\lambda_{22}\lambda_{23}-2\lambda_{32}\lambda_{33})J_1+(2\lambda_{00}\lambda_{02}+2\lambda_{10}\lambda_{12}+2\lambda_{20}\lambda_{22}+2\lambda_{30}\lambda_{32}-2\lambda_{01}\lambda_{03}\nonumber\\
&&-2\lambda_{11}\lambda_{13}-2\lambda_{21}\lambda_{23}-2\lambda_{31}\lambda_{33})J_2+(2\lambda_{00}\lambda_{03}
+2\lambda_{10}\lambda_{13}+2\lambda_{20}\lambda_{23}+2\lambda_{30}\lambda_{33}-2\lambda_{01}\lambda_{02}-2\lambda_{11}\lambda_{12}-2\lambda_{21}\lambda_{22}\nonumber\\
&&-2\lambda_{31}\lambda_{32})J_3],\label{Eq_F00}\\
\mathcal{F}_{01}
&=&\frac{1}{16\pi^2}[4(\lambda_{01}\lambda_{00}-3\lambda_{01}\lambda_{01}-\lambda_{01}\lambda_{02}-\lambda_{01}\lambda_{03}+\lambda_{01}\lambda_{10}+\lambda_{01}\lambda_{11}-\lambda_{01}\lambda_{12}-\lambda_{01}\lambda_{13}+\lambda_{01}\lambda_{20}+\lambda_{01}\lambda_{21}-\lambda_{01}\lambda_{22}-\lambda_{01}\lambda_{23}\nonumber\\
&&+\lambda_{01}\lambda_{30}+\lambda_{01}\lambda_{31}-\lambda_{01}\lambda_{32}-\lambda_{01}\lambda_{33})(I_2+I_3)+(2\lambda_{00}\lambda_{01}+2\lambda_{10}\lambda_{11}+2\lambda_{20}\lambda_{21}+2\lambda_{30}\lambda_{31}+2\lambda_{02}\lambda_{03}+2\lambda_{12}\lambda_{13}+2\lambda_{22}\lambda_{23}\nonumber\\
&&+2\lambda_{32}\lambda_{33})I_0-(\lambda_{00}\lambda_{00}+\lambda_{01}\lambda_{01}+\lambda_{02}\lambda_{02}+\lambda_{03}\lambda_{03}+\lambda_{10}\lambda_{10}+\lambda_{11}\lambda_{11}+\lambda_{12}\lambda_{12}+\lambda_{13}\lambda_{13}+\lambda_{20}\lambda_{20}+\lambda_{21}\lambda_{21}+\lambda_{22}\lambda_{22}\nonumber\\
&&+\lambda_{23}\lambda_{23}+\lambda_{30}\lambda_{30}+\lambda_{31}\lambda_{31}+\lambda_{32}\lambda_{32}+\lambda_{33}\lambda_{33})I_1-(2\lambda_{01}\lambda_{02}+2\lambda_{11}\lambda_{12}+2\lambda_{21}\lambda_{22}+2\lambda_{31}\lambda_{32}+2\lambda_{00}\lambda_{03}+2\lambda_{10}\lambda_{13}+2\lambda_{20}\lambda_{23}\nonumber\\
&&+2\lambda_{30}\lambda_{33})I_2-(2\lambda_{01}\lambda_{03}+2\lambda_{11}\lambda_{13}+2\lambda_{21}\lambda_{23}+2\lambda_{31}\lambda_{33}+2\lambda_{02}\lambda_{00}+2\lambda_{12}\lambda_{10}+2\lambda_{22}\lambda_{20}+2\lambda_{32}\lambda_{30})I_3+(2\lambda_{00}\lambda_{01}+2\lambda_{10}\lambda_{11}\nonumber\\
&&+2\lambda_{20}\lambda_{21}+2\lambda_{30}\lambda_{31}-2\lambda_{02}\lambda_{03}-2\lambda_{12}\lambda_{13}-2\lambda_{22}\lambda_{23}-2\lambda_{32}\lambda_{33})J_0+(\lambda_{00}\lambda_{00}+\lambda_{01}\lambda_{01}+\lambda_{02}\lambda_{02}+\lambda_{03}\lambda_{03}+\lambda_{10}\lambda_{10}+\lambda_{11}\lambda_{11}\nonumber\\
&&+\lambda_{12}\lambda_{12}+\lambda_{13}\lambda_{13}+\lambda_{20}\lambda_{20}+\lambda_{21}\lambda_{21}+\lambda_{22}\lambda_{22}+\lambda_{23}\lambda_{23}+\lambda_{30}\lambda_{30}+\lambda_{31}\lambda_{31}+\lambda_{32}\lambda_{32}+\lambda_{33}\lambda_{33})J_1+(-2\lambda_{01}\lambda_{02}-2\lambda_{11}\lambda_{12}\nonumber\\
&&-2\lambda_{21}\lambda_{22}-2\lambda_{31}\lambda_{32}+2\lambda_{00}\lambda_{03}+2\lambda_{10}\lambda_{13}+2\lambda_{20}\lambda_{23}+2\lambda_{30}\lambda_{33})J_2+(-2\lambda_{01}\lambda_{03}-2\lambda_{11}\lambda_{13}-2\lambda_{21}\lambda_{23}-2\lambda_{31}\lambda_{33}+2\lambda_{02}\lambda_{00}\nonumber\\
&&+2\lambda_{12}\lambda_{10}+2\lambda_{22}\lambda_{20}+2\lambda_{32}\lambda_{30})J_3],\\
\mathcal{F}_{02}
&=&\frac{1}{16\pi^2}[4(\lambda_{02}\lambda_{00}-\lambda_{02}\lambda_{01}-3\lambda_{02}\lambda_{02}-\lambda_{02}\lambda_{03}+\lambda_{02}\lambda_{10}-\lambda_{02}\lambda_{11}+\lambda_{02}\lambda_{12}-\lambda_{02}\lambda_{13}+\lambda_{02}\lambda_{20}-\lambda_{02}\lambda_{21}+\lambda_{02}\lambda_{22}-\lambda_{02}\lambda_{23}\nonumber\\
&&+\lambda_{02}\lambda_{30}-\lambda_{02}\lambda_{31}+\lambda_{02}\lambda_{32}-\lambda_{02}\lambda_{33})(I_1+I_3)+(2\lambda_{00}\lambda_{02}+2\lambda_{10}\lambda_{12}+2\lambda_{20}\lambda_{22}+2\lambda_{30}\lambda_{32}+2\lambda_{01}\lambda_{03}+2\lambda_{11}\lambda_{13}+2\lambda_{21}\lambda_{23}\nonumber\\
&&+2\lambda_{31}\lambda_{33})I_0-(2\lambda_{01}\lambda_{02}+2\lambda_{11}\lambda_{12}+2\lambda_{21}\lambda_{22}+2\lambda_{31}\lambda_{32}+2\lambda_{00}\lambda_{03}+2\lambda_{10}\lambda_{13}+2\lambda_{20}\lambda_{23}+2\lambda_{30}\lambda_{33})I_1-(\lambda_{00}\lambda_{00}+\lambda_{01}\lambda_{01}\nonumber\\
&&+\lambda_{02}\lambda_{02}+\lambda_{03}\lambda_{03}+\lambda_{10}\lambda_{10}+\lambda_{11}\lambda_{11}+\lambda_{12}\lambda_{12}+\lambda_{13}\lambda_{13}+\lambda_{20}\lambda_{20}+\lambda_{21}\lambda_{21}+\lambda_{22}\lambda_{22}+\lambda_{23}\lambda_{23}+\lambda_{30}\lambda_{30}+\lambda_{31}\lambda_{31}+\lambda_{32}\lambda_{32}\nonumber\\
&&+\lambda_{33}\lambda_{33})I_2-(2\lambda_{02}\lambda_{03}+2\lambda_{12}\lambda_{13}+2\lambda_{22}\lambda_{23}+2\lambda_{32}\lambda_{33}+2\lambda_{00}\lambda_{01}+2\lambda_{10}\lambda_{11}+2\lambda_{20}\lambda_{21}+2\lambda_{30}\lambda_{31})I_3+(2\lambda_{00}\lambda_{02}+2\lambda_{10}\lambda_{12}\nonumber\\
&&+2\lambda_{20}\lambda_{22}+2\lambda_{30}\lambda_{32}-2\lambda_{01}\lambda_{03}-2\lambda_{11}\lambda_{13}-2\lambda_{21}\lambda_{23}-2\lambda_{31}\lambda_{33})J_0+(-2\lambda_{01}\lambda_{02}-2\lambda_{11}\lambda_{12}-2\lambda_{21}\lambda_{22}-2\lambda_{31}\lambda_{32}+2\lambda_{00}\lambda_{03}\nonumber\\
&&+2\lambda_{10}\lambda_{13}+2\lambda_{20}\lambda_{23}+2\lambda_{30}\lambda_{33})J_1+(\lambda_{00}\lambda_{00}+\lambda_{01}\lambda_{01}+\lambda_{02}\lambda_{02}+\lambda_{03}\lambda_{03}+\lambda_{10}\lambda_{10}+\lambda_{11}\lambda_{11}+\lambda_{12}\lambda_{12}+\lambda_{13}\lambda_{13}+\lambda_{20}\lambda_{20}\nonumber\\
&&+\lambda_{21}\lambda_{21}+\lambda_{22}\lambda_{22}+\lambda_{23}\lambda_{23}+\lambda_{30}\lambda_{30}+\lambda_{31}\lambda_{31}+\lambda_{32}\lambda_{32}+\lambda_{33}\lambda_{33})J_2+(-2\lambda_{02}\lambda_{03}-2\lambda_{12}\lambda_{13}-2\lambda_{22}\lambda_{23}-2\lambda_{32}\lambda_{33}+2\lambda_{00}\lambda_{01}\nonumber\\
&&+2\lambda_{10}\lambda_{11}+2\lambda_{20}\lambda_{21}+2\lambda_{30}\lambda_{31})J_3],\\
\mathcal{F}_{03}
&=&\frac{1}{16\pi^2}[4(\lambda_{03}\lambda_{00}-\lambda_{03}\lambda_{01}-\lambda_{03}\lambda_{02}-3\lambda_{03}\lambda_{03}+\lambda_{03}\lambda_{10}-\lambda_{03}\lambda_{11}-\lambda_{03}\lambda_{12}+\lambda_{03}\lambda_{13}+\lambda_{03}\lambda_{20}-\lambda_{03}\lambda_{21}-\lambda_{03}\lambda_{22}+\lambda_{03}\lambda_{23}\nonumber\\
&&+\lambda_{03}\lambda_{30}-\lambda_{03}\lambda_{31}-\lambda_{03}\lambda_{32}+\lambda_{03}\lambda_{33})(I_1+I_2)+(2\lambda_{00}\lambda_{03}+2\lambda_{10}\lambda_{13}+2\lambda_{20}\lambda_{23}+2\lambda_{30}\lambda_{33}+2\lambda_{01}\lambda_{02}+2\lambda_{11}\lambda_{12}+2\lambda_{21}\lambda_{22}\nonumber\\
&&+2\lambda_{31}\lambda_{32})I_0-(2\lambda_{01}\lambda_{03}+2\lambda_{11}\lambda_{13}+2\lambda_{21}\lambda_{23}+2\lambda_{31}\lambda_{33}+2\lambda_{00}\lambda_{02}+2\lambda_{10}\lambda_{12}+2\lambda_{20}\lambda_{22}+2\lambda_{30}\lambda_{32})I_1-(2\lambda_{02}\lambda_{03}+2\lambda_{12}\lambda_{13}\nonumber\\
&&+2\lambda_{22}\lambda_{23}+2\lambda_{32}\lambda_{33}+2\lambda_{00}\lambda_{01}+2\lambda_{10}\lambda_{11}+2\lambda_{20}\lambda_{21}+2\lambda_{30}\lambda_{31})I_2-(\lambda_{00}\lambda_{00}+\lambda_{01}\lambda_{01}+\lambda_{02}\lambda_{02}+\lambda_{03}\lambda_{03}+\lambda_{10}\lambda_{10}+\lambda_{11}\lambda_{11}\nonumber\\
&&+\lambda_{12}\lambda_{12}+\lambda_{13}\lambda_{13}+\lambda_{20}\lambda_{20}+\lambda_{21}\lambda_{21}+\lambda_{22}\lambda_{22}+\lambda_{23}\lambda_{23}+\lambda_{30}\lambda_{30}+\lambda_{31}\lambda_{31}+\lambda_{32}\lambda_{32}+\lambda_{33}\lambda_{33})I_3+(2\lambda_{00}\lambda_{03}+2\lambda_{10}\lambda_{13}\nonumber\\
&&+2\lambda_{20}\lambda_{23}+2\lambda_{30}\lambda_{33}-2\lambda_{01}\lambda_{02}-2\lambda_{11}\lambda_{12}-2\lambda_{21}\lambda_{22}-2\lambda_{31}\lambda_{32})J_0+(-2\lambda_{01}\lambda_{03}-2\lambda_{11}\lambda_{13}-2\lambda_{21}\lambda_{23}-2\lambda_{31}\lambda_{33}+2\lambda_{00}\lambda_{02}\nonumber\\
&&+2\lambda_{10}\lambda_{12}+2\lambda_{20}\lambda_{22}+2\lambda_{30}\lambda_{32})J_1+(-2\lambda_{02}\lambda_{03}-2\lambda_{12}\lambda_{13}-2\lambda_{22}\lambda_{23}-2\lambda_{32}\lambda_{33}+2\lambda_{00}\lambda_{01}+2\lambda_{10}\lambda_{11}+2\lambda_{20}\lambda_{21}+2\lambda_{30}\lambda_{31})J_2\nonumber\\
&&+(\lambda_{00}\lambda_{00}+\lambda_{01}\lambda_{01}+\lambda_{02}\lambda_{02}+\lambda_{03}\lambda_{03}+\lambda_{10}\lambda_{10}+\lambda_{11}\lambda_{11}+\lambda_{12}\lambda_{12}+\lambda_{13}\lambda_{13}+\lambda_{20}\lambda_{20}+\lambda_{21}\lambda_{21}+\lambda_{22}\lambda_{22}+\lambda_{23}\lambda_{23}+\lambda_{30}\lambda_{30}\nonumber\\
&&+\lambda_{31}\lambda_{31}+\lambda_{32}\lambda_{32}+\lambda_{33}\lambda_{33})J_3],\\
\mathcal{F}_{10}
&=&\frac{1}{16\pi^2}[(2\lambda_{00}\lambda_{10}+2\lambda_{01}\lambda_{11}+2\lambda_{02}\lambda_{12}+2\lambda_{03}\lambda_{13}+2\lambda_{20}\lambda_{30}+2\lambda_{21}\lambda_{31}+2\lambda_{22}\lambda_{32}+2\lambda_{23}\lambda_{33})I_0-(2\lambda_{01}\lambda_{10}+2\lambda_{11}\lambda_{00}+2\lambda_{22}\lambda_{33}\nonumber\\
&&+2\lambda_{32}\lambda_{23}+2\lambda_{12}\lambda_{03}+2\lambda_{13}\lambda_{02}+2\lambda_{31}\lambda_{20}+2\lambda_{30}\lambda_{21})I_1-(2\lambda_{02}\lambda_{10}+2\lambda_{12}\lambda_{00}+2\lambda_{13}\lambda_{01}+2\lambda_{11}\lambda_{03}+2\lambda_{22}\lambda_{30}+2\lambda_{32}\lambda_{20}\nonumber\\
&&+2\lambda_{21}\lambda_{33}+2\lambda_{23}\lambda_{31})I_2-(2\lambda_{03}\lambda_{10}+2\lambda_{13}\lambda_{00}+2\lambda_{12}\lambda_{01}+2\lambda_{02}\lambda_{11}+2\lambda_{23}\lambda_{30}+2\lambda_{20}\lambda_{33}+2\lambda_{21}\lambda_{32}+2\lambda_{22}\lambda_{31})I_3+(2\lambda_{00}\lambda_{10}\nonumber\\
&&+2\lambda_{01}\lambda_{11}+2\lambda_{02}\lambda_{12}+2\lambda_{03}\lambda_{13}-2\lambda_{20}\lambda_{30}-2\lambda_{21}\lambda_{31}-2\lambda_{22}\lambda_{32}-2\lambda_{23}\lambda_{33})J_0+(2\lambda_{01}\lambda_{10}+2\lambda_{11}\lambda_{00}+2\lambda_{22}\lambda_{33}+2\lambda_{32}\lambda_{23}\nonumber\\
&&-2\lambda_{12}\lambda_{03}-2\lambda_{13}\lambda_{02}-2\lambda_{31}\lambda_{20}-2\lambda_{30}\lambda_{21})J_1+(2\lambda_{02}\lambda_{10}+2\lambda_{12}\lambda_{00}-2\lambda_{13}\lambda_{01}-2\lambda_{11}\lambda_{03}-2\lambda_{22}\lambda_{30}-2\lambda_{32}\lambda_{20}+2\lambda_{21}\lambda_{33}\nonumber\\
&&+2\lambda_{23}\lambda_{31})J_2+(2\lambda_{03}\lambda_{10}+2\lambda_{13}\lambda_{00}-2\lambda_{12}\lambda_{01}-2\lambda_{02}\lambda_{11}
-2\lambda_{23}\lambda_{30}-2\lambda_{20}\lambda_{33}+2\lambda_{21}\lambda_{32}+2\lambda_{22}\lambda_{31})J_3],\\
\mathcal{F}_{11}
&=&\frac{1}{16\pi^2}[4(\lambda_{11}\lambda_{00}+\lambda_{11}\lambda_{01}-\lambda_{11}\lambda_{02}-\lambda_{11}\lambda_{03}+\lambda_{11}\lambda_{10}-3\lambda_{11}\lambda_{11}-\lambda_{11}\lambda_{12}-\lambda_{11}\lambda_{13}-\lambda_{11}\lambda_{20}-\lambda_{11}\lambda_{21}+\lambda_{11}\lambda_{22}+\lambda_{11}\lambda_{23}\nonumber\\
&&-\lambda_{11}\lambda_{30}-\lambda_{11}\lambda_{31}+\lambda_{11}\lambda_{32}+\lambda_{11}\lambda_{33})(I_2+I_3)+(2\lambda_{00}\lambda_{11}+2\lambda_{01}\lambda_{10}+2\lambda_{20}\lambda_{31}+2\lambda_{21}\lambda_{30}+2\lambda_{12}\lambda_{03}+2\lambda_{13}\lambda_{02}+2\lambda_{22}\lambda_{33}\nonumber\\
&&+2\lambda_{23}\lambda_{32})I_0-(2\lambda_{10}\lambda_{00}+2\lambda_{20}\lambda_{30}+2\lambda_{11}\lambda_{01}+2\lambda_{12}\lambda_{02}+2\lambda_{13}\lambda_{03}+2\lambda_{21}\lambda_{31}+2\lambda_{22}\lambda_{32}+2\lambda_{23}\lambda_{33})I_1-(2\lambda_{03}\lambda_{10}+2\lambda_{13}\lambda_{00}\nonumber\\
&&+2\lambda_{23}\lambda_{30}+2\lambda_{20}\lambda_{33}+2\lambda_{11}\lambda_{02}+2\lambda_{12}\lambda_{01}+2\lambda_{21}\lambda_{32}+2\lambda_{22}\lambda_{31})I_2-(2\lambda_{03}\lambda_{11}+2\lambda_{13}\lambda_{01}+2\lambda_{23}\lambda_{31}+2\lambda_{21}\lambda_{33}+2\lambda_{12}\lambda_{00}\nonumber\\
&&+2\lambda_{10}\lambda_{02}+2\lambda_{20}\lambda_{32}+2\lambda_{22}\lambda_{30})I_3+(2\lambda_{00}\lambda_{11}+2\lambda_{01}\lambda_{10}-2\lambda_{20}\lambda_{31}-2\lambda_{21}\lambda_{30}-2\lambda_{12}\lambda_{03}-2\lambda_{13}\lambda_{02}+2\lambda_{22}\lambda_{33}+2\lambda_{23}\lambda_{32})J_0\nonumber\\
&&+(2\lambda_{10}\lambda_{00}-2\lambda_{20}\lambda_{30}+2\lambda_{11}\lambda_{01}+2\lambda_{12}\lambda_{02}+2\lambda_{13}\lambda_{03}-2\lambda_{21}\lambda_{31}-2\lambda_{22}\lambda_{32}-2\lambda_{23}\lambda_{33})J_1+(2\lambda_{03}\lambda_{10}+2\lambda_{13}\lambda_{00}-2\lambda_{23}\lambda_{30}\nonumber\\
&&-2\lambda_{20}\lambda_{33}-2\lambda_{11}\lambda_{02}-2\lambda_{12}\lambda_{01}+2\lambda_{21}\lambda_{32}+2\lambda_{22}\lambda_{31})J_2+(-2\lambda_{03}\lambda_{11}-2\lambda_{13}\lambda_{01}+2\lambda_{23}\lambda_{31}+2\lambda_{21}\lambda_{33}+2\lambda_{12}\lambda_{00}+2\lambda_{10}\lambda_{02}\nonumber\\
&&-2\lambda_{20}\lambda_{32}-2\lambda_{22}\lambda_{30})J_3],\\
\mathcal{F}_{12}
&=&\frac{1}{16\pi^2}[4(\lambda_{12}\lambda_{00}-\lambda_{12}\lambda_{01}+\lambda_{12}\lambda_{02}-\lambda_{12}\lambda_{03}+\lambda_{12}\lambda_{10}-\lambda_{12}\lambda_{11}-3\lambda_{12}\lambda_{12}-\lambda_{12}\lambda_{13}-\lambda_{12}\lambda_{20}+\lambda_{12}\lambda_{21}-\lambda_{12}\lambda_{22}+\lambda_{12}\lambda_{23}\nonumber\\
&&-\lambda_{12}\lambda_{30}+\lambda_{12}\lambda_{31}-\lambda_{12}\lambda_{32}+\lambda_{12}\lambda_{33})(I_1+I_3)+(2\lambda_{00}\lambda_{12}+2\lambda_{02}\lambda_{10}+2\lambda_{20}\lambda_{32}+2\lambda_{22}\lambda_{30}+2\lambda_{11}\lambda_{03}+2\lambda_{13}\lambda_{01}+2\lambda_{21}\lambda_{33}\nonumber\\
&&+2\lambda_{23}\lambda_{31})I_0-(2\lambda_{12}\lambda_{01}+2\lambda_{11}\lambda_{02}+2\lambda_{13}\lambda_{00}+2\lambda_{10}\lambda_{03}+2\lambda_{21}\lambda_{32}+2\lambda_{22}\lambda_{31}+2\lambda_{23}\lambda_{30}+2\lambda_{20}\lambda_{33})I_1-(2\lambda_{10}\lambda_{00}+2\lambda_{11}\lambda_{01}\nonumber\\
&&+2\lambda_{12}\lambda_{02}+2\lambda_{13}\lambda_{03}+2\lambda_{20}\lambda_{30}+2\lambda_{21}\lambda_{31}+2\lambda_{22}\lambda_{32}+2\lambda_{23}\lambda_{33})I_2-(2\lambda_{11}\lambda_{00}+2\lambda_{10}\lambda_{01}+2\lambda_{12}\lambda_{03}+2\lambda_{13}\lambda_{02}+2\lambda_{22}\lambda_{33}\nonumber\\
&&+2\lambda_{23}\lambda_{32}+2\lambda_{21}\lambda_{30}+2\lambda_{20}\lambda_{31})I_3+(2\lambda_{00}\lambda_{12}+2\lambda_{02}\lambda_{10}-2\lambda_{20}\lambda_{32}-2\lambda_{22}\lambda_{30}-2\lambda_{11}\lambda_{03}-2\lambda_{13}\lambda_{01}+2\lambda_{21}\lambda_{33}+2\lambda_{23}\lambda_{31})J_0\nonumber\\
&&+(-2\lambda_{12}\lambda_{01}-2\lambda_{11}\lambda_{02}+2\lambda_{13}\lambda_{00}+2\lambda_{10}\lambda_{03}+2\lambda_{21}\lambda_{32}+2\lambda_{22}\lambda_{31}-2\lambda_{23}\lambda_{30}-2\lambda_{20}\lambda_{33})J_1+(2\lambda_{10}\lambda_{00}+2\lambda_{11}\lambda_{01}+2\lambda_{12}\lambda_{02}\nonumber\\
&&+2\lambda_{13}\lambda_{03}-2\lambda_{20}\lambda_{30}-2\lambda_{21}\lambda_{31}-2\lambda_{22}\lambda_{32}-2\lambda_{23}\lambda_{33})J_2+(2\lambda_{11}\lambda_{00}+2\lambda_{10}\lambda_{01}-2\lambda_{12}\lambda_{03}-2\lambda_{13}\lambda_{02}+2\lambda_{22}\lambda_{33}+2\lambda_{23}\lambda_{32}\nonumber\\
&&-2\lambda_{21}\lambda_{30}-2\lambda_{20}\lambda_{31})J_3],\\
\mathcal{F}_{13}
&=&\frac{1}{16\pi^2}[4(\lambda_{13}\lambda_{00}-\lambda_{13}\lambda_{01}-\lambda_{13}\lambda_{02}+\lambda_{13}\lambda_{03}+\lambda_{13}\lambda_{10}-\lambda_{13}\lambda_{11}-\lambda_{13}\lambda_{12}-3\lambda_{13}\lambda_{13}-\lambda_{13}\lambda_{20}+\lambda_{13}\lambda_{21}+\lambda_{13}\lambda_{22}-\lambda_{13}\lambda_{23}\nonumber\\
&&-\lambda_{13}\lambda_{30}+\lambda_{13}\lambda_{31}+\lambda_{13}\lambda_{32}-\lambda_{13}\lambda_{33})(I_1+I_2)+(2\lambda_{00}\lambda_{13}+2\lambda_{03}\lambda_{10}+2\lambda_{23}\lambda_{30}+2\lambda_{20}\lambda_{33}+2\lambda_{12}\lambda_{01}+2\lambda_{11}\lambda_{02}+2\lambda_{22}\lambda_{31}\nonumber\\
&&+2\lambda_{21}\lambda_{32})I_0-(2\lambda_{13}\lambda_{01}+2\lambda_{11}\lambda_{03}+2\lambda_{12}\lambda_{00}+2\lambda_{10}\lambda_{02}+2\lambda_{21}\lambda_{33}+2\lambda_{23}\lambda_{31}+2\lambda_{22}\lambda_{30}+2\lambda_{20}\lambda_{32})I_1-(2\lambda_{13}\lambda_{02}+2\lambda_{12}\lambda_{03}\nonumber\\
&&+2\lambda_{11}\lambda_{00}+2\lambda_{10}\lambda_{01}+2\lambda_{22}\lambda_{33}+2\lambda_{23}\lambda_{32}+2\lambda_{21}\lambda_{30}+2\lambda_{20}\lambda_{31})I_2-(2\lambda_{10}\lambda_{00}+2\lambda_{11}\lambda_{01}+2\lambda_{12}\lambda_{02}+2\lambda_{13}\lambda_{03}+2\lambda_{20}\lambda_{30}\nonumber\\
&&+2\lambda_{21}\lambda_{31}+2\lambda_{22}\lambda_{32}+2\lambda_{23}\lambda_{33})I_3+(2\lambda_{00}\lambda_{13}+2\lambda_{03}\lambda_{10}-2\lambda_{23}\lambda_{30}-2\lambda_{20}\lambda_{33}-2\lambda_{12}\lambda_{01}-2\lambda_{11}\lambda_{02}+2\lambda_{22}\lambda_{31}+2\lambda_{21}\lambda_{32})J_0\nonumber\\
&&+(-2\lambda_{13}\lambda_{01}-2\lambda_{11}\lambda_{03}+2\lambda_{12}\lambda_{00}+2\lambda_{10}\lambda_{02}+2\lambda_{21}\lambda_{33}+2\lambda_{23}\lambda_{31}-2\lambda_{22}\lambda_{30}-2\lambda_{20}\lambda_{32})J_1+(-2\lambda_{13}\lambda_{02}-2\lambda_{12}\lambda_{03}+2\lambda_{11}\lambda_{00}\nonumber\\
&&+2\lambda_{10}\lambda_{01}+2\lambda_{22}\lambda_{33}+2\lambda_{23}\lambda_{32}-2\lambda_{21}\lambda_{30}-2\lambda_{20}\lambda_{31})J_2+(2\lambda_{10}\lambda_{00}+2\lambda_{11}\lambda_{01}+2\lambda_{12}\lambda_{02}+2\lambda_{13}\lambda_{03}-2\lambda_{20}\lambda_{30}-2\lambda_{21}\lambda_{31}\nonumber\\
&&-2\lambda_{22}\lambda_{32}-2\lambda_{23}\lambda_{33})J_3],\\
\mathcal{F}_{20}
&=&\frac{1}{16\pi^2}[(2\lambda_{00}\lambda_{20}+2\lambda_{01}\lambda_{21}+2\lambda_{02}\lambda_{22}+2\lambda_{03}\lambda_{23}+2\lambda_{10}\lambda_{30}+2\lambda_{11}\lambda_{31}+2\lambda_{12}\lambda_{32}+2\lambda_{13}\lambda_{33})I_0-(2\lambda_{20}\lambda_{01}+2\lambda_{21}\lambda_{00}+2\lambda_{22}\lambda_{03}\nonumber\\
&&+2\lambda_{23}\lambda_{02}+2\lambda_{11}\lambda_{30}+2\lambda_{10}\lambda_{31}+2\lambda_{12}\lambda_{33}+2\lambda_{13}\lambda_{32})I_1-(2\lambda_{20}\lambda_{02}+2\lambda_{22}\lambda_{00}+2\lambda_{21}\lambda_{03}+2\lambda_{23}\lambda_{01}+2\lambda_{12}\lambda_{30}+2\lambda_{10}\lambda_{32}\nonumber\\
&&+2\lambda_{11}\lambda_{33}+2\lambda_{13}\lambda_{31})I_2-(2\lambda_{20}\lambda_{03}+2\lambda_{23}\lambda_{00}+2\lambda_{21}\lambda_{02}+2\lambda_{22}\lambda_{01}+2\lambda_{10}\lambda_{33}+2\lambda_{13}\lambda_{30}+2\lambda_{11}\lambda_{32}+2\lambda_{12}\lambda_{31})I_3+(2\lambda_{00}\lambda_{20}\nonumber\\
&&+2\lambda_{01}\lambda_{21}+2\lambda_{02}\lambda_{22}+2\lambda_{03}\lambda_{23}-2\lambda_{10}\lambda_{30}-2\lambda_{11}\lambda_{31}-2\lambda_{12}\lambda_{32}-2\lambda_{13}\lambda_{33})J_0+(2\lambda_{20}\lambda_{01}+2\lambda_{21}\lambda_{00}-2\lambda_{22}\lambda_{03}-2\lambda_{23}\lambda_{02}\nonumber\\
&&-2\lambda_{11}\lambda_{30}-2\lambda_{10}\lambda_{31}+2\lambda_{12}\lambda_{33}+2\lambda_{13}\lambda_{32})J_1+(2\lambda_{20}\lambda_{02}+2\lambda_{22}\lambda_{00}-2\lambda_{21}\lambda_{03}-2\lambda_{23}\lambda_{01}-2\lambda_{12}\lambda_{30}-2\lambda_{10}\lambda_{32}+2\lambda_{11}\lambda_{33}\nonumber\\
&&+2\lambda_{13}\lambda_{31})J_2+(2\lambda_{20}\lambda_{03}+2\lambda_{23}\lambda_{00}-2\lambda_{21}\lambda_{02}-2\lambda_{22}\lambda_{01}-2\lambda_{10}\lambda_{33}-2\lambda_{13}\lambda_{30}+2\lambda_{11}\lambda_{32}+2\lambda_{12}\lambda_{31})J_3],\\
\mathcal{F}_{21}
&=&\frac{1}{16\pi^2}[4(\lambda_{21}\lambda_{00}+\lambda_{21}\lambda_{01}-\lambda_{21}\lambda_{02}-\lambda_{21}\lambda_{03}-\lambda_{21}\lambda_{10}-\lambda_{21}\lambda_{11}+\lambda_{21}\lambda_{12}+\lambda_{21}\lambda_{13}+\lambda_{21}\lambda_{20}-3\lambda_{21}\lambda_{21}-\lambda_{21}\lambda_{22}-\lambda_{21}\lambda_{23}\nonumber\\
&&-\lambda_{21}\lambda_{30}-\lambda_{21}\lambda_{31}+\lambda_{21}\lambda_{32}+\lambda_{21}\lambda_{33})(I_2+I_3)+(2\lambda_{00}\lambda_{21}+2\lambda_{01}\lambda_{20}+2\lambda_{03}\lambda_{22}+2\lambda_{02}\lambda_{23}+2\lambda_{11}\lambda_{30}+2\lambda_{10}\lambda_{31}+2\lambda_{12}\lambda_{33}\nonumber\\
&&+2\lambda_{13}\lambda_{32})I_0-(2\lambda_{00}\lambda_{20}+2\lambda_{01}\lambda_{21}+2\lambda_{02}\lambda_{22}+2\lambda_{03}\lambda_{23}+2\lambda_{10}\lambda_{30}+2\lambda_{11}\lambda_{31}+2\lambda_{12}\lambda_{32}+2\lambda_{13}\lambda_{33})I_1-(2\lambda_{02}\lambda_{21}+2\lambda_{01}\lambda_{22}\nonumber\\
&&+2\lambda_{00}\lambda_{23}+2\lambda_{03}\lambda_{20}+2\lambda_{11}\lambda_{32}+2\lambda_{12}\lambda_{31}+2\lambda_{13}\lambda_{30}+2\lambda_{10}\lambda_{33})I_2-(2\lambda_{21}\lambda_{03}+2\lambda_{23}\lambda_{01}+2\lambda_{22}\lambda_{00}+2\lambda_{20}\lambda_{02}+2\lambda_{11}\lambda_{33}\nonumber\\
&&+2\lambda_{13}\lambda_{31}+2\lambda_{12}\lambda_{30}+2\lambda_{10}\lambda_{32})I_3+(2\lambda_{00}\lambda_{21}+2\lambda_{01}\lambda_{20}-2\lambda_{03}\lambda_{22}-2\lambda_{02}\lambda_{23}-2\lambda_{11}\lambda_{30}-2\lambda_{10}\lambda_{31}+2\lambda_{12}\lambda_{33}+2\lambda_{13}\lambda_{32})J_0\nonumber\\
&&+(2\lambda_{00}\lambda_{20}+2\lambda_{01}\lambda_{21}+2\lambda_{02}\lambda_{22}+2\lambda_{03}\lambda_{23}-2\lambda_{10}\lambda_{30}-2\lambda_{11}\lambda_{31}-2\lambda_{12}\lambda_{32}-2\lambda_{13}\lambda_{33})J_1+(-2\lambda_{02}\lambda_{21}-2\lambda_{01}\lambda_{22}+2\lambda_{00}\lambda_{23}\nonumber\\
&&+2\lambda_{03}\lambda_{20}+2\lambda_{11}\lambda_{32}+2\lambda_{12}\lambda_{31}-2\lambda_{13}\lambda_{30}-2\lambda_{10}\lambda_{33})J_2+(-2\lambda_{21}\lambda_{03}-2\lambda_{23}\lambda_{01}+2\lambda_{22}\lambda_{00}+2\lambda_{20}\lambda_{02}+2\lambda_{11}\lambda_{33}+2\lambda_{13}\lambda_{31}\nonumber\\
&&-2\lambda_{12}\lambda_{30}-2\lambda_{10}\lambda_{32})J_3],\\
\mathcal{F}_{22}
&=&\frac{1}{16\pi^2}[4(\lambda_{22}\lambda_{00}-\lambda_{22}\lambda_{01}+\lambda_{22}\lambda_{02}-\lambda_{22}\lambda_{03}-\lambda_{22}\lambda_{10}+\lambda_{22}\lambda_{11}-\lambda_{22}\lambda_{12}+\lambda_{22}\lambda_{13}+\lambda_{22}\lambda_{20}-\lambda_{22}\lambda_{21}-3\lambda_{22}\lambda_{22}-\lambda_{22}\lambda_{23}\nonumber\\
&&-\lambda_{22}\lambda_{30}+\lambda_{22}\lambda_{31}-\lambda_{22}\lambda_{32}+\lambda_{22}\lambda_{33})(I_1+I_3)+(2\lambda_{00}\lambda_{22}+2\lambda_{02}\lambda_{20}+2\lambda_{03}\lambda_{21}+2\lambda_{01}\lambda_{23}+2\lambda_{10}\lambda_{32}+2\lambda_{12}\lambda_{30}+2\lambda_{11}\lambda_{33}\nonumber\\
&&+2\lambda_{13}\lambda_{31})I_0-(2\lambda_{01}\lambda_{22}+2\lambda_{02}\lambda_{21}+2\lambda_{00}\lambda_{23}+2\lambda_{03}\lambda_{20}+2\lambda_{12}\lambda_{31}+2\lambda_{11}\lambda_{32}+2\lambda_{13}\lambda_{30}+2\lambda_{10}\lambda_{33})I_1-(2\lambda_{00}\lambda_{20}+2\lambda_{01}\lambda_{21}\nonumber\\
&&+2\lambda_{02}\lambda_{22}+2\lambda_{03}\lambda_{23}+2\lambda_{10}\lambda_{30}+2\lambda_{11}\lambda_{31}+2\lambda_{12}\lambda_{32}+2\lambda_{13}\lambda_{33})I_2-(2\lambda_{03}\lambda_{22}+2\lambda_{02}\lambda_{23}+2\lambda_{00}\lambda_{21}+2\lambda_{01}\lambda_{20}+2\lambda_{12}\lambda_{33}\nonumber\\
&&+2\lambda_{13}\lambda_{32}+2\lambda_{11}\lambda_{30}+2\lambda_{10}\lambda_{31})I_3+(2\lambda_{00}\lambda_{22}+2\lambda_{02}\lambda_{20}-2\lambda_{03}\lambda_{21}-2\lambda_{01}\lambda_{23}-2\lambda_{10}\lambda_{32}-2\lambda_{12}\lambda_{30}+2\lambda_{11}\lambda_{33}+2\lambda_{13}\lambda_{31})J_0\nonumber\\
&&+(-2\lambda_{01}\lambda_{22}-2\lambda_{02}\lambda_{21}+2\lambda_{00}\lambda_{23}+2\lambda_{03}\lambda_{20}+2\lambda_{12}\lambda_{31}+2\lambda_{11}\lambda_{32}-2\lambda_{13}\lambda_{30}-2\lambda_{10}\lambda_{33})J_1+(2\lambda_{00}\lambda_{20}+2\lambda_{01}\lambda_{21}+2\lambda_{02}\lambda_{22}\nonumber\\
&&+2\lambda_{03}\lambda_{23}-2\lambda_{10}\lambda_{30}-2\lambda_{11}\lambda_{31}-2\lambda_{12}\lambda_{32}-2\lambda_{13}\lambda_{33})J_2+(-2\lambda_{03}\lambda_{22}-2\lambda_{02}\lambda_{23}+2\lambda_{00}\lambda_{21}+2\lambda_{01}\lambda_{20}+2\lambda_{12}\lambda_{33}+2\lambda_{13}\lambda_{32}\nonumber\\
&&-2\lambda_{11}\lambda_{30}-2\lambda_{10}\lambda_{31})J_3],\\
\mathcal{F}_{23}
&=&\frac{1}{16\pi^2}[4(\lambda_{23}\lambda_{00}-\lambda_{23}\lambda_{01}-\lambda_{23}\lambda_{02}+\lambda_{23}\lambda_{03}-\lambda_{23}\lambda_{10}+\lambda_{23}\lambda_{11}+\lambda_{23}\lambda_{12}-\lambda_{23}\lambda_{13}+\lambda_{23}\lambda_{20}-\lambda_{23}\lambda_{21}-\lambda_{23}\lambda_{22}-3\lambda_{23}\lambda_{23}\nonumber\\
&&-\lambda_{23}\lambda_{30}+\lambda_{23}\lambda_{31}+\lambda_{23}\lambda_{32}-\lambda_{23}\lambda_{33})(I_1+I_2)+(2\lambda_{00}\lambda_{23}+2\lambda_{03}\lambda_{20}+2\lambda_{02}\lambda_{21}+2\lambda_{01}\lambda_{22}+2\lambda_{10}\lambda_{33}+2\lambda_{13}\lambda_{30}+2\lambda_{11}\lambda_{32}\nonumber\\
&&+2\lambda_{12}\lambda_{31})I_0-(2\lambda_{01}\lambda_{23}+2\lambda_{03}\lambda_{21}+2\lambda_{00}\lambda_{22}+2\lambda_{02}\lambda_{20}+2\lambda_{13}\lambda_{31}+2\lambda_{11}\lambda_{33}+2\lambda_{12}\lambda_{30}+2\lambda_{10}\lambda_{32})I_1-(2\lambda_{02}\lambda_{23}+2\lambda_{03}\lambda_{22}\nonumber\\
&&+2\lambda_{00}\lambda_{21}+2\lambda_{01}\lambda_{20}+2\lambda_{13}\lambda_{32}+2\lambda_{12}\lambda_{33}+2\lambda_{11}\lambda_{30}+2\lambda_{10}\lambda_{31})I_2-(2\lambda_{00}\lambda_{20}+2\lambda_{01}\lambda_{21}+2\lambda_{02}\lambda_{22}+2\lambda_{03}\lambda_{23}+2\lambda_{10}\lambda_{30}\nonumber\\
&&+2\lambda_{11}\lambda_{31}+2\lambda_{12}\lambda_{32}+2\lambda_{13}\lambda_{33})I_3+(2\lambda_{00}\lambda_{23}+2\lambda_{03}\lambda_{20}-2\lambda_{02}\lambda_{21}-2\lambda_{01}\lambda_{22}-2\lambda_{10}\lambda_{33}-2\lambda_{13}\lambda_{30}+2\lambda_{11}\lambda_{32}+2\lambda_{12}\lambda_{31})J_0\nonumber\\
&&+(-2\lambda_{01}\lambda_{23}-2\lambda_{03}\lambda_{21}+2\lambda_{00}\lambda_{22}+2\lambda_{02}\lambda_{20}+2\lambda_{13}\lambda_{31}+2\lambda_{11}\lambda_{33}-2\lambda_{12}\lambda_{30}-2\lambda_{10}\lambda_{32})J_1+(-2\lambda_{02}\lambda_{23}-2\lambda_{03}\lambda_{22}+2\lambda_{00}\lambda_{21}\nonumber\\
&&+2\lambda_{01}\lambda_{20}+2\lambda_{13}\lambda_{32}+2\lambda_{12}\lambda_{33}-2\lambda_{11}\lambda_{30}-2\lambda_{10}\lambda_{31})J_2+(2\lambda_{00}\lambda_{20}+2\lambda_{01}\lambda_{21}+2\lambda_{02}\lambda_{22}+2\lambda_{03}\lambda_{23}-2\lambda_{10}\lambda_{30}-2\lambda_{11}\lambda_{31}\nonumber\\
&&-2\lambda_{12}\lambda_{32}-2\lambda_{13}\lambda_{33})J_3],\\
\mathcal{F}_{30}
&=&\frac{1}{16\pi^2}[(2\lambda_{00}\lambda_{30}+2\lambda_{01}\lambda_{31}+2\lambda_{02}\lambda_{32}+2\lambda_{03}\lambda_{33}+2\lambda_{10}\lambda_{20}+2\lambda_{11}\lambda_{21}+2\lambda_{12}\lambda_{22}+2\lambda_{13}\lambda_{23})I_0-(2\lambda_{30}\lambda_{01}+2\lambda_{31}\lambda_{00}+2\lambda_{32}\lambda_{03}\nonumber\\
&&+2\lambda_{33}\lambda_{02}+2\lambda_{10}\lambda_{21}+2\lambda_{11}\lambda_{20}+2\lambda_{12}\lambda_{23}+2\lambda_{13}\lambda_{22})I_1-(2\lambda_{30}\lambda_{02}+2\lambda_{32}\lambda_{00}+2\lambda_{31}\lambda_{03}+2\lambda_{33}\lambda_{01}+2\lambda_{10}\lambda_{22}+2\lambda_{12}\lambda_{20}\nonumber\\
&&+2\lambda_{11}\lambda_{23}+2\lambda_{13}\lambda_{21})I_2-(2\lambda_{30}\lambda_{03}+2\lambda_{33}\lambda_{00}+2\lambda_{31}\lambda_{02}+2\lambda_{32}\lambda_{01}+2\lambda_{10}\lambda_{23}+2\lambda_{13}\lambda_{20}+2\lambda_{11}\lambda_{22}+2\lambda_{12}\lambda_{21})I_3+(2\lambda_{00}\lambda_{30}\nonumber\\
&&+2\lambda_{01}\lambda_{31}+2\lambda_{02}\lambda_{32}+2\lambda_{03}\lambda_{33}-2\lambda_{10}\lambda_{20}-2\lambda_{11}\lambda_{21}-2\lambda_{12}\lambda_{22}-2\lambda_{13}\lambda_{23})J_0+(2\lambda_{30}\lambda_{01}+2\lambda_{31}\lambda_{00}-2\lambda_{32}\lambda_{03}-2\lambda_{33}\lambda_{02}\nonumber\\
&&-2\lambda_{10}\lambda_{21}-2\lambda_{11}\lambda_{20}+2\lambda_{12}\lambda_{23}+2\lambda_{13}\lambda_{22})J_1+(2\lambda_{30}\lambda_{02}+2\lambda_{32}\lambda_{00}-2\lambda_{31}\lambda_{03}-2\lambda_{33}\lambda_{01}-2\lambda_{10}\lambda_{22}-2\lambda_{12}\lambda_{20}+2\lambda_{11}\lambda_{23}\nonumber\\
&&+2\lambda_{13}\lambda_{21})J_2+(2\lambda_{30}\lambda_{03}+2\lambda_{33}\lambda_{00}-2\lambda_{31}\lambda_{02}-2\lambda_{32}\lambda_{01}-2\lambda_{10}\lambda_{23}-2\lambda_{13}\lambda_{20}+2\lambda_{11}\lambda_{22}+2\lambda_{12}\lambda_{21})J_3],\\
\mathcal{F}_{31}
&=&\frac{1}{16\pi^2}[4(\lambda_{31}\lambda_{00}+\lambda_{31}\lambda_{01}-\lambda_{31}\lambda_{02}-\lambda_{31}\lambda_{03}-\lambda_{31}\lambda_{10}-\lambda_{31}\lambda_{11}+\lambda_{31}\lambda_{12}+\lambda_{31}\lambda_{13}-\lambda_{31}\lambda_{20}-\lambda_{31}\lambda_{21}+\lambda_{31}\lambda_{22}+\lambda_{31}\lambda_{23}\nonumber\\
&&+\lambda_{31}\lambda_{30}-3\lambda_{31}\lambda_{31}-\lambda_{31}\lambda_{32}-\lambda_{31}\lambda_{33})(I_2+I_3)+(2\lambda_{00}\lambda_{31}+2\lambda_{01}\lambda_{30}+2\lambda_{03}\lambda_{32}+2\lambda_{02}\lambda_{33}+2\lambda_{10}\lambda_{21}+2\lambda_{11}\lambda_{20}+2\lambda_{12}\lambda_{23}\nonumber\\
&&+2\lambda_{13}\lambda_{22})I_0-(2\lambda_{00}\lambda_{30}+2\lambda_{01}\lambda_{31}+2\lambda_{02}\lambda_{32}+2\lambda_{03}\lambda_{33}+2\lambda_{10}\lambda_{20}+2\lambda_{11}\lambda_{21}+2\lambda_{12}\lambda_{22}+2\lambda_{13}\lambda_{23})I_1-(2\lambda_{02}\lambda_{31}+2\lambda_{01}\lambda_{32}\nonumber\\
&&+2\lambda_{00}\lambda_{33}+2\lambda_{03}\lambda_{30}+2\lambda_{11}\lambda_{22}+2\lambda_{12}\lambda_{21}+2\lambda_{13}\lambda_{20}+2\lambda_{10}\lambda_{23})I_2-(2\lambda_{03}\lambda_{31}+2\lambda_{01}\lambda_{33}+2\lambda_{00}\lambda_{32}+2\lambda_{02}\lambda_{30}+2\lambda_{11}\lambda_{23}\nonumber\\
&&+2\lambda_{13}\lambda_{21}+2\lambda_{12}\lambda_{20}+2\lambda_{10}\lambda_{22})I_3+(2\lambda_{00}\lambda_{31}+2\lambda_{01}\lambda_{30}-2\lambda_{03}\lambda_{32}-2\lambda_{02}\lambda_{33}-2\lambda_{10}\lambda_{21}-2\lambda_{11}\lambda_{20}+2\lambda_{12}\lambda_{23}+2\lambda_{13}\lambda_{22})J_0\nonumber\\
&&+(2\lambda_{00}\lambda_{30}+2\lambda_{01}\lambda_{31}+2\lambda_{02}\lambda_{32}+2\lambda_{03}\lambda_{33}-2\lambda_{10}\lambda_{20}-2\lambda_{11}\lambda_{21}-2\lambda_{12}\lambda_{22}-2\lambda_{13}\lambda_{23})J_1+(-2\lambda_{02}\lambda_{31}-2\lambda_{01}\lambda_{32}+2\lambda_{00}\lambda_{33}\nonumber\\
&&+2\lambda_{03}\lambda_{30}+2\lambda_{11}\lambda_{22}+2\lambda_{12}\lambda_{21}-2\lambda_{13}\lambda_{20}-2\lambda_{10}\lambda_{23})J_2+(-2\lambda_{03}\lambda_{31}-2\lambda_{01}\lambda_{33}+2\lambda_{00}\lambda_{32}+2\lambda_{02}\lambda_{30}+2\lambda_{11}\lambda_{23}+2\lambda_{13}\lambda_{21}\nonumber\\
&&-2\lambda_{12}\lambda_{20}-2\lambda_{10}\lambda_{22})J_3],\\
\mathcal{F}_{32}
&=&\frac{1}{16\pi^2}[4(\lambda_{32}\lambda_{00}-\lambda_{32}\lambda_{01}+\lambda_{32}\lambda_{02}-\lambda_{32}\lambda_{03}-\lambda_{32}\lambda_{10}+\lambda_{32}\lambda_{11}-\lambda_{32}\lambda_{12}+\lambda_{32}\lambda_{13}-\lambda_{32}\lambda_{20}+\lambda_{32}\lambda_{21}-\lambda_{32}\lambda_{22}+\lambda_{32}\lambda_{23}\nonumber\\
&&+\lambda_{32}\lambda_{30}-\lambda_{32}\lambda_{31}-3\lambda_{32}\lambda_{32}-\lambda_{32}\lambda_{33})(I_1+I_3)+(2\lambda_{00}\lambda_{32}+2\lambda_{02}\lambda_{30}+2\lambda_{03}\lambda_{31}+2\lambda_{01}\lambda_{33}+2\lambda_{10}\lambda_{22}+2\lambda_{12}\lambda_{20}+2\lambda_{11}\lambda_{23}\nonumber\\
&&+2\lambda_{13}\lambda_{21})I_0-(2\lambda_{01}\lambda_{32}+2\lambda_{02}\lambda_{31}+2\lambda_{00}\lambda_{33}+2\lambda_{03}\lambda_{30}+2\lambda_{11}\lambda_{22}+2\lambda_{12}\lambda_{21}+2\lambda_{13}\lambda_{20}+2\lambda_{10}\lambda_{23})I_1-(2\lambda_{00}\lambda_{30}+2\lambda_{01}\lambda_{31}\nonumber\\
&&+2\lambda_{02}\lambda_{32}+2\lambda_{03}\lambda_{33}+2\lambda_{10}\lambda_{20}+2\lambda_{11}\lambda_{21}+2\lambda_{12}\lambda_{22}+2\lambda_{13}\lambda_{23})I_2-(2\lambda_{03}\lambda_{32}+2\lambda_{02}\lambda_{33}+2\lambda_{00}\lambda_{31}+2\lambda_{01}\lambda_{30}+2\lambda_{12}\lambda_{23}\nonumber\\
&&+2\lambda_{13}\lambda_{22}+2\lambda_{11}\lambda_{20}+2\lambda_{10}\lambda_{21})I_3+(2\lambda_{00}\lambda_{32}+2\lambda_{02}\lambda_{30}-2\lambda_{03}\lambda_{31}-2\lambda_{01}\lambda_{33}-2\lambda_{10}\lambda_{22}-2\lambda_{12}\lambda_{20}+2\lambda_{11}\lambda_{23}+2\lambda_{13}\lambda_{21})J_0\nonumber\\
&&+(-2\lambda_{01}\lambda_{32}-2\lambda_{02}\lambda_{31}+2\lambda_{00}\lambda_{33}+2\lambda_{03}\lambda_{30}+2\lambda_{11}\lambda_{22}+2\lambda_{12}\lambda_{21}-2\lambda_{13}\lambda_{20}-2\lambda_{10}\lambda_{23})J_1+(2\lambda_{00}\lambda_{30}+2\lambda_{01}\lambda_{31}+2\lambda_{02}\lambda_{32}\nonumber\\
&&+2\lambda_{03}\lambda_{33}-2\lambda_{10}\lambda_{20}-2\lambda_{11}\lambda_{21}-2\lambda_{12}\lambda_{22}-2\lambda_{13}\lambda_{23})J_2+(-2\lambda_{03}\lambda_{32}-2\lambda_{02}\lambda_{33}+2\lambda_{00}\lambda_{31}+2\lambda_{01}\lambda_{30}+2\lambda_{12}\lambda_{23}+2\lambda_{13}\lambda_{22}\nonumber\\
&&-2\lambda_{11}\lambda_{20}-2\lambda_{10}\lambda_{21})J_3],\\
\mathcal{F}_{33}
&=&\frac{1}{16\pi^2}[4(\lambda_{33}\lambda_{00}-\lambda_{33}\lambda_{01}-\lambda_{33}\lambda_{02}+\lambda_{33}\lambda_{03}-\lambda_{33}\lambda_{10}+\lambda_{33}\lambda_{11}+\lambda_{33}\lambda_{12}-\lambda_{33}\lambda_{13}-\lambda_{33}\lambda_{20}+\lambda_{33}\lambda_{21}+\lambda_{33}\lambda_{22}-\lambda_{33}\lambda_{23}\nonumber\\
&&+\lambda_{33}\lambda_{30}-\lambda_{33}\lambda_{31}-\lambda_{33}\lambda_{32}-3\lambda_{33}\lambda_{33})(I_1+I_2)+(2\lambda_{00}\lambda_{33}+2\lambda_{03}\lambda_{30}+2\lambda_{02}\lambda_{31}+2\lambda_{01}\lambda_{32}+2\lambda_{10}\lambda_{23}+2\lambda_{13}\lambda_{20}+2\lambda_{11}\lambda_{22}\nonumber\\
&&+2\lambda_{12}\lambda_{21})I_0-(2\lambda_{01}\lambda_{33}+2\lambda_{03}\lambda_{31}+2\lambda_{00}\lambda_{32}+2\lambda_{02}\lambda_{30}+2\lambda_{13}\lambda_{21}+2\lambda_{11}\lambda_{23}+2\lambda_{12}\lambda_{20}+2\lambda_{10}\lambda_{22})I_1-(2\lambda_{02}\lambda_{33}+2\lambda_{03}\lambda_{32}\nonumber\\
&&+2\lambda_{00}\lambda_{31}+2\lambda_{01}\lambda_{30}+2\lambda_{13}\lambda_{22}+2\lambda_{12}\lambda_{23}+2\lambda_{11}\lambda_{20}+2\lambda_{10}\lambda_{21})I_2-(2\lambda_{00}\lambda_{30}+2\lambda_{01}\lambda_{31}+2\lambda_{02}\lambda_{32}+2\lambda_{03}\lambda_{33}+2\lambda_{10}\lambda_{20}\nonumber\\
&&+2\lambda_{11}\lambda_{21}+2\lambda_{12}\lambda_{22}+2\lambda_{13}\lambda_{23})I_3+(2\lambda_{00}\lambda_{33}+2\lambda_{03}\lambda_{30}-2\lambda_{02}\lambda_{31}-2\lambda_{01}\lambda_{32}-2\lambda_{10}\lambda_{23}-2\lambda_{13}\lambda_{20}+2\lambda_{11}\lambda_{22}+2\lambda_{12}\lambda_{21})J_0\nonumber\\
&&+(-2\lambda_{01}\lambda_{33}-2\lambda_{03}\lambda_{31}+2\lambda_{00}\lambda_{32}+2\lambda_{02}\lambda_{30}+2\lambda_{13}\lambda_{21}+2\lambda_{11}\lambda_{23}-2\lambda_{12}\lambda_{20}-2\lambda_{10}\lambda_{22})J_1+(-2\lambda_{02}\lambda_{33}-2\lambda_{03}\lambda_{32}+2\lambda_{00}\lambda_{31}\nonumber\\
&&+2\lambda_{01}\lambda_{30}+2\lambda_{13}\lambda_{22}+2\lambda_{12}\lambda_{23}-2\lambda_{11}\lambda_{20}-2\lambda_{10}\lambda_{21})J_2+(2\lambda_{00}\lambda_{30}+2\lambda_{01}\lambda_{31}+2\lambda_{02}\lambda_{32}+2\lambda_{03}\lambda_{33}-2\lambda_{10}\lambda_{20}-2\lambda_{11}\lambda_{21}\nonumber\\
&&-2\lambda_{12}\lambda_{22}-2\lambda_{13}\lambda_{23})J_3],\label{Eq_F33}
\end{eqnarray}
\end{small}
where the related coefficients are designated as
\begin{small}
\begin{eqnarray}
I_0&=&\int_0^{2\pi}\mathrm{d}\theta\frac{1}{\sqrt{d_1^2\cos^2{2\theta}+\frac{1}{4}d_2^2\sin^2{2\theta}+d_3^2}},\hspace{0.8cm}
I_1=\int_0^{2\pi}\mathrm{d}\theta\frac{d_1^2\cos^2{2\theta}}{(d_1^2\cos^2{2\theta}
+\frac{1}{4}d_2^2\sin^2{2\theta}+d_3^2)^{\frac{3}{2}}},\label{Eq_I_0}\\
I_2&=&\int_0^{2\pi}\mathrm{d}\theta\frac{\frac{1}{4}d_2^2\sin^2{2\theta}}{(d_1^2\cos^2{2\theta}+\frac{1}{4}d_2^2\sin^2{2\theta}
+d_3^2)^{\frac{3}{2}}},\hspace{0.7cm}
I_3=\int_0^{2\pi}\mathrm{d}\theta\frac{d_3^2}{(d_1^2\cos^2{2\theta}+\frac{1}{4}d_2^2\sin^2{2\theta}+d_3^2)^{\frac{3}{2}}},\\
J_0&=&\int_0^{2\pi}\mathrm{d}\theta\frac{\sqrt{d_1^2\cos^2{2\theta}+\frac{1}{4}d_2^2\sin^2{2\theta}+d_3^2}}
{d_0^2-d_1^2\cos^2{2\theta}-\frac{1}{4}d_2^2\sin^2{2\theta}-d_3^2},\hspace{0.35cm}
J_1=\int_0^{2\pi}\mathrm{d}\theta\frac{d_1^2\cos^2{2\theta}/
\sqrt{d_1^2\cos^2{2\theta}+\frac{1}{4}d_2^2\sin^2{2\theta}+d_3^2}}
{(d_0^2-d_1^2\cos^2{2\theta}-\frac{1}{4}d_2^2\sin^2{2\theta}-d_3^2)},\\
J_2&=&\int_0^{2\pi}\mathrm{d}\theta\frac{d_2^2\sin^2{2\theta}/\sqrt{d_1^2\cos^2{2\theta}+\frac{1}{4}d_2^2\sin^2{2\theta}+d_3^2}}
{4(d_0^2-d_1^2\cos^2{2\theta}-\frac{1}{4}d_2^2\sin^2{2\theta}-d_3^2)},\hspace{0.1cm}
J_3=\int_0^{2\pi}\mathrm{d}\theta\frac{d_3^2/\sqrt{d_1^2\cos^2{2\theta}+\frac{1}{4}d_2^2\sin^2{2\theta}+d_3^2}}
{(d_0^2-d_1^2\cos^2{2\theta}-\frac{1}{4}d_2^2\sin^2{2\theta}-d_3^2)}.\label{Eq_J_3}
\end{eqnarray}
\end{small}
\end{widetext}

\section{Boundary conditions of FPs and stable subregions of FPs}\label{Sec_Appendix_BC_flows}

For Case II, stable subregions in the $d_1$-$d_2$-$d_3$ space corresponding to $\mathrm{FP}_{1,4,5,6}$ are shown in Fig.~\ref{Fig_FP1-4-II} and Fig.~\ref{Fig_FP5-6-II}. For Case III, stable subregions for $\mathrm{FP}_{1,7,8,9,10}$ appear in Figs.~\ref{Fig_FP1-7-III}-\ref{Fig_FP10-III}. Defining $u \equiv \log{d_1}$, $v \equiv \log{d_2}$, $w \equiv \log{d_3}$, we perform plane-fitting analysis to obtain boundary conditions (BC) for fixed points in Table~\ref{table:fixpoint}. Specifically, $\mathrm{BC}$-$\mathrm{FP}_1$-$\mathrm{II}$ is approximately bounded by seven planes,
\begin{small}
\begin{eqnarray}
0.011975u - 0.005987v - 0.999910w + 4.981889 = 0,& \nonumber \\
0.008666u + 0.002654v - 0.999959w - 0.930985 = 0,& \nonumber \\
0.999919u - 0.005983v - 0.011207w + 4.978154 = 0,& \nonumber \\
0.998376u + 0.054420v + 0.016822w + 5.103586 = 0,& \ \ \label{Eq_condition_3} \\
0.706989u - 0.707124v + 0.011912w + 0.258573 = 0,& \nonumber \\
0.017034u + 0.717703v - 0.696141w + 0.659630 = 0,& \nonumber \\
0.031311u + 0.707179v - 0.706341w + 0.702609 = 0, \nonumber
\end{eqnarray}
\end{small}
$\mathrm{BC}$-$\mathrm{FP}_4$-$\mathrm{case}$ $\mathrm{II}$ is approximately
bounded by 10 planes,
\begin{small}
\begin{eqnarray}
0.741962u - 0.068425v - 0.666941w - 0.149299 = 0,& \nonumber \\
0.749986u - 0.105301v - 0.653018w - 0.060643 = 0,& \nonumber \\
0.029404u - 0.030030v + 0.999116w + 5.026644 = 0,& \nonumber \\
0.013940u + 0.001608v - 0.999902w - 4.989536 = 0,& \nonumber \\
0.950634u + 0.304046v - 0.062066w + 0.311431 = 0,& \ \ \label{Eq_condition_5} \\
0.993333u + 0.114615v + 0.012376w + 0.772497 = 0,& \nonumber \\
0.129089u + 0.991258v + 0.027272w - 5.342738 = 0,& \nonumber \\
0.028733u - 0.999569v + 0.006070w + 4.911828 = 0,& \nonumber \\
0.701747u - 0.712150v - 0.019834w + 1.812948 = 0,& \nonumber \\
0.714783u - 0.698942v - 0.023782w + 0.480929 = 0, \nonumber
\end{eqnarray}
\end{small}
$\mathrm{BC}$-$\mathrm{FP}_5$-$\mathrm{case}$ $\mathrm{II}$
is approximately bounded by 20 planes,
\begin{small}
\begin{eqnarray}
1.000000u - 0.000000v + 0.000000w + 5.000000 = 0,& \nonumber \\
0.999233u + 0.016352v - 0.035582w + 4.924962 = 0,& \nonumber \\
0.018891u + 0.999185v - 0.035669w - 4.841747 = 0,& \nonumber \\
0.007860u + 0.997304v - 0.072963w - 4.823509 = 0,& \nonumber \\
0.013378u - 0.796091v + 0.605029w + 3.841906 = 0,& \nonumber \\
0.008072u + 0.828593v - 0.559794w - 3.907926 = 0,& \nonumber \\
0.010864u + 0.120547v + 0.992648w + 0.541192 = 0,& \nonumber \\
0.004567u + 0.102742v + 0.994698w + 0.582930 = 0,& \nonumber \\
0.010307u + 0.722226v - 0.691581w - 2.383820 = 0,& \nonumber \\
0.007290u - 0.745816v + 0.666112w + 2.512607 = 0,& \ \ \label{Eq_condition_6} \\
0.572951u - 0.819320v + 0.021008w + 3.883559 = 0,& \nonumber \\
0.655750u - 0.754934v - 0.008213w + 3.564708 = 0,& \nonumber \\
0.998836u - 0.030585v + 0.037289w + 1.110053 = 0,& \nonumber \\
0.997625u + 0.068863v - 0.001766w + 0.667084 = 0,& \nonumber \\
0.075465u - 0.996978v - 0.018447w + 4.827712 = 0,& \nonumber \\
0.641838u - 0.766769v - 0.010512w + 2.626501 = 0,& \nonumber \\
0.685591u - 0.726727v + 0.042808w + 2.611468 = 0,& \nonumber \\
0.109621u - 0.026703v + 0.993615w + 5.078327 = 0,& \nonumber \\
0.032285u - 0.034222v - 0.998893w - 4.831522 = 0,& \nonumber \\
0.031474u - 0.999475v - 0.007694w + 4.902638 = 0, \nonumber
\end{eqnarray}
\end{small}
$\mathrm{BC}$-$\mathrm{FP}_6$-$\mathrm{case}$ $\mathrm{II}$
is approximately bounded by 12 planes,
\begin{small}
\begin{eqnarray}
0.995466u + 0.049349v - 0.081320w + 4.649462 = 0,& \nonumber \\
0.009807u - 0.773276v + 0.633994w + 3.770791 = 0,& \nonumber \\
0.038601u - 0.829495v + 0.557179w + 4.061968 = 0,& \nonumber \\
0.001753u - 0.993560v + 0.113293w + 4.950718 = 0,& \nonumber \\
0.036898u + 0.999314v - 0.003014w - 4.815088 = 0,& \nonumber \\
0.037874u + 0.007110v + 0.999257w + 1.150643 = 0,& \ \ \label{Eq_condition_7} \\
0.029923u + 0.044394v + 0.998566w + 0.961467 = 0,& \nonumber \\
0.991067u - 0.130907v + 0.025501w + 1.605737 = 0,& \nonumber \\
0.998209u - 0.058718v + 0.011439w + 1.246236 = 0,& \nonumber \\
0.622078u - 0.782792v + 0.016007w + 3.880454 = 0,& \nonumber \\
0.046721u - 0.998587v - 0.025331w + 4.858400 = 0,& \nonumber \\
0.129661u - 0.007627v - 0.991529w - 4.803195 = 0, \nonumber
\end{eqnarray}
\end{small}
$\mathrm{BC}$-$\mathrm{FP}_1$-$\mathrm{case}$ $\mathrm{III}$
is approximately bounded by eight planes,
\begin{small}
\begin{eqnarray}
0.000070u + 0.005917v - 0.999982w + 4.928352 = 0,& \nonumber \\
0.008169u - 0.001378v - 0.999966w - 0.952217 = 0,& \nonumber \\
0.705163u + 0.000023v - 0.709045w + 0.865328 = 0,& \nonumber \\
0.707716u + 0.010348v - 0.706421w + 0.893102 = 0,& \ \ \label{Eq_condition_8} \\
0.070469u + 0.992911v - 0.095717w + 5.023896 = 0,& \nonumber \\
0.011763u + 0.999891v - 0.008893w + 4.849523 = 0,& \nonumber \\
0.698304u - 0.714804v - 0.037767w + 0.141599 = 0,& \nonumber \\
0.719338u - 0.694413v - 0.018542w + 0.245219 = 0, \nonumber
\end{eqnarray}
\end{small}
$\mathrm{BC}$-$\mathrm{FP}_7$-$\mathrm{case}$ $\mathrm{III}$
is approximately bounded by nine planes,
\begin{small}
\begin{eqnarray}
0.999964u - 0.005298v + 0.006660w - 5.005087 = 0,& \nonumber \\
0.999567u + 0.000004v + 0.029414w - 5.077264 = 0,& \nonumber\\
0.001506u + 0.001165v - 0.999998w - 0.899148 = 0,& \nonumber \\
0.095254u + 0.033716v + 0.994882w + 0.925149 = 0,& \nonumber \\
0.084919u + 0.994810v - 0.056050w + 4.700842 = 0,& \ \ \label{Eq_condition_9} \\
0.058563u - 0.997535v - 0.038654w - 5.007731 = 0,& \nonumber \\
0.190707u + 0.611098v - 0.768238w - 0.345928 = 0,& \nonumber \\
0.723903u - 0.000179v - 0.689902w - 0.548629 = 0,& \nonumber \\
0.706971u - 0.000086v - 0.707242w - 1.625096 = 0, \nonumber
\end{eqnarray}
\end{small}
$\mathrm{BC}$-$\mathrm{FP}_8$-$\mathrm{case}$ $\mathrm{III}$ is approximately
bounded by 11 planes,
\begin{small}
\begin{eqnarray}
0.028661u + 0.752563v - 0.657896w - 0.968019 = 0,& \nonumber \\
0.174571u + 0.599057v - 0.781445w - 0.991093 = 0,& \nonumber \\
0.024605u - 0.024967v + 0.999385w + 4.948749 = 0,& \nonumber \\
0.000700u + 0.008638v - 0.999962w - 4.996042 = 0,& \nonumber \\
0.396250u + 0.918115v - 0.007149w - 0.085531 = 0,& \nonumber \\
0.079535u + 0.980934v - 0.177319w - 0.458429 = 0,& \ \ \label{Eq_condition_10} \\
0.999717u + 0.023803v + 0.000000w - 4.916464 = 0,& \nonumber \\
0.998722u - 0.047940v + 0.016022w - 4.627061 = 0,& \nonumber \\
0.653248u - 0.757071v + 0.010480w + 0.565797 = 0,& \nonumber \\
0.696523u - 0.704314v + 0.137103w + 0.717224 = 0,& \nonumber \\
0.691457u - 0.722418v + 0.000000w - 1.294676 = 0, \nonumber
\end{eqnarray}
\end{small}
$\mathrm{BC}$-$\mathrm{FP}_9$-$\mathrm{case}$ $\mathrm{III}$ is approximately
bounded by 19 planes,
\begin{small}
\begin{eqnarray}
0.000000u - 1.000000v + 0.000000w - 5.000000 = 0,& \nonumber \\
0.054456u + 0.994710v - 0.087100w + 4.768431 = 0,& \nonumber \\
0.999405u + 0.034354v - 0.003008w - 4.723184 = 0,& \nonumber \\
0.989205u - 0.000013v - 0.146537w - 4.575986 = 0,& \nonumber \\
0.787909u - 0.007061v - 0.615751w - 3.459752 = 0,& \nonumber \\
0.808080u + 0.006682v - 0.589034w - 3.508806 = 0,& \nonumber \\
0.110380u - 0.010286v - 0.993836w - 1.165536 = 0,& \nonumber \\
0.106929u - 0.011401v - 0.994201w - 1.158251 = 0,& \nonumber \\
0.734397u - 0.000059v - 0.678720w - 2.342787 = 0,& \nonumber \\
0.743296u - 0.006923v - 0.668926w - 2.390099 = 0,& \ \ \label{Eq_condition_11} \\
0.826746u - 0.562444v + 0.012183w - 3.441840 = 0,& \nonumber \\
0.827091u - 0.561939v + 0.012052w - 3.443800 = 0,& \nonumber \\
0.039617u - 0.998533v - 0.036916w - 0.727425 = 0,& \nonumber \\
0.149619u - 0.986156v - 0.071493w - 1.164556 = 0,& \nonumber \\
0.981427u - 0.188421v + 0.036046w - 4.479992 = 0,& \nonumber \\
0.725693u - 0.687842v + 0.015615w - 2.140314 = 0,& \nonumber \\
0.706809u - 0.707347v - 0.008967w - 2.128821 = 0,& \nonumber \\
0.062813u - 0.002792v + 0.998021w + 4.682116 = 0,& \nonumber \\
0.998641u - 0.049932v + 0.014943w - 4.816128 = 0, \nonumber
\end{eqnarray}
\end{small}
and $\mathrm{BC}$-$\mathrm{FP}_{10}$-$\mathrm{case}$ $\mathrm{III}$ is approximately
bounded by 13 planes,
\begin{small}
\begin{eqnarray}
0.000000u + 1.000000v + 0.000000w + 5.000000 = 0,& \nonumber \\
1.000000u + 0.000000v + 0.000000w - 5.000000 = 0,& \nonumber \\
0.995686u + 0.018664v - 0.090894w - 4.975993 = 0,& \nonumber \\
0.831274u - 0.016989v - 0.555603w - 4.062835 = 0,& \nonumber \\
0.826289u - 0.020423v - 0.562877w - 4.052740 = 0,& \nonumber \\
0.000040u - 0.018805v - 0.999823w - 1.093949 = 0,& \nonumber \\
0.184322u + 0.055767v + 0.981283w + 0.338412 = 0,& \ \ \label{Eq_condition_12} \\
0.061317u - 0.997749v - 0.027132w - 1.081940 = 0,& \nonumber \\
0.005858u - 0.999258v - 0.038077w - 0.864799 = 0,& \nonumber \\
0.997410u - 0.071894v - 0.001955w - 4.943622 = 0,& \nonumber \\
0.874151u - 0.485640v + 0.003716w - 4.012230 = 0,& \nonumber \\
0.858597u - 0.512555v - 0.009953w - 3.983959 = 0,& \nonumber \\
0.151518u - 0.084176v - 0.984864w - 5.622983 = 0. \nonumber
\end{eqnarray}
\end{small}

\begin{figure}[H]
\centering
\subfigure[]{
\includegraphics[width=0.5in]{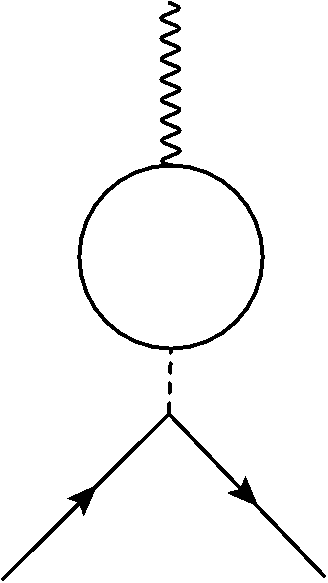}
}
\hspace{0.55cm}
\subfigure[]{
\includegraphics[width=0.5in]{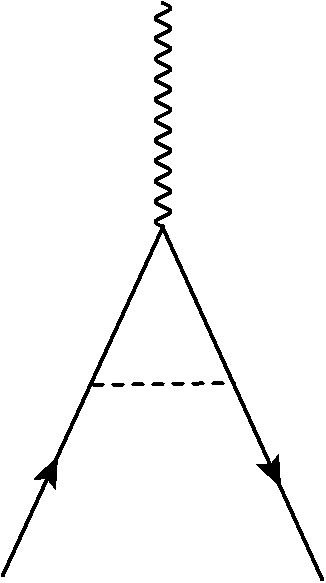}
}
\hspace{0.55cm}
\subfigure[]{
\includegraphics[width=0.5in]{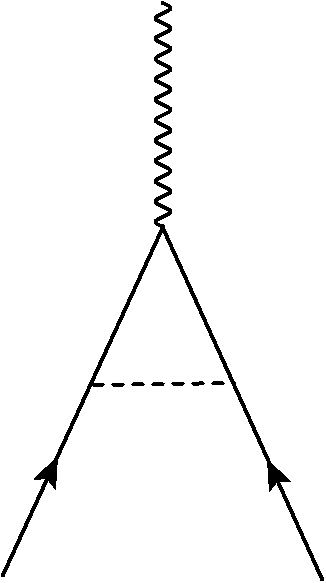}
}
\caption{One-loop Feynman diagrams for the bilinear fermionic
source terms: (a)-(b) the particle-hole channel and (c) the
particle-particle channel. The solid, dash,
and wave lines correspond to the fermionic propagator, fermion-fermion
interaction and source term, respectively.}\label{Fig_source_terms}
\end{figure}

\section{Flow equations of source terms}\label{Appendix_1L-Delta_i}

On the basis of the effective action~(\ref{Eq_S_eff}) and the source terms~(\ref{Eq_source_term}), fermion-fermion interactions generate  one-loop corrections to the sources~\cite{Vafek2012PRB,Vafek2014PRB,Wang2017PRB} as illustrated in Fig.~\ref{Fig_source_terms}. Combining the one-loop calculations in Appendix~\ref{Appendix_1L_corrections} with the RG scalings in Sec.~\ref{Sec_RG_analysis}, we obtain the energy-dependent evolutions
of all source terms,
\begin{widetext}
\begin{small}
\begin{eqnarray}
\frac{d\Delta^c_{1}}{dl}
&=&\Bigl\{2+\frac{1}{32\pi}\Bigl[-7\lambda_{00}(I_0-I_1-I_2-I_3)+\lambda_{01}(I_0-I_1-I_2-I_3)
+\lambda_{02}(I_0-I_1-I_2-I_3)+\lambda_{03}(I_0-I_1-I_2-I_3)\nonumber\\
&&+\lambda_{10}(I_0-I_1-I_2-I_3)+\lambda_{11}(I_0-I_1-I_2-I_3)+\lambda_{12}(I_0-I_1-I_2-I_3)+\lambda_{13}(I_0-I_1-I_2-I_3)\nonumber\\
&&+\lambda_{20}(I_0-I_1-I_2-I_3)+\lambda_{21}(I_0-I_1-I_2-I_3)+\lambda_{22}(I_0-I_1-I_2-I_3)+\lambda_{23}(I_0-I_1-I_2-I_3)\nonumber\\
&&+\lambda_{30}(I_0-I_1-I_2-I_3)+\lambda_{31}(I_0-I_1-I_2-I_3)+\lambda_{32}(I_0-I_1-I_2-I_3)
+\lambda_{33}(I_0-I_1-I_2-I_3)\Bigr]\Bigr\}\Delta^c_{1},\label{Eq_source_1}\\
\frac{d\Delta^{c}_{2}}{dl}
&=&\Bigl\{2+\frac{1}{32\pi}\Bigl[\lambda_{00}(I_0-I_1+I_2+I_3)-7\lambda_{01}(I_0-I_1+I_2+I_3)
-\lambda_{02}(I_0-I_1+I_2+I_3)-\lambda_{03}(I_0-I_1+I_2+I_3)\nonumber\\
&&+\lambda_{10}(I_0-I_1+I_2+I_3)+\lambda_{11}(I_0-I_1+I_2+I_3)-\lambda_{12}(I_0-I_1+I_2+I_3)
-\lambda_{13}(I_0-I_1+I_2+I_3)\nonumber\\
&&+\lambda_{20}(I_0-I_1+I_2+I_3)+\lambda_{21}(I_0-I_1+I_2+I_3)-\lambda_{22}(I_0-I_1+I_2+I_3)-\lambda_{23}(I_0-I_1+I_2+I_3)\nonumber\\
&&+\lambda_{30}(I_0-I_1+I_2+I_3)+\lambda_{31}(I_0-I_1+I_2+I_3)-\lambda_{32}(I_0-I_1+I_2+I_3)
-\lambda_{33}(I_0-I_1+I_2+I_3)\Bigr]\Bigr\}\Delta^{c}_{2},\\
\frac{d\Delta_{3}^{c}}{dl}
&=&\Bigl\{2+\frac{1}{32\pi}\Bigl[\lambda_{00}(I_0+I_1-I_2+I_3)-\lambda_{01}(I_0+I_1-I_2+I_3)-7\lambda_{02}(I_0+I_1-I_2+I_3)
-\lambda_{03}(I_0+I_1-I_2+I_3)\nonumber\\
&&+\lambda_{10}(I_0+I_1-I_2+I_3)-\lambda_{11}(I_0+I_1-I_2+I_3)
+\lambda_{12}(I_0+I_1-I_2+I_3)-\lambda_{13}(I_0+I_1-I_2+I_3)\nonumber\\
&&+\lambda_{20}(I_0+I_1-I_2+I_3)-\lambda_{21}(I_0+I_1-I_2+I_3)+\lambda_{22}(I_0+I_1-I_2+I_3)-\lambda_{23}(I_0+I_1-I_2+I_3)\nonumber\\
&&+\lambda_{30}(I_0+I_1-I_2+I_3)-\lambda_{31}(I_0+I_1-I_2+I_3)+\lambda_{32}(I_0+I_1-I_2+I_3)
-\lambda_{33}(I_0+I_1-I_2+I_3)\Bigr]\Bigr\}\Delta_{3}^{c},\\
\frac{d\Delta_{4}^{c}}{dl}
&=&\Bigl\{2+\frac{1}{32\pi}\Bigl[\lambda_{00}(I_0+I_1+I_2-I_3)-\lambda_{01}(I_0+I_1+I_2-I_3)-\lambda_{02}(I_0+I_1+I_2-I_3)
-7\lambda_{03}(I_0+I_1+I_2-I_3)\nonumber\\
&&+\lambda_{10}(I_0+I_1+I_2-I_3)-\lambda_{11}(I_0+I_1+I_2-I_3)-\lambda_{12}(I_0+I_1+I_2-I_3)+\lambda_{13}(I_0+I_1+I_2-I_3)\nonumber\\
&&+\lambda_{20}(I_0+I_1+I_2-I_3)-\lambda_{21}(I_0+I_1+I_2-I_3)-\lambda_{22}(I_0+I_1+I_2-I_3)+\lambda_{23}(I_0+I_1+I_2-I_3)\nonumber\\
&&+\lambda_{30}(I_0+I_1+I_2-I_3)-\lambda_{31}(I_0+I_1+I_2-I_3)-\lambda_{32}(I_0+I_1+I_2-I_3)
+\lambda_{33}(I_0+I_1+I_2-I_3)\Bigr]\Bigr\}\Delta_{4}^{c},\\
\frac{d\Delta_{1x}^{s}}{dl}
&=&\Big\{2+\frac{1}{32\pi}\Bigl[\lambda_{00}(I_0-I_1-I_2-I_3)+\lambda_{01}(I_0-I_1-I_2-I_3)+\lambda_{02}(I_0-I_1-I_2-I_3)
+\lambda_{03}(I_0-I_1-I_2-I_3)\nonumber\\
&&-7\lambda_{10}(I_0-I_1-I_2-I_3)+\lambda_{11}(I_0-I_1-I_2-I_3)+\lambda_{12}(I_0-I_1-I_2-I_3)+\lambda_{13}(I_0-I_1-I_2-I_3)\nonumber\\
&&-\lambda_{20}(I_0-I_1-I_2-I_3)-\lambda_{21}(I_0-I_1-I_2-I_3)-\lambda_{22}(I_0-I_1-I_2-I_3)-\lambda_{23}(I_0-I_1-I_2-I_3)\nonumber\\
&&-\lambda_{30}(I_0-I_1-I_2-I_3)-\lambda_{31}(I_0-I_1-I_2-I_3)-\lambda_{32}(I_0-I_1-I_2-I_3)
-\lambda_{33}(I_0-I_1-I_2-I_3)\Bigr]\Bigr\}\Delta_{1x}^{s},\\
\frac{d\Delta_{1y}^{s}}{dl}
&=&\Big\{2+\frac{1}{32\pi}\Bigl[\lambda_{00}(I_0-I_1-I_2-I_3)+\lambda_{01}(I_0-I_1-I_2-I_3)+\lambda_{02}(I_0-I_1-I_2-I_3)
+\lambda_{03}(I_0-I_1-I_2-I_3)\nonumber\\
&&-\lambda_{10}(I_0-I_1-I_2-I_3)-\lambda_{11}(I_0-I_1-I_2-I_3)-\lambda_{12}(I_0-I_1-I_2-I_3)-\lambda_{13}(I_0-I_1-I_2-I_3)\nonumber\\
&&-7\lambda_{20}(I_0-I_1-I_2-I_3)+\lambda_{21}(I_0-I_1-I_2-I_3)+\lambda_{22}(I_0-I_1-I_2-I_3)+\lambda_{23}(I_0-I_1-I_2-I_3)\nonumber\\
&&-\lambda_{30}(I_0-I_1-I_2-I_3)-\lambda_{31}(I_0-I_1-I_2-I_3)-\lambda_{32}(I_0-I_1-I_2-I_3)
-\lambda_{33}(I_0-I_1-I_2-I_3)\Bigr]\Bigr\}\Delta_{1y}^{s},\\
\frac{d\Delta_{1z}^{s}}{dl}
&=&\Big\{2+\frac{1}{32\pi}\Bigl[\lambda_{00}(I_0-I_1-I_2-I_3)+\lambda_{01}(I_0-I_1-I_2-I_3)+\lambda_{02}(I_0-I_1-I_2-I_3)
+\lambda_{03}(I_0-I_1-I_2-I_3)\nonumber\\
&&-\lambda_{10}(I_0-I_1-I_2-I_3)-\lambda_{11}(I_0-I_1-I_2-I_3)-\lambda_{12}(I_0-I_1-I_2-I_3)-\lambda_{13}(I_0-I_1-I_2-I_3)\nonumber\\
&&-\lambda_{20}(I_0-I_1-I_2-I_3)-\lambda_{21}(I_0-I_1-I_2-I_3)-\lambda_{22}(I_0-I_1-I_2-I_3)-\lambda_{23}(I_0-I_1-I_2-I_3)\nonumber\\
&&-7\lambda_{30}(I_0-I_1-I_2-I_3)+\lambda_{31}(I_0-I_1-I_2-I_3)+\lambda_{32}(I_0-I_1-I_2-I_3)
+\lambda_{33}(I_0-I_1-I_2-I_3)\Bigr]\Bigr\}\Delta_{1z}^{s},\\
\frac{d\Delta_{2x}^{s}}{dl}
&=&\Big\{2+\frac{1}{32\pi}\Bigl[\lambda_{00}(I_0-I_1+I_2+I_3)+\lambda_{01}(I_0-I_1+I_2+I_3)-\lambda_{02}(I_0-I_1+I_2+I_3)
-\lambda_{03}(I_0-I_1+I_2+I_3)\nonumber\\
&&+\lambda_{10}(I_0-I_1+I_2+I_3)-7\lambda_{11}(I_0-I_1+I_2+I_3)-\lambda_{12}(I_0-I_1+I_2+I_3)-\lambda_{13}(I_0-I_1+I_2+I_3)\nonumber\\
&&-\lambda_{20}(I_0-I_1+I_2+I_3)-\lambda_{21}(I_0-I_1+I_2+I_3)+\lambda_{22}(I_0-I_1+I_2+I_3)+\lambda_{23}(I_0-I_1+I_2+I_3)\nonumber\\
&&-\lambda_{30}(I_0-I_1+I_2+I_3)-\lambda_{31}(I_0-I_1+I_2+I_3)+\lambda_{32}(I_0-I_1+I_2+I_3)
+\lambda_{33}(I_0-I_1+I_2+I_3)\Bigr]\Bigr\}\Delta_{2x}^{s},\\
\frac{d\Delta_{2y}^{s}}{dl}
&=&\Big\{2+\frac{1}{32\pi}\Bigl[\lambda_{00}(I_0-I_1+I_2+I_3)+\lambda_{01}(I_0-I_1+I_2+I_3)-\lambda_{02}(I_0-I_1+I_2+I_3)
-\lambda_{03}(I_0-I_1+I_2+I_3)\nonumber\\
&&-\lambda_{10}(I_0-I_1+I_2+I_3)-\lambda_{11}(I_0-I_1+I_2+I_3)+\lambda_{12}(I_0-I_1+I_2+I_3)+\lambda_{13}(I_0-I_1+I_2+I_3)\nonumber\\
&&+\lambda_{20}(I_0-I_1+I_2+I_3)-7\lambda_{21}(I_0-I_1+I_2+I_3)-\lambda_{22}(I_0-I_1+I_2+I_3)-\lambda_{23}(I_0-I_1+I_2+I_3)\nonumber\\
&&-\lambda_{30}(I_0-I_1+I_2+I_3)-\lambda_{31}(I_0-I_1+I_2+I_3)+\lambda_{32}(I_0-I_1+I_2+I_3)
+\lambda_{33}(I_0-I_1+I_2+I_3)\Bigr]\Bigr\}\Delta_{2y}^{s},\\
\frac{d\Delta_{2z}^{s}}{dl}
&=&\Big\{2+\frac{1}{32\pi}\Bigl[\lambda_{00}(I_0-I_1+I_2+I_3)+\lambda_{01}(I_0-I_1+I_2+I_3)-\lambda_{02}(I_0-I_1+I_2+I_3)
-\lambda_{03}(I_0-I_1+I_2+I_3)\nonumber\\
&&-\lambda_{10}(I_0-I_1+I_2+I_3)-\lambda_{11}(I_0-I_1+I_2+I_3)+\lambda_{12}(I_0-I_1+I_2+I_3)+\lambda_{13}(I_0-I_1+I_2+I_3)\nonumber\\
&&-\lambda_{20}(I_0-I_1+I_2+I_3)-\lambda_{21}(I_0-I_1+I_2+I_3)+\lambda_{22}(I_0-I_1+I_2+I_3)+\lambda_{23}(I_0-I_1+I_2+I_3)\nonumber\\
&&+\lambda_{30}(I_0-I_1+I_2+I_3)-7\lambda_{31}(I_0-I_1+I_2+I_3)-\lambda_{32}(I_0-I_1+I_2+I_3)
-\lambda_{33}(I_0-I_1+I_2+I_3)\Bigr]\Bigr\}\Delta_{2z}^{s},\\
\frac{d\Delta_{3x}^{s}}{dl}
&=&\Big\{2+\frac{1}{32\pi}\Bigl[\lambda_{00}(I_0+I_1-I_2+I_3)-\lambda_{01}(I_0+I_1-I_2+I_3)+\lambda_{02}(I_0+I_1-I_2+I_3)
-\lambda_{03}(I_0+I_1-I_2+I_3)\nonumber\\
&&+\lambda_{10}(I_0+I_1-I_2+I_3)-\lambda_{11}(I_0+I_1-I_2+I_3)-7\lambda_{12}(I_0+I_1-I_2+I_3)-\lambda_{13}(I_0+I_1-I_2+I_3)\nonumber\\
&&-\lambda_{20}(I_0+I_1-I_2+I_3)+\lambda_{21}(I_0+I_1-I_2+I_3)-\lambda_{22}(I_0+I_1-I_2+I_3)+\lambda_{23}(I_0+I_1-I_2+I_3)\nonumber\\
&&-\lambda_{30}(I_0+I_1-I_2+I_3)+\lambda_{31}(I_0+I_1-I_2+I_3)-\lambda_{32}(I_0+I_1-I_2+I_3)
+\lambda_{33}(I_0+I_1-I_2+I_3)\Bigr]\Bigr\}\Delta_{3x}^{s},\\
\frac{d\Delta_{3y}^{s}}{dl}
&=&\Big\{2+\frac{1}{32\pi}\Bigl[\lambda_{00}(I_0+I_1-I_2+I_3)-\lambda_{01}(I_0+I_1-I_2+I_3)+\lambda_{02}(I_0+I_1-I_2+I_3)
-\lambda_{03}(I_0+I_1-I_2+I_3)\nonumber\\
&&-\lambda_{10}(I_0+I_1-I_2+I_3)+\lambda_{11}(I_0+I_1-I_2+I_3)-\lambda_{12}(I_0+I_1-I_2+I_3)+\lambda_{13}(I_0+I_1-I_2+I_3)\nonumber\\
&&+\lambda_{20}(I_0+I_1-I_2+I_3)-\lambda_{21}(I_0+I_1-I_2+I_3)-7\lambda_{22}(I_0+I_1-I_2+I_3)-\lambda_{23}(I_0+I_1-I_2+I_3)\nonumber\\
&&-\lambda_{30}(I_0+I_1-I_2+I_3)+\lambda_{31}(I_0+I_1-I_2+I_3)-\lambda_{32}(I_0+I_1-I_2+I_3)
+\lambda_{33}(I_0+I_1-I_2+I_3)\Bigr]\Bigr\}\Delta_{3y}^{s},\\
\frac{d\Delta_{3z}^{s}}{dl}
&=&\Big\{2+\frac{1}{32\pi}\Bigl[\lambda_{00}(I_0+I_1-I_2+I_3)-\lambda_{01}(I_0+I_1-I_2+I_3)+\lambda_{02}(I_0+I_1-I_2+I_3)
-\lambda_{03}(I_0+I_1-I_2+I_3)\nonumber\\
&&-\lambda_{10}(I_0+I_1-I_2+I_3)+\lambda_{11}(I_0+I_1-I_2+I_3)-\lambda_{12}(I_0+I_1-I_2+I_3)+\lambda_{13}(I_0+I_1-I_2+I_3)\nonumber\\
&&-\lambda_{20}(I_0+I_1-I_2+I_3)+\lambda_{21}(I_0+I_1-I_2+I_3)-\lambda_{22}(I_0+I_1-I_2+I_3)+\lambda_{23}(I_0+I_1-I_2+I_3)\nonumber\\
&&+\lambda_{30}(I_0+I_1-I_2+I_3)-\lambda_{31}(I_0+I_1-I_2+I_3)-7\lambda_{32}(I_0+I_1-I_2+I_3)
-\lambda_{33}(I_0+I_1-I_2+I_3)\Bigr]\Bigr\}\Delta_{3z}^{s},\\
\frac{d\Delta_{4x}^{s}}{dl}
&=&\Big\{2+\frac{1}{32\pi}\Bigl[\lambda_{00}(I_0+I_1+I_2-I_3)-\lambda_{01}(I_0+I_1+I_2-I_3)-\lambda_{02}(I_0+I_1+I_2-I_3)
+\lambda_{03}(I_0+I_1+I_2-I_3)\nonumber\\
&&+\lambda_{10}(I_0+I_1+I_2-I_3)-\lambda_{11}(I_0+I_1+I_2-I_3)-\lambda_{12}(I_0+I_1+I_2-I_3)-7\lambda_{13}(I_0+I_1+I_2-I_3)\nonumber\\
&&-\lambda_{20}(I_0+I_1+I_2-I_3)+\lambda_{21}(I_0+I_1+I_2-I_3)+\lambda_{22}(I_0+I_1+I_2-I_3)-\lambda_{23}(I_0+I_1+I_2-I_3)\nonumber\\
&&-\lambda_{30}(I_0+I_1+I_2-I_3)+\lambda_{31}(I_0+I_1+I_2-I_3)+\lambda_{32}(I_0+I_1+I_2-I_3)
-\lambda_{33}(I_0+I_1+I_2-I_3)\Bigr]\Bigr\}\Delta_{4x}^{s},\\
\frac{d\Delta_{4y}^{s}}{dl}
&=&\Big\{2+\frac{1}{32\pi}\Bigl[\lambda_{00}(I_0+I_1+I_2-I_3)-\lambda_{01}(I_0+I_1+I_2-I_3)-\lambda_{02}(I_0+I_1+I_2-I_3)
+\lambda_{03}(I_0+I_1+I_2-I_3)\nonumber\\
&&-\lambda_{10}(I_0+I_1+I_2-I_3)+\lambda_{11}(I_0+I_1+I_2-I_3)+\lambda_{12}(I_0+I_1+I_2-I_3)-\lambda_{13}(I_0+I_1+I_2-I_3)\nonumber\\
&&+\lambda_{20}(I_0+I_1+I_2-I_3)-\lambda_{21}(I_0+I_1+I_2-I_3)-\lambda_{22}(I_0+I_1+I_2-I_3)-7\lambda_{23}(I_0+I_1+I_2-I_3)\nonumber\\
&&-\lambda_{30}(I_0+I_1+I_2-I_3)+\lambda_{31}(I_0+I_1+I_2-I_3)+\lambda_{32}(I_0+I_1+I_2-I_3)
-\lambda_{33}(I_0+I_1+I_2-I_3)\Bigr]\Bigr\}\Delta_{4y}^{s},\\
\frac{d\Delta_{4z}^{s}}{dl}
&=&\Big\{2+\frac{1}{32\pi}\Bigl[\lambda_{00}(I_0+I_1+I_2-I_3)-\lambda_{01}(I_0+I_1+I_2-I_3)-\lambda_{02}(I_0+I_1+I_2-I_3)
+\lambda_{03}(I_0+I_1+I_2-I_3)\nonumber\\
&&-\lambda_{10}(I_0+I_1+I_2-I_3)+\lambda_{11}(I_0+I_1+I_2-I_3)+\lambda_{12}(I_0+I_1+I_2-I_3)-\lambda_{13}(I_0+I_1+I_2-I_3)\nonumber\\
&&-\lambda_{20}(I_0+I_1+I_2-I_3)+\lambda_{21}(I_0+I_1+I_2-I_3)+\lambda_{22}(I_0+I_1+I_2-I_3)-\lambda_{23}(I_0+I_1+I_2-I_3)\nonumber\\
&&+\lambda_{30}(I_0+I_1+I_2-I_3)-\lambda_{31}(I_0+I_1+I_2-I_3)-\lambda_{32}(I_0+I_1+I_2-I_3)
-7\lambda_{33}(I_0+I_1+I_2-I_3)\Bigr]\Bigr\}\Delta_{4z}^{s},\\
\frac{d\Delta_{1}^{\mathrm{pp}}}{dl}
&=&\Bigl\{2+\frac{1}{32\pi}\Bigl[\lambda_{00}(J_0-J_1-J_2+J_3)-\lambda_{01}(J_0-J_1-J_2+J_3)+\lambda_{02}(J_0-J_1-J_2+J_3)
+\lambda_{03}(J_0-J_1-J_2+J_3)\nonumber\\
&&-\lambda_{10}(J_0-J_1-J_2+J_3)+\lambda_{11}(J_0-J_1-J_2+J_3)-\lambda_{12}(J_0-J_1-J_2+J_3)-\lambda_{13}(J_0-J_1-J_2+J_3)\nonumber\\
&&-\lambda_{20}(J_0-J_1-J_2+J_3)+\lambda_{21}(J_0-J_1-J_2+J_3)-\lambda_{22}(J_0-J_1-J_2+J_3)-\lambda_{23}(J_0-J_1-J_2+J_3)\nonumber\\
&&-\lambda_{30}(J_0-J_1-J_2+J_3)+\lambda_{31}(J_0-J_1-J_2+J_3)-\lambda_{32}(J_0-J_1-J_2+J_3)
-\lambda_{33}(J_0-J_1-J_2+J_3)\!\Bigr]\!\Bigr\}\Delta_{1}^{\mathrm{pp}},\\
\frac{d\Delta_{2}^{\mathrm{pp}}}{dl}
&=&\Bigl\{2+\frac{1}{32\pi}\Bigl[\lambda_{00}(J_0+J_1-J_2-J_3)+\lambda_{01}(J_0+J_1-J_2-J_3)+\lambda_{02}(J_0+J_1-J_2-J_3)
-\lambda_{03}(J_0+J_1-J_2-J_3)\nonumber\\
&&-\lambda_{10}(J_0+J_1-J_2-J_3)-\lambda_{11}(J_0+J_1-J_2-J_3)-\lambda_{12}(J_0+J_1-J_2-J_3)+\lambda_{13}(J_0+J_1-J_2-J_3)\nonumber\\
&&-\lambda_{20}(J_0+J_1-J_2-J_3)-\lambda_{21}(J_0+J_1-J_2-J_3)-\lambda_{22}(J_0+J_1-J_2-J_3)+\lambda_{23}(J_0+J_1-J_2-J_3)\nonumber\\
&&-\lambda_{30}(J_0+J_1-J_2-J_3)-\lambda_{31}(J_0+J_1-J_2-J_3)-\lambda_{32}(J_0+J_1-J_2-J_3)
+\lambda_{33}(J_0+J_1-J_2-J_3)\!\Bigr]\!\Bigr\}\Delta_{2}^{\mathrm{pp}},\\
\frac{d\Delta_{3}^{\mathrm{pp}}}{dl}
&=&\Bigl\{2+\frac{1}{32\pi}\Bigl[\lambda_{00}(J_0+J_1+J_2+J_3)+\lambda_{01}(J_0+J_1+J_2+J_3)-\lambda_{02}(J_0+J_1+J_2+J_3)
+\lambda_{03}(J_0+J_1+J_2+J_3)\nonumber\\
&&-\lambda_{10}(J_0+J_1+J_2+J_3)-\lambda_{11}(J_0+J_1+J_2+J_3)+\lambda_{12}(J_0+J_1+J_2+J_3)-\lambda_{13}(J_0+J_1+J_2+J_3)\nonumber\\
&&-\lambda_{20}(J_0+J_1+J_2+J_3)-\lambda_{21}(J_0+J_1+J_2+J_3)+\lambda_{22}(J_0+J_1+J_2+J_3)-\lambda_{23}(J_0+J_1+J_2+J_3)\nonumber\\
&&-\lambda_{30}(J_0+J_1+J_2+J_3)-\lambda_{31}(J_0+J_1+J_2+J_3)+\lambda_{32}(J_0+J_1+J_2+J_3)
-\lambda_{33}(J_0+J_1+J_2+J_3)\!\Bigr]\!\Bigr\}\Delta_{3}^{\mathrm{pp}},\\
\frac{d\Delta_{4x}^{\mathrm{pp}}}{dl}
&=&\Bigl\{2+\frac{1}{32\pi}\Bigl[(\lambda_{00}(J_0-J_1+J_2-J_3)-\lambda_{01}(J_0-J_1+J_2-J_3)-\lambda_{02}(J_0-J_1+J_2-J_3)
-\lambda_{03}(J_0-J_1+J_2-J_3)\nonumber\\
&&+\lambda_{10}(J_0-J_1+J_2-J_3)-\lambda_{11}(J_0-J_1+J_2-J_3)-\lambda_{12}(J_0-J_1+J_2-J_3)-\lambda_{13}(J_0-J_1+J_2-J_3)\nonumber\\
&&-\lambda_{20}(J_0-J_1+J_2-J_3)+\lambda_{21}(J_0-J_1+J_2-J_3)+\lambda_{22}(J_0-J_1+J_2-J_3)+\lambda_{23}(J_0-J_1+J_2-J_3)\nonumber\\
&&+\lambda_{30}(J_0-J_1+J_2-J_3)-\lambda_{31}(J_0-J_1+J_2-J_3)-\lambda_{32}(J_0-J_1+J_2-J_3)
-\lambda_{33}(J_0-J_1+J_2-J_3)\!\Bigr]\!\Bigr\}\Delta_{4x}^{\mathrm{pp}},\\
\frac{d\Delta_{4y}^{\mathrm{pp}}}{dl}
&=&\Bigl\{2+\frac{1}{32\pi}\Bigl[\lambda_{00}(J_0-J_1+J_2-J_3)-\lambda_{01}(J_0-J_1+J_2-J_3)-\lambda_{02}(J_0-J_1+J_2-J_3)
-\lambda_{03}(J_0-J_1+J_2-J_3)\nonumber\\
&&+\lambda_{10}(J_0-J_1+J_2-J_3)-\lambda_{11}(J_0-J_1+J_2-J_3)-\lambda_{12}(J_0-J_1+J_2-J_3)-\lambda_{13}(J_0-J_1+J_2-J_3)\nonumber\\
&&+\lambda_{20}(J_0-J_1+J_2-J_3)-\lambda_{21}(J_0-J_1+J_2-J_3)-\lambda_{22}(J_0-J_1+J_2-J_3)-\lambda_{23}(J_0-J_1+J_2-J_3)\nonumber\\
&&-\lambda_{30}(J_0-J_1+J_2-J_3)+\lambda_{31}(J_0-J_1+J_2-J_3)+\lambda_{32}(J_0-J_1+J_2-J_3)
+\lambda_{33}(J_0-J_1+J_2-J_3)\!\Bigr]\!\Bigr\}\Delta_{4y}^{\mathrm{pp}},\\
\frac{d\Delta_{4z}^{\mathrm{pp}}}{dl}
&=&\Bigl\{2+\frac{1}{32\pi}\Bigl[\lambda_{00}(J_0-J_1+J_2-J_3)-\lambda_{01}(J_0-J_1+J_2-J_3)-\lambda_{02}(J_0-J_1+J_2-J_3)
-\lambda_{03}(J_0-J_1+J_2-J_3)\nonumber\\
&&-\lambda_{10}(J_0-J_1+J_2-J_3)+\lambda_{11}(J_0-J_1+J_2-J_3)+\lambda_{12}(J_0-J_1+J_2-J_3)+\lambda_{13}(J_0-J_1+J_2-J_3)\nonumber\\
&&+\lambda_{20}(J_0-J_1+J_2-J_3)-\lambda_{21}(J_0-J_1+J_2-J_3)-\lambda_{22}(J_0-J_1+J_2-J_3)-\lambda_{23}(J_0-J_1+J_2-J_3)\nonumber\\
&&+\lambda_{30}(J_0-J_1+J_2-J_3)-\lambda_{31}(J_0-J_1+J_2-J_3)-\lambda_{32}(J_0-J_1+J_2-J_3)
-\lambda_{33}(J_0-J_1+J_2-J_3)\!\Bigr]\!\Bigr\}\Delta_{4z}^{\mathrm{pp}},\label{Eq_source_2}
\end{eqnarray}
\end{small}
where the related coefficients are defined in Eqs.~(\ref{Eq_I_0})-(\ref{Eq_J_3}) and the subscripts $c$ and $s$
label charge, spin, and particle-particle channels, respectively.
\end{widetext}

%%%%%%%%%%%%%%%%%%%%%%%%%%%%%%%%%%%%%%%%%%%%%%%%%%%%%%%%%%%%%%%%%%%%%%%%%%%%%%%%%%%%%%%%%
%%%%%%%%%%%%%%%%%%%%%%%%%%%%%%%%%%%%%%%%%%%%%%%%%%%%%%%%%%%%%%%%%%%%%%%%%%%%%%%%%%%%%%%%%

\end{document}